\documentclass[onecolumn]{emulateapj}
\usepackage{amsmath}
\usepackage{rotating}

\slugcomment{January 6th, 2012}

\def\bfnabla{{\mbox{\boldmath $\nabla$}}}

\def\msun{M_\odot}

\renewcommand\bv{{\mbox{\boldmath $v$}}}
\newcommand\bb{{\mbox{\boldmath $B$}}}
\newcommand\bP{{\mbox{\boldmath $P$}}}
\newcommand\bF{{\mbox{\boldmath $F$}}}
\newcommand\bfr{{\sf\boldmath f}}
\newcommand\bI{{{\sf\boldmath I}}}
\newcommand\bW{{\mbox{\boldmath $W$}}}
\newcommand\bU{{\mbox{\boldmath $U$}}}

\newcommand\Crat{{\mathbb{C}}}
\newcommand\Prat{{\mathbb{P}}}
\def\<{\,\langle\langle}
\def\>{\,\rangle\rangle}

\begin{document}

\shortauthors{Y.-F. Jiang et al.}
\author{Yan-Fei Jiang\altaffilmark{1}, James M. Stone\altaffilmark{1} \& Shane W. Davis\altaffilmark{2}}
\affil{$^1$Department of Astrophysical Sciences, Princeton
University, Princeton, NJ 08544, USA} 
\affil{$^2$Canadian Institute for Theoretical Astrophysics. Toronto, ON M5S3H4, Canada}

\title{A Godunov Method for Multidimensional Radiation Magnetohydrodynamics
based on a variable Eddington tensor}
 
\begin{abstract}
We describe a numerical algorithm to integrate the equations of
radiation magnetohydrodynamics in multidimensions using Godunov methods.
This algorithm solves the radiation moment equations in the mixed
frame, without invoking any diffusion-like approximations.  The moment
equations are closed using a variable Eddington tensor whose components
are calculated from a formal solution of the transfer equation at a large
number of angles using the method of short characteristics.  We use a
comprehensive test suite to verify the algorithm, including convergence
tests of radiation-modified linear acoustic and magnetosonic waves,
the structure of
radiation modified shocks, and two-dimensional tests of photon bubble
instability and the ablation of dense clouds by an intense radiation
field. These tests cover a very wide range of regimes, including both optically 
thick and thin flows, and ratios of the
radiation to gas pressure of at least $10^{-4}$ to $10^{4}$.  
Across most of the parameter space, we find the method is accurate. 
However, the tests also reveal there are regimes where the method needs 
improvement,
for example when both the radiation pressure 
and absorption opacity are very large. We suggest modifications to the algorithm 
that will improve accuracy in this case.
We discuss
the advantages of this method over those based on flux-limited diffusion.
In particular, we find the method is not only substantially more
accurate, but often no more expensive than the diffusion approximation 
for our intended applications.

\end{abstract}

\keywords{(magnetohydrodynamics:) MHD $-$ methods: numerical $-$  radiative transfer}

\section{Introduction}
\label{sec:intro}
Moving fluids can absorb, emit and scatter photons, and via these
processes energy and momentum are exchanged between the radiation field
and the rest of the flow.  When the fluxes of energy and
momentum carried by photons are significant compared to those
carried by particles or magnetic field, the fluid dynamics
will be significantly affected (and perhaps even controlled)
by the radiation field.  Radiation has been realized to play an
important role in the dynamics of many different astrophysical
systems, such as the formation of stars in different environments
\citep[e.g.,][]{McKeeOstriker2007,Krumholzetal2009,Offneretal2009,JiangGoodman2011},
supernovae \citep[e.g.,][]{Arnettetal1989,Jankaetal2007,Nordhausetal2010},
accretion flows around supermassive massive black holes
\citep[e.g.,][]{ShakuraSunyaev1973,Hiroseetal2009,Spruit2010}, and the
evolution of galaxies with feedback from a central massive black hole
\citep[e.g.,][]{CiottiOstriker2007,Proga2007}, to name but a few.

The dynamics of radiating fluids can be divided into quite different
regimes, depending on whether the photons dominate energy and/or
momentum transport, and whether the optical depth across the region of
interest is small or large (an excellent introduction to all aspects
of radiation hydrodynamics is given in the monographs of
\citealt[][]{MihalasMihalas1984} and \citealt[][]{Castor2004}).
In this work, we are motivated by the properties of accretion flows around
compact objects, the dynamics of accretion disk boundary layers, and the
production of radiation-driven winds and outflows from stars, accretion
disks, and galaxies.  In these systems, radiation often dominates {\em
both} the energy and momentum transport, and the optical depth can vary
widely between different regions in the flow.  Since magnetohydrodynamic
(MHD) effects are crucial in accretion flows \cite[see the review][and references therein]{BalbusHawley1998}, 
they must be included.
Thus, for these applications the appropriate equations are the those
of gas dynamics, Maxwell's equations, and the radiation transfer equation.
Our goal is to develop numerical algorithms to solve this system of
equations in order to investigate our motivating applications.

Although the basic equations of radiation MHD are well known, there is
still considerable uncertainty regarding even fundamental issues of how to
solve them.  For example, it is not clear whether to treat the radiation
field in the co-moving (fluid) or laboratory frame  in order to obtain
the simplest and most accurate solutions.  \cite{Castor2009} has provided a
detailed critique of both approaches.  In this work, we adopt the mixed
frame \citep[e.g.,][]{MihalasMihalas1984}, in which both the radiation and fluid variables are integrated
in the laboratory frame, but the radiation-material interaction terms
are treated in the co-moving frame, with $\mathcal{O}(v/c)$ expansions
used to transform these terms back to the lab frame.
A proper accounting of all $\mathcal{O}(v/c)$ terms in these transformations
is necessary in order to correctly account for all effects \citep[e.g.,][]{Lowrieetal1999,Krumholzetal2007}.
Although there are limitations to this approach (it is not suitable for
treating line transport), we find it is well suited to our applications.

Rather than integrating the time-dependent transfer equation directly
(which, even in the case of frequency independent transport, requires
solving an integro-differential equation in 6 dimensions), we instead
integrate a hierarchy of angular moments, using a variable Eddington tensor
(VET) that relates the zeroth and second moments in order to close the
system of equations.  The VET is computed directly from a formal solution
of the {\em time-independent} transfer equation at every time step.
This approach has been implemented previously in both 2D and 3D versions
of the ZEUS code \citep[][]{Stoneetal1992,HayesNorman2003},
the adaptive mesh refinement code TITAN \citep[][]{GehmeyrMihalas1994} 
as well as a Soften Lagrangian Hydrodynamic code described in \cite{GnedinAbel2001}.
However, because of the expense of
solving the transfer equation in multidimensions, the VET method has not
been widely used for astrophysical applications.  Advances in algorithms and
computer hardware now make the VET feasible even in 3D, and as we show
in this paper, the VET method has considerable advantages over other
methods.  In a companion paper \citep[][]{Davisetal2012}, we provide a detailed description
of the algorithms we have implemented to solve the transfer equation.

A popular alternative to the VET method for closing the
radiation moment equations is to adopt the flux-limited diffusion (FLD)
approximation \citep[e.g.,][]{LevermorePomraning1981}.  In this approach,
the radiation flux is assumed to be in the direction of the gradient of
the radiation energy density, with a value that is limited to prevent
superluminal transport.  In fact, the majority of multidimensional
radiation hydrodynamic codes used in astrophysics to date adopt FLD
\citep[e.g.,][]{TurnerStone2001,Krumholzetal2007,Gittingsetal2008,SwestyMyra2009,
Holstetal2011,Commerconetal2011,Zhangetal2011},
However, there are a number of well-known and potentially serious
limitations to FLD \citep[e.g.,][]{MihalasMihalas1984,HayesNorman2003}.  
For example, the fact that the direction of
the flux is assumed rather than computed from the transfer equation
can introduce serious error in optically thin regions, such as the
inability to follow shadows, or a net force on the fluid which is in the
wrong direction.  Dropping the time-dependent term in the radiation flux
eliminates radiation inertia, which can be important when the radiation
field carries a significant amount of the total momentum in the flow.
Finally, in FLD the off-diagonal components of the radiation pressure
tensor are dropped, which eliminates effects such as radiation viscosity.
The VET method overcomes all of these limitations.  We compare and
contrast these two methods throughout this paper.

A significant difference between the algorithm developed in this
paper and previous implementations of the VET method is the algorithm
used to integrate the hydrodynamic (as well as the MHD) equations.
Most previous implementations of VET used methods based on operator
splitting, with artificial viscosity required for shock capturing
as implemented in, for example, the ZEUS code \citep[][]{Stoneetal1992}.
In this work we combine the VET approach with a dimensionally-unsplit
higher-order Godunov method for hydrodynamics and MHD (as implemented
in the Athena code, see \citealt{Stoneetal2008}).  By adopting a Riemann
solver to compute fluxes of the conserved variables, Godunov methods
do not require any artificial viscosity for shock capturing.  However,
a significant challenge to this approach is how to treat the stiff source
terms associated with the interaction of the radiation and material.
With Godunov methods, simply splitting these
terms from the flux differences (the simplest and most often adopted
approach) is known to be problematic when the source terms are stiff
\citep[e.g.,][]{LowrieMorel2001}.  Instead in this work we adopt the modified Godunov method of
\cite{MiniatiColella2007}, which provides a stable second-order accurate algorithm.  In a
previous paper \cite{SekoraStone2010} (hereafter SS10), we introduced
a one-dimensional version of this algorithm.  This paper represents
the multidimensional extension of the SS10 algorithm.

In addition to the modified Godunov method, a second crucial ingredient to
the algorithm is the method by which the VET is computed.  We use a formal
solution of the radiative transfer (RT) equation based on the method of
short characteristics, and including multi-frequency, scattering, and
non-LTE effects.  Angular quadratures of the specific intensity are then
used to compute the VET from first principles.  A complete description of
our algorithm for solving the RT equation, including tests, is given in a
companion paper \citep[][]{Davisetal2012}.  The RT module can also
be used to compute images and spectra of MHD simulations for diagnostic
purposes, and can even by used to compute the radiation source terms in
the MHD equations for problems where only energy transport via photons is
important.  In effect, the algorithm described in this paper is a marriage
between the modified Godunov method of SS10 (extended to multidimensions)
and a modern non-LTE stellar atmospheres code to compute the VET.

Of course, the final measure of any algorithm is provided by quantitative
testing.  Unfortunately, there are very few analytic solutions available
in radiation hydrodynamics and MHD useful for such tests.  In this paper,
we introduce a comprehensive test suite
that includes error convergence tests of radiation-modified
acoustic waves in a variety of regimes, quantitative comparison of
the structure of radiating shocks in the non-equilibrium diffusion
limit to semi-analytic solutions over a wide range of Mach numbers,
quantitative comparison to previous numerical solutions of the structure
of subcritical radiating shocks computed with full-transport, the linear
growth rate and nonlinear regime of the photon bubble instability, and
shadowing and ablation of dense clumps by an intense radiation field.
It is likely that many of these tests will be useful for others developing
radiation hydrodynamic codes.

The outline of this paper is as follows.  In the next section, we catalog
the equations of motion we solve.  In section \ref{sec:numerical},
we describe in detail our numerical algorithms, highlighting the
extensions we have made to the 1D method described by SS10. In section
\ref{sec:EnergyError}, we discuss the importance of exact energy
conservation, and present test problems to measure the energy error.
We present the results of our comprehensive test suite in  section
\ref{sec:tests}.  Finally, we compare and contrast our numerical
algorithms to those based on the diffusion approximation in section
\ref{sec:discussFLD}, and summarize in section \ref{sec:summary}.

\section{Equations}
\label{sec:equations}

Following SS10, we write the equations of radiation MHD in the
mixed frame, and solve the radiation and material energy and momentum
conservation laws separately, including the $\mathcal{O}(v/c)$  source
terms as given by \cite{Lowrieetal1999}. Similar equations have 
also been derived by \cite{Krumholzetal2007} in the flux-limited 
diffusion approximation, and the source terms used here are identical to 
those in \cite{Krumholzetal2007} 
to $\mathcal{O}(v/c)$. Some important $(v/c)^2$ terms are also included in 
our formula, such as the advective flux of radiation enthalpy. See the 
discussions in \cite{Lowrieetal1999} on the importance of keeping these terms.
We find it convenient to
solve the system of equations in a dimensionless form by adopting the
two ratios $\Crat=c/a_0$ and $\Prat=a_rT_0^4/P_0$.   Here $a_0$, $T_0$,
and $P_0$ are the characteristic values of sound speed, gas temperature,
and gas pressure respectively, and $a_r$ is the radiation constant.  Thus,
$\Crat$ is the dimensionless speed of light, and $\Prat$ the dimensionless
radiation pressure.  With these scalings, the units of the radiation
energy density and flux are $a_rT_0^4$ and $ca_rT_0^4$ respectively.

To further simplify the equations, we assume frequency-independent (gray)
opacities, and local thermodynamic equilibrium (LTE).  The scattering
opacity is assumed to be isotropic and coherent in the co-moving
frame.  Thus, we only consider the equations for frequency-integrated
quantities, and do not need to distinguish between Planck and
flux-mean opacities.  Extension of our method to frequency dependent
transport problems, for example using multigroup methods \citep[e.g.,][]{Vaytetetal2011}, is
straightforward but beyond the scope of this paper.  The equations
of radiation MHD are then:
\begin{eqnarray}
\frac{\partial\rho}{\partial t}+\bfnabla\cdot(\rho \bv)&=&0, \nonumber \\
\frac{\partial( \rho\bv)}{\partial t}+\bfnabla\cdot({\rho \bv\bv-\bb\bb+{{\sf P}^{\ast}}}) &=&-\mathbb{P}{\bf S_r}(\bP),\  \nonumber \\
\frac{\partial{E}}{\partial t}+\bfnabla\cdot\left[(E+P^{\ast})\bv-\bb(\bb\cdot\bv)\right]&=&-\mathbb{PC}S_r(E),  \nonumber \\
\frac{\partial\bb}{\partial t}-\bfnabla\times(\bv\times\bb)&=&0, \nonumber \\
\frac{\partial E_r}{\partial t}+\mathbb{C}\bfnabla\cdot \bF_r&=&\mathbb{C}S_r(E), \nonumber \\
\frac{\partial \bF_r}{\partial t}+\mathbb{C}\bfnabla\cdot{\sf P}_r&=&\mathbb{C}{\bf S_r}(\bP),
\label{equations}
\end{eqnarray}
where the source terms are,
\begin{eqnarray}
{\bf S_r}(\bP)&=&-\sigma_t\left(\bF_r-\frac{\bv E_r+\bv\cdot{\sf P} _r}{\mathbb{C}}\right)+\sigma_a\frac{\bv}{\mathbb{C}}(T^4-E_r),\nonumber\\
S_r(E)&=&\sigma_a(T^4-E_r)+(\sigma_a-\sigma_s)\frac{\bv}{\mathbb{C}}\cdot\left(\bF _r-\frac{\bv E_r+\bv\cdot{\sf  P} _r}{\mathbb{C}}\right).
\label{sources}
\end{eqnarray}
In the above, $\rho$ is density, ${\sf P}^{\ast}\equiv(P+B^2/2)\bI$ (with $\bI$
the unit tensor), $\sigma_a$ and $\sigma_s$ are the absorption and
scattering opacities\footnote{Flux mean, Planck mean and energy mean opacities 
are treated as the same here. But they can be easily extended to be different values.} 
(which can be functions of both density and temperature),
and the magnetic permeability $\mu=1$.  The total gas energy density is
\begin{eqnarray}
E=E_g+\frac{1}{2}\rho v^2+\frac{B^2}{2},
\end{eqnarray}
where $E_g$ is the internal gas energy density.   We adopt an equation of state
for an ideal gas with adiabatic index $\gamma$, thus
$E_g=P/(\gamma-1)$ for $\gamma\neq 1$ and $T=P/R_{\text{ideal}}\rho$, where
$R_{\text{ideal}}$ is the ideal gas constant. 
The radiation pressure ${\sf P}_r$ is related to the radiation energy density $E_r$
by the closure relation
\begin{eqnarray}
{\sf P}_r=\bfr E_r.
\end{eqnarray}
where $\bfr$ is the VET.
It is straightforward to convert the dimensionless radiation MHD equations
given above to their more familiar dimensional form by setting $\Prat=1$,
and replacing $\Crat$ with the speed of light $c$, $T^4$ with $a_rT^4$,
and $\bF_r$ with $\bF_r/c$.

The VET used to close the hierarchy of radiation moment equations
is calculated directly from angular quadratures of the frequency 
averaged specific intensity $I_r$
\begin{eqnarray}
\bfr=\frac{{\sf P}_r}{E_r}=\frac{\oint I_r \hat{\bold{n}}\hat{\bold{n}}d\omega}{\oint I_rd\omega},
\label{Edd}
\end{eqnarray}
where $\omega$ is solid angle and $\hat{\bold{n}}$ is the unit vector.
The specific intensity $I_r$ is calculated from a formal solution of the
time-independent transfer equation
\begin{eqnarray}
\frac{\partial I_r}{\partial s}=\kappa_t (S-I_r),
\label{formalsolution}
\end{eqnarray}
where $\kappa_t$ is the total specific opacity and $S$ the source function.
In a companion paper \citep[][]{Davisetal2012}, we describe in detail the algorithm
we use to solve the transfer equation \ref{formalsolution},
which is based on the method
of short characteristics.  In section \ref{sec:Eddtensor}, we describe how
the angular quadratures of the specific intensity returned by the transfer
solver are computed to give the Eddington tensor.

In the above equations, 
radiation quantities are always defined in the Eulerian (lab) frame.
To order $\mathcal{O}(1/\Crat)$, the co-moving radiation energy density
$E_{r,0}$ and radiation flux $\bF_{r,0}$ are related to the lab frame
values $E_r$ and $\bF_{r}$ by \citep[e.g.,][]{Castor2004}
\begin{eqnarray}
E_{r,0}&=&E_r-2\frac{\bv}{\mathbb{C}}\cdot\bF_r,\nonumber\\
\bF_{r,0}&=&\bF_r-\frac{1}{\mathbb{C}}\left(\bv E_r+\bv \cdot{\sf P}_r\right), \nonumber\\
E_r&=&E_{r,0}+2\frac{\bv}{\mathbb{C}}\cdot\bF_{r,0},\nonumber\\
\bF_r&=&\bF_{r,0}+\frac{1}{\mathbb{C}}\left(\bv E_{r,0}+\bv \cdot{\sf P}_{r,0}\right).
\label{Comoving}
\end{eqnarray}

The source terms (equation \ref{sources}) in the mixed frame
representation were originally developed by \cite{MihalasKlein1982},
and extended by \cite{Lowrieetal1999} to include an extra term
$(\sigma_a-\sigma_s)\bv\cdot(\bv E_r+\bv\cdot{\sf P}_r)/\Crat^2$ in the
energy source term $S_r(E)$.  This term is necessary to ensure the
correct thermal equilibrium state in moving fluids, and is especially
important when scattering opacity is dominant (see a full discussion
in \citealt{Lowrieetal1999}).  Further discussion
of the physical interpretation of the source terms can be found in Section
\ref{sec:CompareDiffusion}.

We do not solve the equations in strong conservation form, which
means that total (radiation plus gas) energy and momentum are not
conserved exactly (to round-off error).  However, exact conservation
is easily enforced only for explicit algorithms, in which case the time step
is limited by the light crossing time across a cell.  Once implicit
differencing methods are adopted, then conservation to round-off error
is generally not possible due to the much larger error tolerance used
when inverting
the matrix representing the difference equations.   Since the use of
implicit methods is crucial in our application domain, we choose to
solve the radiation and material conservation laws with source terms as
accurately as possible, and monitor the energy error as a diagnostic.
The issue of the importance of exact energy conservation is discussed
in more detail, along with results of tests of energy conservation in
our methods, in section \ref{sec:EnergyError}.

\section{Numerical Algorithm}
\label{sec:numerical}

In a previous paper (SS10) we presented a one-dimensional modified Godunov
algorithm for radiation hydrodynamics.  In this paper our focus will be
on the additional extensions and improvements to the SS10 method required
in multidimensions.  The numerical algorithms described in this paper
have been implemented in the Athena MHD code \citep[][]{Stoneetal2008}.
Athena implements higher-order reconstruction in the primitive variables,
an extension of a dimensionally unsplit integrator  to MHD,
the constrained transport (CT) algorithm to enforce the divergence-free
constraint on the magnetic field, and a variety of approximate and exact
Riemann solvers.  Since comprehensive descriptions of the algorithms
in Athena has been given previously \citep{Stoneetal2008}, in this section we only
describe the extensions to these algorithms necessary for radiation MHD.

One challenge to any numerical algorithm for radiation MHD
is the large range of timescales.  In our applications, we are interested in
evolving the system on the sound crossing time, which can be many
orders of magnitude larger than the light crossing time.  Thus, implicit
differencing methods are essential.  In this work, to improve both stability 
and accuracy, we split an implicit solution of the radiation subsystem
from a modified explicit update of the rest of the equations.  That is,
we update the radiation energy density $E_r$ and flux $\bF_r$
by solving the moment equations
\begin{eqnarray}
\frac{\partial E_r}{\partial t}+\mathbb{C}\bfnabla\cdot \bF_r&=&\mathbb{C}S_r(E), \nonumber \\
\frac{\partial \bF_r}{\partial t}+\mathbb{C}\bfnabla\cdot{\sf P}_r&=&\mathbb{C}{\bf S_r}(\bP).
\label{radmomeqs}
\end{eqnarray}
using fully implicit, backward Euler differencing.  As discussed in SS10,
higher-order implicit time integration schemes can lead to oscillatory
solutions with large time steps.  Thus, to ensure a non-oscillatory
method, we restrict the update to first-order backward Euler.  During this
step, the gas variables in the source terms $S_r(E)$ and ${\bf S_r}(\bP)$
are held constant.

On the other hand, the gas quantities are updated by solving the 
ideal MHD equations using a time-explicit
modified Godunov algorithm for the stiff source terms
\begin{eqnarray}
\frac{\partial\rho}{\partial t}+\bfnabla\cdot(\rho \bv)&=&0, \nonumber \\
\frac{\partial( \rho\bv)}{\partial t}+\bfnabla\cdot({\rho \bv\bv-\bb\bb+{\sf P}^{\ast}}) &=&-\mathbb{P}{\bf S_r}(\bP),\  \nonumber \\
\frac{\partial{E}}{\partial t}+\bfnabla\cdot\left[(E+P^{\ast})\bv-\bb(\bb\cdot\bv)\right]&=&-\mathbb{PC}S_r(E),  \nonumber \\
\frac{\partial\bb}{\partial t}-\bfnabla\times(\bv\times\bb)&=&0.
\label{gasconslaws}
\end{eqnarray}
In this step, the radiation variables are held constant.  Note that
the source terms themselves are {\em not split} from the flux-divergence
terms.  This is critical for achieving stable and accurate solutions when
the source terms are stiff.  

The order in which we do these updates
is arbitrary.  In most cases, we update the gas dynamical variables 
(that is, equations \ref{gasconslaws}) first.  However, when the
the radiation pressure is completely negligible, we have found
switching the order of the update is more robust. 

\subsection{Basic Steps in the Algorithm}
\label{sec:steps}

To begin with, it is useful to summarize all of the individual steps in the
overall algorithm:

{\it Step 1.---}  Using the gas variables to compute the opacities
and source function, solve the transfer equation over a large set of angles
using short characteristics, and compute the Eddington tensor $\bfr$.

{\it Step 2.---} Reconstruct the left- and right-states in the primitive
variables at cell interfaces along each of the $x$, $y$, and $z$ directions
independently.  We reconstruct the primitive variables, instead of the
conserved variables as in SS10, since we have found it to be less oscillatory
and more accurate.  Either second- or third-order spatial reconstruction is
possible.

{\it Step 3.---} Add the radiation energy and momentum source terms
to the left and right states (see equations 79 and 80 in SS10). 
Source terms for the left (right) states are calculated by using cell-centered
quantities from the cell to the left (right) of the interface, and use
the modified Godunov method.

{\it Step 4.---} Compute 1D fluxes of the conserved variables with the
appropriate Riemann solver. For radiation hydrodynamic simulations, we
use the HLLC solver, while for radiation MHD simulations we generally
use HLLD.  We use the adiabatic sound speed (instead of the radiation
modified sound speed as was used by SS10) to calculate the fluxes.
This adds extra dissipation and makes the algorithm more robust.
We do not calculate fluxes for $E_r$ and $\bF_r$ because they
are not updated in this step; therefore the Riemann solvers are the same
as those described in \cite{Stoneetal2008}.

{\it Step 5.---} For radiation MHD calculations,
use constrained transport (CT) algorithm, 
(step 3 of the Athena 3D algorithm) to calculate 
the electric field at cell corners and update the face centered magnetic field. 

{\it Step 6.---} Evolve the left- and right-states at each interface
with the transverse flux gradients 
for half time step $\Delta t/2$, as required for the unsplit CTU
integrator in Athena. 

{\it Step 7.---} For radiation MHD calculations,
calculate the cell-centered electric field at time 
$t^{n+1/2}$ to use as a reference state in CT algorithm
(step 6 of the Athena 3D algorithm).

{\it Step 8.---} Calculate new fluxes along each direction with the corrected
left- and right-states and the appropriate Riemann solver. 

{\it Step 9.---} Update the area averaged magnetic field at cell faces
from time step $n$ to $n+1$ using CT (step 8 of Athena 3D algorithm).

{\it Step 10.---} Update the gas variables from time $n$ to $n+1$ by adding
the divergence of the flux gradient and the source terms 
using the modified Godunov method.

{\it Step 11.---}  Estimate the value of radiation energy source term added 
to the gas total energy using the method 
described in Section \ref{sec:TErterm}.  Then update the radiation 
variables $E_r,\ \bF_r$ using fully implicit, backward Euler differencing
of the radiation moment equations \ref{radmomeqs}.  This requires solving
a large linear system in multidimensions.
Gas variables do not change during this step. 

{\it Step 12.---} Update the time $t^{n+1}=t^n+\Delta t$, calculate the new time 
step according to the CFL condition with the adiabatic sound speed (fast
magnetosonic speed) for radiation hydrodynamics (MHD). 

In comparison to the algorithm introduced in SS10, the primary changes
we have made to extend the scheme to multidimensions are in Step 1
(compute the Eddington tensor in multidimensions), Step 2 (reconstruction in the primitive
rather than conserved variables), Step 11 (using
an estimate of the radiation energy source term, and solving the implicit moment
equations in multidimension), and Step 12 (choose a timestep based
on the adiabatic sound speed).

\subsection{Modified Godunov Method in Multidimensions}
In this subsection, we describe in detail the modified Godunov method to 
calculate the flux, especially the way to add the stiff source terms 
to the left and right states. 

\subsubsection{Reconstruction Step}

To compute the left- and right-states of the vector of conserved
variables required to calculate the fluxes via a Riemann solver, 
we reconstruct the primitive variables as originally proposed by 
\cite{MiniatiColella2007}, instead of the conserved 
variables (SS10).  In hydrodynamics,
the primitive and conserved variables are related via
 \begin{equation}
 \footnotesize
\left(
\begin{array}{c}
  \rho  \\
  v_x   \\
  v_y  \\
  v_z \\
   P   
\end{array}
\right)
=
\left[
\begin{array}{ccccc}
  1 & 0 & 0 & 0 & 0\\
  0 & 1/\rho &  0 & 0 & 0  \\
  0 &  0   & 1/\rho & 0 & 0\\  
  0 &  0   & 0 & 1/\rho & 0\\  
   (\gamma-1)(v_x^2+v_y^2+v_z^2)/2 & -(\gamma-1)v_x & -(\gamma - 1)v_y  & -(\gamma - 1)v_z & \gamma-1   
\end{array}
\right]
\left(
\begin{array}{c}
  \rho  \\
  \rho v_x   \\
  \rho v_y  \\
  \rho v_z \\
   E   
\end{array}
\right).
\end{equation}
The radiation source terms for velocity and pressure are
\begin{eqnarray}
 {\bf S_r}(\bv)&=&-\Prat {\bf S_r}(\bP)/\rho \\ \nonumber
 S_r(P) &=& (\gamma-1)\Prat \bv\cdot {\bf S_r}(\bP) -(\gamma-1)\Prat\Crat S_r(E),
\end{eqnarray}
where ${\bf S_r}(\bP)$ and $S_r(E)$ are given by equation \ref{sources}.
These source terms are the same in the case of radiation MHD, even
though in this case the definition of total gas energy $E$ includes the
magnetic energy.  Either second- or third-order reconstruction schemes
can be used \citep[][]{Stoneetal2008}.

The radiation source terms for the primitive variables must be added
to the left- and right-states for one half time step.  This requires
calculating the propagation operator $\mathcal{I}$ to project off
the unstable mode \citep[e.g.,][SS10]{MiniatiColella2007}, using the
gradient of radiation source terms on the plane of primitive variables
$ \bfnabla_W{\bf S_r}(\bW)$.  As discussed in SS10, in most cases it is the
energy source term $S_r(E)$ (or equivalently $S_r(P)$) that defines the
stiffness of the problem.  The momentum source term can be added as a
normal body force.  However, in the extreme case of radiation pressure
so large that $\Prat \gtrsim \Crat$, the momentum source term may also
become stiff.  For example, this can happen in the inner regions of an
accretion disk around a supermassive black hole, where $\Crat\sim 10^2$
while $\Prat\sim 10^2$ \citep[e.g.,][]{ShakuraSunyaev1973, Turneretal2003}.  
For this reason, we keep the leading term
in the momentum source term, so $\bfnabla_WS_r(\bW)$ can be written as
\begin{equation} \bfnabla_W{\bf S_r}(\bW)= \left( \begin{array}{ccccc}
  0 & 0 & 0  & 0 & 0\\ 0 & S^{v_x}_{v_x} &  0 & 0 & 0  \\
    0 & 0 &  S^{v_y}_{v_y} & 0 & 0  \\
      0 & 0 &  0 & S^{v_z}_{v_z} & 0  \\
   S^P_{\rho} & S^P_{v_x}  & S^P_{v_y} & S^P_{v_z} & S^P_P
\end{array} \right), \label{dSWdW} 
\end{equation}
where $S^x_y\equiv\partial S_r(x)/\partial y$ for any quantity $x$ and $y$.

For example, in the $x$-direction 
$S^{P}_{v_x}/S^P_P\sim \mathcal{O}(v/\Crat^2)$ and
$S^{P}_{v_x}/S^P_{\rho}\sim \mathcal{O}(v/\Crat^2)$. 
To order $v/\Crat$, we can take $S^P_{v_x}=S^P_{v_y}=S^P_{v_z}=0$, which
significantly simplifies the analysis (a significant advantage of using
the primitive variables).  Then the propagation operator is
\begin{eqnarray}
 {\sf \mathcal{I}}\left(\frac{\Delta t}{2}\right)&=&\frac{1}{\Delta t/2}\int_0^{\Delta t/2}e^{\tau \bfnabla_W {\bf S_r}\left(\bW\right)}d\tau \nonumber\\
 &=&
\left(
\begin{array}{ccccc}
  1 & 0 & 0  & 0  & 0  \\
  0  & \beta_x &  0 & 0  & 0  \\
  0 &  0   & \beta_y &  0  & 0  \\
  0 & 0   &   0   &   \beta_z &  0   \\
   (\alpha-1)S^P_{\rho}/S^P_P & 0  &  0   &  0  & \alpha   
\end{array}
\right)\nonumber\\
&\approx&\left(
\begin{array}{ccccc}
  1 & 0 & 0  & 0  & 0  \\
  0  & \beta_x &  0 & 0  & 0  \\
  0 &  0   & \beta_y &  0  & 0  \\
  0 & 0   &   0   &   \beta_z &  0   \\
   (1-\alpha)P/\rho & 0  &  0   &  0  & \alpha   
\end{array}
\right),
\end{eqnarray}
where we define the following quantities
\begin{eqnarray}
 \beta_x&=&\left[\text{exp}(S^{v_x}_{v_x}\Delta t/2)-1\right]/(S^{v_x}_{v_x}\Delta t/2), \nonumber \\
 \beta_y&=&\left[\text{exp}(S^{v_y}_{v_y}\Delta t/2)-1\right]/(S^{v_y}_{v_y}\Delta t/2), \nonumber \\
 \beta_z&=&\left[\text{exp}(S^{v_z}_{v_z}\Delta t/2)-1\right]/(S^{v_z}_{v_z}\Delta t/2), \nonumber \\
 \alpha  &=&\left[\text{exp}(S^P_P\Delta t/2)-1\right]/(S^P_P\Delta t/2). 
\end{eqnarray}

Note that $S^{v_x}_{v_x}$, $S^{v_y}_{v_y}$, $S^{v_z}_{v_z}$
$\sim\mathcal{O}(-\sigma_t\Prat/\Crat)$ and 
when $\sigma_t\Prat/\Crat\ll 1$, then $\beta_x,\ \beta_y,\ \beta_z\approx 1$.
In this case the propagation operator is reduced to the form used in SS10.
When $\sigma_t\Prat/\Crat\gg 1$, then $\beta_x$, $\beta_y$ and $\beta_z$ 
ensure that the algorithm is still stable in this regime.  With this
propagation operator, the source terms added to the left- and right-states
are $0.5\Delta t\times {\sf \mathcal{I}}(\Delta t/2){\bf S_r}(\bW)$.
 
\subsubsection{Fluxes from the Riemann Solver}
\label{RiemannFlux}

Once the left- and right-states are calculated with the appropriate
radiation source terms added, the fluxes of the conserved variables
in each direction can be calculated from any of the Riemann solvers
currently implemented in Athena (e.g., HLLC for radiation hydro, and HLLD
for radiation MHD).  Unlike SS10, the characteristic speed uses in those
solvers is not the radiation modified sound speed (equation 73 of SS10),
but the adiabatic sound speed.  We have found this is necessary to make
the multidimensional algorithm stable at a CFL number of 1.0 in 2D,
and 0.5 in 3D (the same stability limits for the extension of the CTU
integrator to MHD without radiation).

In radiation MHD, the cell-centered electromotive force (EMF) must be
calculated as a reference state in the CT algorithm after we get the
above fluxes (step 5 of the 2D integrator and step 6 of 3D integrator
in \citealt{Stoneetal2008}). In this step, momentum needs to be evolved
for a half time step and radiation source terms ${\bf S_r}(\bP)$ need to be added.

In the multidimensional CTU integrator, the left- and right-states must be
corrected with the transverse flux gradients.  Because the radiation 
subsystem is only first order accurate, transverse gradients of the 
source terms do not need to be added.
This is done in the same
way as in Athena (equation 86 and 87 of \citealt[][]{Stoneetal2008}).
Using the corrected left- and right-states, we call the Riemann solver
again to calculate the fluxes, which are then used in the following
predict-correct step.

\subsubsection{Updating the conserved variables with the predict-correct scheme}
\label{sec:addsource}

To achieve second-order accuracy, the conserved variables in this step
are updated with a predict-correct scheme, similar to the approach
taken by \cite{MiniatiColella2007} and SS10.  The radiation quantities
(energy and flux) have already been updated by the implicit solution of
the radiation subsystem, so they are kept constant during this step.
To be consistent with our reconstruction step, the stiffness of the
momentum source terms in some regions must also be taken into account.

Given the divergence of the fluxes (computed as described above) 
$\bfnabla\cdot \bF^{n+1/2}$ and the radiation source term ${\bf S_r}(\bU^{n})$, 
a predict solution is estimated as
\begin{eqnarray}
\hat{\bU}=\bU^n+\Delta t(\bI-\Delta t \bfnabla_U{\bf S_r}(\bU^n))^{-1}\left({\bf S_r}(\bU^n)-\bfnabla\cdot\bF^{n+1/2}\right),
\end{eqnarray}
where $\bfnabla_U{\bf S_r}(\bU^n)$ is the gradient of the radiation source term
${\bf S_r}(\bU)$ with respect to the conserved variables $\bU^n$ as
\begin{eqnarray}
\bfnabla_U{\bf S_r}(\bU)=
\left(
\begin{array}{ccccc}
  0 & 0 & 0  & 0 & 0\\
 0 & S^{M_x}_{M_x} &  0 & 0 & 0  \\
  0 & 0 &  S^{M_y}_{M_y} & 0 & 0  \\
  0 & 0 &  0 & S^{M_z}_{M_z} & 0  \\
   S^E_{\rho} & S^E_{M_x}  & S^E_{M_y} & S^E_{M_z} & S^E_E   
\end{array}
\right).
\label{dSUdU}
\end{eqnarray}
  
The error in this predicted solution, $\epsilon(\Delta t)$, is estimated as
\begin{eqnarray}
 \epsilon(\Delta t)=\bU^n+\frac{\Delta t}{2}\left({\bf S_r}(\bU^n) + {\bf S_r}(\hat{\bU})\right) -\Delta t \bfnabla\cdot\bF^{n+1/2}-\hat{\bU}.
\end{eqnarray}
Then the correction step is
\begin{eqnarray}
\bU^{n+1}=\hat{\bU}+(\bI-\Delta t \bfnabla_U{\bf S_r}(\bU^n))^{-1}\epsilon(\Delta t).
\end{eqnarray}
At the end of this step, all gas and radiation quantities are updated
to time step $n+1$, and the entire algorithm can be repeated in the
next cycle.

\subsection{Radiation Subsystem in Multidimensions}
\label{sec:Radsystem}

In this subsection, we describe the extensions to SS10 required
to integrate the radiation subsystem equations \ref{radmomeqs}
in multidimensions. 

\subsubsection{Special Treatment of the energy source term}
\label{sec:TErterm}

The radiation energy density $E_r$ and flux $\bF_r$ are updated from
time step $n$ to $n+1$ based on the updated gas
quantities at time step $n+1$, using first-order backward Euler time
differencing to make the method stable with a time step much larger
than the light crossing time.  However, in SS10, the source terms on the
right hand side of equations \ref{radmomeqs} were calculated using the
temperature at time step $n$.  We have found this can introduce
significant error in the total energy when the gas and radiation are
far from thermal equilibrium, and the time step is much bigger than
the thermalization time.  The source of this error is the assumption
that the gas temperature is constant during the update and thus the 
energy source terms added to the gas energy density and radiation 
energy density are not the same. This will be worse if there is 
continuous heating or cooling source terms and the energy error 
may accumulate. In reality,
the gas temperature should evolve simultaneously with the radiation
energy density until thermal equilibrium is reached. 

To reduce this 
energy error, we first estimate the change of gas energy density 
due to the radiation energy source term in the modified Godunov step. 
Let $\bfnabla\cdot \bF^{n+1/2}$ be the Riemann flux calculated in 
Section \ref{RiemannFlux}. The updated gas total energy after the 
Godunov step is 
$E^{n+1}$ and the gas total energy at the beginning of the step is 
$E^n$. Then the change of gas energy due to the radiation energy source 
term in this step $dS_r(E)$ can be estimated as
\begin{eqnarray}
dS_r(E)=E^{n+1}-(E^{n}-dt\bfnabla\cdot\bF^{n+1/2}).
\label{dEsource}
\end{eqnarray} 
To conserve total energy, the energy source term added to the 
radiation subsystem is then $\tilde{S}_r(E)=\mathcal{R}\left(\Crat\sigma_a(\tilde{T}^4-E_r)
+(\sigma_a-\sigma_s)\bv\cdot\left(\bF _r-(\bv E_r+\bv\cdot{\sf  P} _r)/\mathbb{C}\right)\right)-(1-\mathcal{R})dS_r(E)/(dt\Prat)$, 
where the estimated temperature $\tilde{T}^4=-(1/(dt\Crat\sigma_a)+1)dS_r(E)/\Prat+E_r^n$. 
Here $\mathcal{R}$ is a parameter we can choose to get the best 
energy conservation. In practice, we find $\mathcal{R}=0.05$ can 
give the best results for the tests we have done.

With this special treatment of the energy source term in the radiation
moment equations, we can reduce the energy error significantly, especially
for states that are initially far from thermal equilibrium in optically
thick regions that are evolved with a time step much larger than the
thermalization time (in practice, such states are extremely rare in any
dynamical evolution, and usually occur only if set up specifically in
the initial conditions).  However, we still cannot conserve total energy
to round-off error.  We present tests in section \ref{sec:EnergyError}
that show in practice, the error in total energy conservation is small.

\subsubsection{The 3D matrix}

The generalization of the fully implicit, backward Euler update of the
radiation moment equations to 3D is straightforward.
For this step, the vector of conserved quantities at time step
$n$ is ${\bf U}^{n}_{i,j,k} =  (E_r,F_{r,x},F_{r,y},F_{r,z})$ .  The radiation flux 
$F_{r,x},F_{r,y},F_{r,z}$ are advanced to time step $n+1$ by solving
\begin{eqnarray}
{\bf U}_{i,j,k}^{n+1}&=&{\bf U}_{i,j,k}^n-\frac{\Delta t}{\Delta x}\left[{\bf F}^{HLLE}_{i+1/2,j,k}-{\bf F}^{HLLE}_{i-1/2,j,k}\right]
-\frac{\Delta t}{\Delta y}\left[{\bf F}^{HLLE}_{i,j+1/2,k}-{\bf F}^{HLLE}_{i,j-1/2,k}\right]\nonumber\\
&-&\frac{\Delta t}{\Delta z}\left[{\bf F}^{HLLE}_{i,j,k+1/2}-{\bf F}^{HLLE}_{i,j,k-1/2}\right]
+\Delta t {\bf S}({\bf U}^{n+1}_{i,j,k}),
\label{3DImplicit}
\end{eqnarray}
while the radiation energy density $E_r$ is updated as
\begin{eqnarray}
E_{r,i,j,k}^{n+1}&=&E_{r,i,j,k}^n-\frac{\Delta t}{\Delta x}\left[{\bf F}^{HLLE}_{i+1/2,j,k}-{\bf F}^{HLLE}_{i-1/2,j,k}\right]
-\frac{\Delta t}{\Delta y}\left[{\bf F}^{HLLE}_{i,j+1/2,k}-{\bf F}^{HLLE}_{i,j-1/2,k}\right]\nonumber\\
&-&\frac{\Delta t}{\Delta z}\left[{\bf F}^{HLLE}_{i,j,k+1/2}-{\bf F}^{HLLE}_{i,j,k-1/2}\right]
+\tilde{S}_r(E),
\label{3DImplicitEr}
\end{eqnarray}
where the ${\bf F}^{HLLE}$ are the vector of fluxes for the conserved quantities
at each cell interface computed by a HLLE Riemann solver (equation 39
in SS10) and $dS_r(E)$ is the estimate energy source term (equation \ref{dEsource}). 
Since the backward Euler differencing is only first-order in
time, first-order spatial reconstruction is used to compute the left- and
right-states for the HLLE solver.  Moreover, since both
the HLLE fluxes and the source
term ${\bf S}({\bf U}^{n+1})$ in equation \ref{3DImplicit} and \ref{3DImplicitEr} are {\em linear} in
the unknowns ${\bf U}^{n+1}_{i,j,k}$, then only one matrix solve is required
per time step, which can represent a significant savings compared to
the implicit solution of nonlinear equations that can result from other
splittings \citep[e.g.,][]{TurnerStone2001,Gonzalezetal2007}.  
The matrix of coefficients that must be inverted to solve
the linear system in this step is given in Appendix \ref{3Dmatrix}.

\subsection{Computing the Eddington tensor}
\label{sec:Eddtensor}

In order to calculate the VET, we use a formal solution of
time-independent transfer equation \ref{formalsolution}.  The methods
for solving the radiative transfer equation are described in \cite{Davisetal2012}, 
and the reader is referred to that work for further
details. At each time step, and for each grid cell, the specific
intensity $I_{r}$ must be integrated along many different rays  $\hat{n}$.  The
angular discretization and corresponding quadratures are chosen to
cover the unit sphere as uniformly as possible, but still be invariant
under 90 degree rotations about the coordinate axes
\citep[e.g.,][]{Brulsetal1999}.  Along each ray, the method of short
characteristics \citep[e.g.,][]{Mihalasetal1978,OlsonKunasz1987} is
used to integrate the transfer equation across the entire mesh.  For
multidimensional problems, this requires the interpolation of
intensities, opacities, and emissivities from (only) neighboring grid
zones.  Simulations with scattering opacity (non-LTE problems) are
solved iteratively with accelerated lambda iteration, with methods
similar to those described in \cite{TrujilloFabiani1995}.  Intensity
boundary conditions and parallelization are handled via ghost zones,
as described in section 3.5 of \cite{Davisetal2012}.

During the integration along each angle, the zeroth moment ($J$) and
second moment (${\sf K}$) of the specific intensity are summed into
running quadratures. This saves having to allocate an array to store
$I_{r}$ over all angles. For each discreet ray $\hat{n}_k$, there is a
vector of direction cosines $(\mu_{0k},\mu_{1k},\mu_{2k})$ with
$\mu_{ik}=\hat{n}_k \cdot \hat{x}_i$ and quadrature weights $w_k$.
The moments are then given by
\begin{eqnarray}
J &=& \sum_{k=0}^{n_a-1} w_k I_k \\ K_{ij} &=&
\sum_{k=0}^{n_a-1} w_k I_k \mu_{ik} \mu_{jk},
\end{eqnarray}
where $I_k \equiv I_{r}(\hat{n}_k)$.  Up to a common factor of
$4\pi/c$, $J$ and ${\sf K}$ are equivalent to radiation energy density
and pressure, respectively.  Since they are computed by the radiation
transfer solver, they will (in general) differ from the $E_r$ and
${\bf P_r}$ of the integrator. As defined in equation (6), the VET is
then simply the ratio of these quadratures ${\sf f} = {\sf K}/J$,
evaluated in each grid cell.

\section{Evaluating the Importance of Exact Energy Conservation}
\label{sec:EnergyError}

As discussed in section \ref{sec:Radsystem}, our algorithm does not
conserve energy exactly (to round-off error), in part because we separate
the implicit solution of the radiation subsystem from the modified Godunov
update of the material conservation laws.  In fact, even if the strong
conservation form of the equations is adopted, it is in general not
possible to conserve energy to round-off error with implicit differencing
if an iterative method is used to solve the resulting linear system to
some error criterion.  Therefore, our philosophy is to monitor energy
conservation as a measure of the accuracy of the solution, rather than
adopting ad-hoc strategies to enforce conservation.

We have found that one important modification to the original algorithm
described in SS10 that improves energy conservation is to estimate
the energy source term added in the modified Godunov step and 
then use the same value of energy source term in the 
radiation subsystem (section \ref{sec:TErterm}). 
A test which demonstrates the importance of this improvement is thermal
relaxation in a uniform, stationary medium.
Consider an infinite uniform
gas with density $\rho$, temperature $T$, and constant absorption
opacity $\sigma_a$.  The radiation energy density $E_r$ is also uniform
everywhere and the radiation flux is zero.  If the gas and radiation are
not in thermal equilibrium, i.e. $E_r\neq T^4$ in our dimensionless units,
then they will evolve towards the equilibrium state $\tilde{E_r}=\tilde{T}^4$ 
on the thermalization time scale $\sim
P/(\Prat E_r \sigma_a \Crat)$ \citep[e.g.,][]{BlaesSocrates2003}. 
The exact solution is set by the condition of energy conservation 
$\Prat E_r+\rho R_{\text{ideal}}T/(\gamma-1)=\Prat \tilde{E_r}+\rho R_{\text{ideal}} \tilde{T}/(\gamma-1)$.
We test
this evolution using $\Crat=10^4, \Prat=1, \rho=1, R_{\text{ideal}}=1,
\gamma=5/3$ and $\sigma_a=100$, with different initial $T$ and $E_r$,
in a domain of size $L=2$ and 128 grid points, and periodic boundary
conditions.  In our algorithm the time step is determined by the CFL
condition using the adiabatic sound speed, which gives about $\sim
5\times10^{-3}$, much larger than the thermalization time $\sim 10^{-6}$.
This problem has also been used to test other codes \citep[e.g., SS10,][]{Zhangetal2011}, 
but
here we choose extreme values of the parameters in order to demonstrate
potential problems.  These values are almost certainly unrealistic in
that no dynamical evolution could result in these conditions in any cell,
but nevertheless it is useful.

\begin{figure}
\centering 
\includegraphics[width=0.9\hsize]{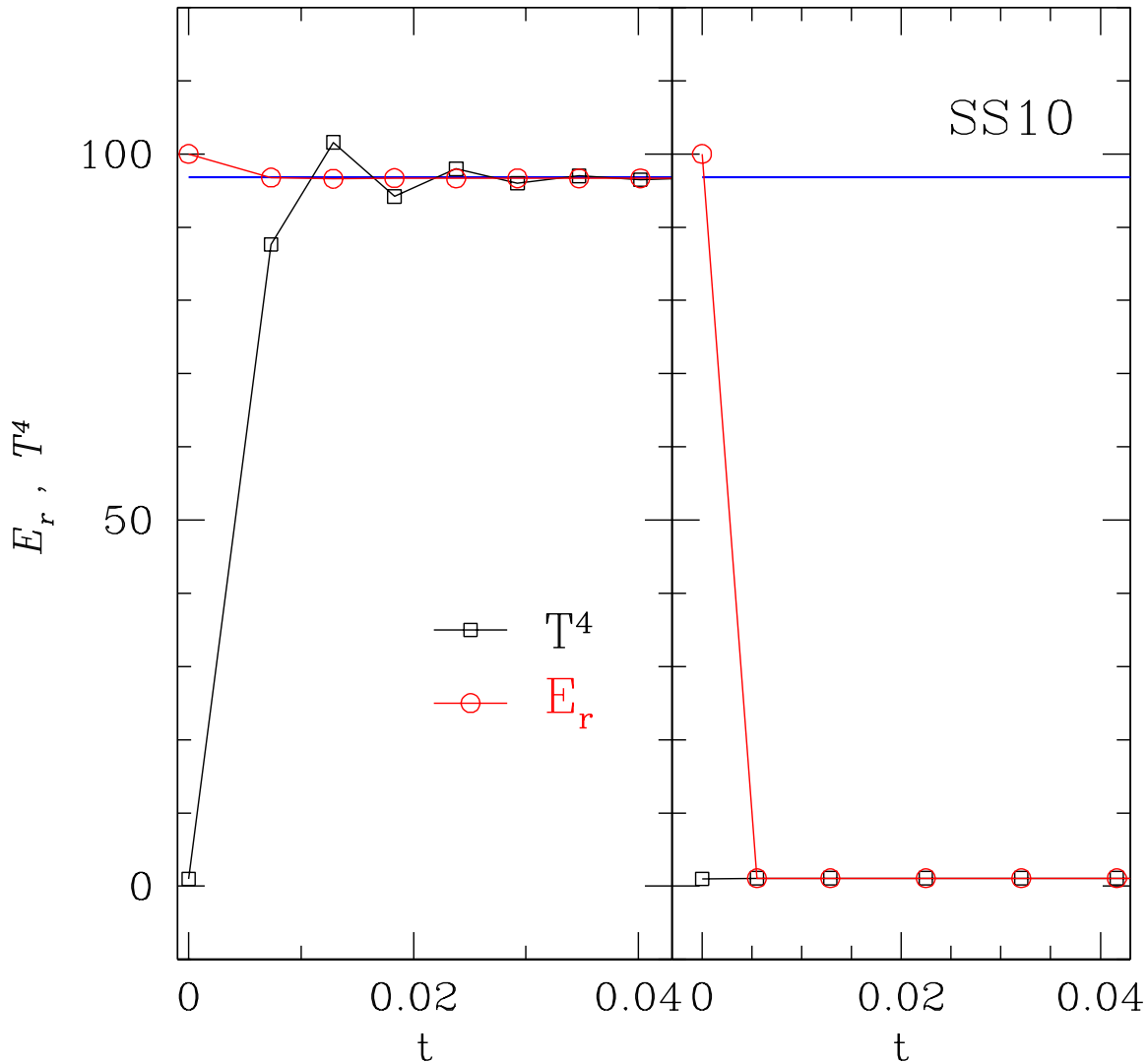}
\caption{Evolution of the radiation and gas energies for
relaxation to thermal equilibrium using a timestep $\sim 10^{3}$ larger
than the thermalization time.  The left 
panel is the result with our special treatment of the 
energy source term (section \ref{sec:TErterm}) while the right column 
is the result with the original method used by 
SS10.  The red (black) lines are for $E_r$ ($T^4$)
while the blue lines are the thermal equilibrium state.}
\label{EnergyError1}
\end{figure}

\begin{figure}
\centering 
\includegraphics[width=0.9\hsize]{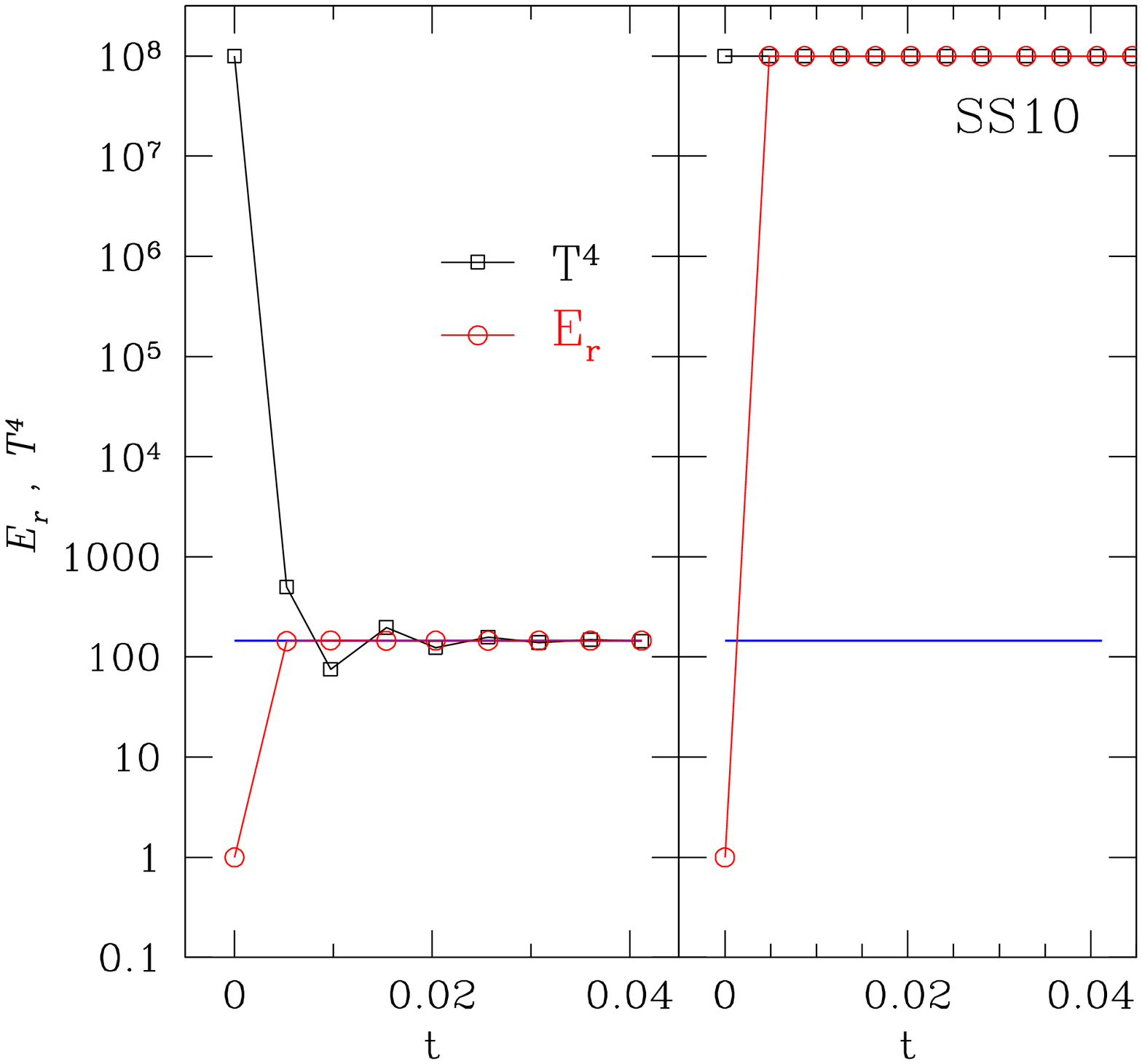}
\caption{Same as figure \ref{EnergyError1}, but with a different
initial condition.  Although the initial difference of $T^4-E_r$ is $\sim 10^8$,
our improved algorithm relaxes to the equilibrium state accurately.}
\label{EnergyError2}
\end{figure}

We show the solutions for two different initial conditions: $E_r=100, T=1$
(Figure \ref{EnergyError1}) and $E_r=1, T=100$ (Figure \ref{EnergyError2})
respectively.  Results with our modifications are shown in the left panels
of these figures, while the solutions given by the method used in SS10
are shown in the right panels.  It is clear the our special treatment of the 
radiation energy source term
reduces the error in the energy significantly, especially
in the case when $E_r-T^4$ is very large.  The right columns confirm the
analysis in section \ref{sec:TErterm}, in that with the original method
proposed by SS10, when the time step is much larger than thermalization
time, $E_r$ approaches $T^4$ in a single step.  Unfortunately, this is the
wrong solution in the case that $E_r=1$ and $T=100$.  Because our
algorithm adds the same energy source term to the gas 
and radiation energy density, we conserve total energy much better. 
In this case, the energy error in our algorithm is determined by the 
tolerance level we set in the matrix solver.
These tests also demonstrate
stability even when the time step is much larger than the
thermalization time.

The tests shown above are done with extreme conditions: a very large
time step compared to the thermalization time and initial conditions
far away from thermal equilibrium. The velocity and radiation flux are
always zero in the tests. As a test of energy conservation in fully
dynamical problems, we have studied the conservation of total energy
for magneto-rotational instability \citep[e.g.,][]{BalbusHawley1998}
in an unstratified shearing-box simulations of a radiation dominated
black hole accretion disk.  The initial condition and shearing
periodic boundary condition are the same as the fiducial model in
\cite{Turneretal2003} for location I (see their table 1). The only
difference is that the ratio between gas pressure and magnetic
pressure is $1600$ in our simulation.  In our units, the initial
parameters are $\rho=1,P=1,E_r=1,B_x=0,B_y=0,B_z=0.035$.  The angular
frequency is $\Omega=1$. The dimensionless parameters $\Prat=376.26$
and $\Crat=4895.61$. A resolution of $32\times 128\times 32$ grids is
used for a box size $L_x=1,L_y=4,L_z=1$.  The gas and radiation field are
continuously heated up due to the MRI turbulence. The energy source comes
from the differential rotation of the disk.  We calculate the difference
between the work done on the simulation box and the total energy change
according to equation 8 of \cite{HGB1995}, which is the energy error. Over
$100$ period, the energy error is only about $0.6\%$ of the final total
energy. If measured with respect to the total work done on the simulation
box, the energy error is only about $1\%$ for $100$ periods.


\section{Tests of the radiation subsystem}
Our implicit Backward Euler scheme to solve the radiation subsystem
(equation \ref{radmomeqs}) is only first-order accurate.  One might be
concerned that a first-order integration algorithm may be too diffusive,
especially when there are steep spatial gradients.  In this section,
we provide further tests of the numerical solution of the radiation
subsystem, and show that it can in fact capture sharp features.

\subsection{Marshak Wave Test}
\begin{figure}
\centering
\includegraphics[width=0.9\hsize]{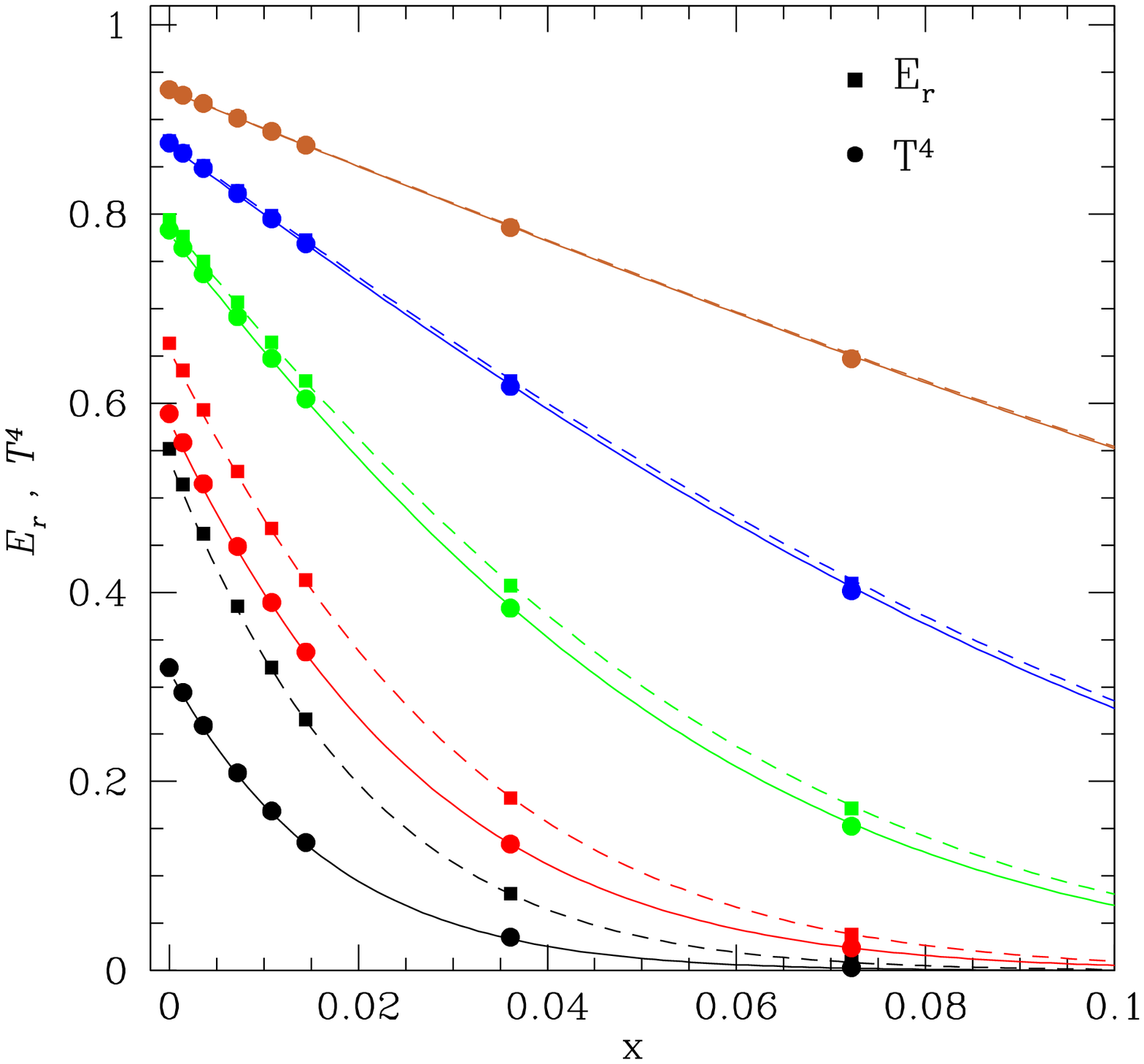}
\caption{Marshak wave test. The filled circles and squares
are the semi-analytic solution of \cite{SuOlson1996}, while the solid and
dashed lines are numerical solutions, for the fourth power of the gas temperature
and the radiation energy density respectively. 
Each pair of lines with the same color are at a different time,
with time increasing from the bottom to the top pair.}
\label{Marshak}
\end{figure}

Evolution of a Marshak wave is a 1D non-equilibrium
diffusion process originally proposed by \cite{Marshak1958}.  
A semi-analytic solution is given
by \cite{SuOlson1996}, and we compare the results from our code with
this solution. This test has also been used by many other authors
\citep[][]{Gonzalezetal2007,Gittingsetal2008,Holstetal2011,Zhangetal2011}. 
It consists of a cold uniform medium with constant absorption
opacity $\sigma_a$.  A constant radiation flux $F_r^{\text{inc}}$ is
applied at the left boundary at $t=0$ and allowed to diffuse through
the medium.  The cold gas is heated up by the radiation field, and
eventually the gas and radiation field reach thermal equilibrium.
This is not a dynamical test, so the hydrodynamic algorithm is
not used, and only the radiation subsystem is solved, with the gas
temperature updated according to equation (124) of SS10.  All of the
other parameters are the same as in SS10.  In particular, the time step
is limited by the light crossing time of each cell in a domain of
size $L=0.1$ and 512 grid points.  As shown in Figure \ref{Marshak}, our
numerical solution matches the semi-analytic solution extremely well.

\subsection{Tophat Test}
\label{sec:Tophat}
The ``tophat" or ``crooked pipe" test  \citep[e.g.,][]{Gentile2001} is 
designed to study the propagation of a radiation field in a low opacity 
pipe with a complex shape.  Surrounding the pipe is a high opacity material.
It is a challenging problem 
because it requires following the propagation of radiation through
complex shapes bounded by discontinuities in opacities. 
We find this test useful to show that our
first-order Backward-Euler differencing with VET can capture sharp gradients
in the radiation field, as well as follow the propagation of radiation along
the pipe properly.

The problem is initialized following the description in
\cite{Gentile2001}, except that we solve the problem in Cartesian instead
of cylindrical coordinates.  Because of this difference, we cannot
reproduce the solution given by \cite{Gentile2001} exactly.  Nonetheless,
the test is still useful to demonstrate we achieve qualitatively the
same solution, and that the VET method can solve this problem properly.

\begin{figure}
\centering
\includegraphics[width=0.9\hsize]{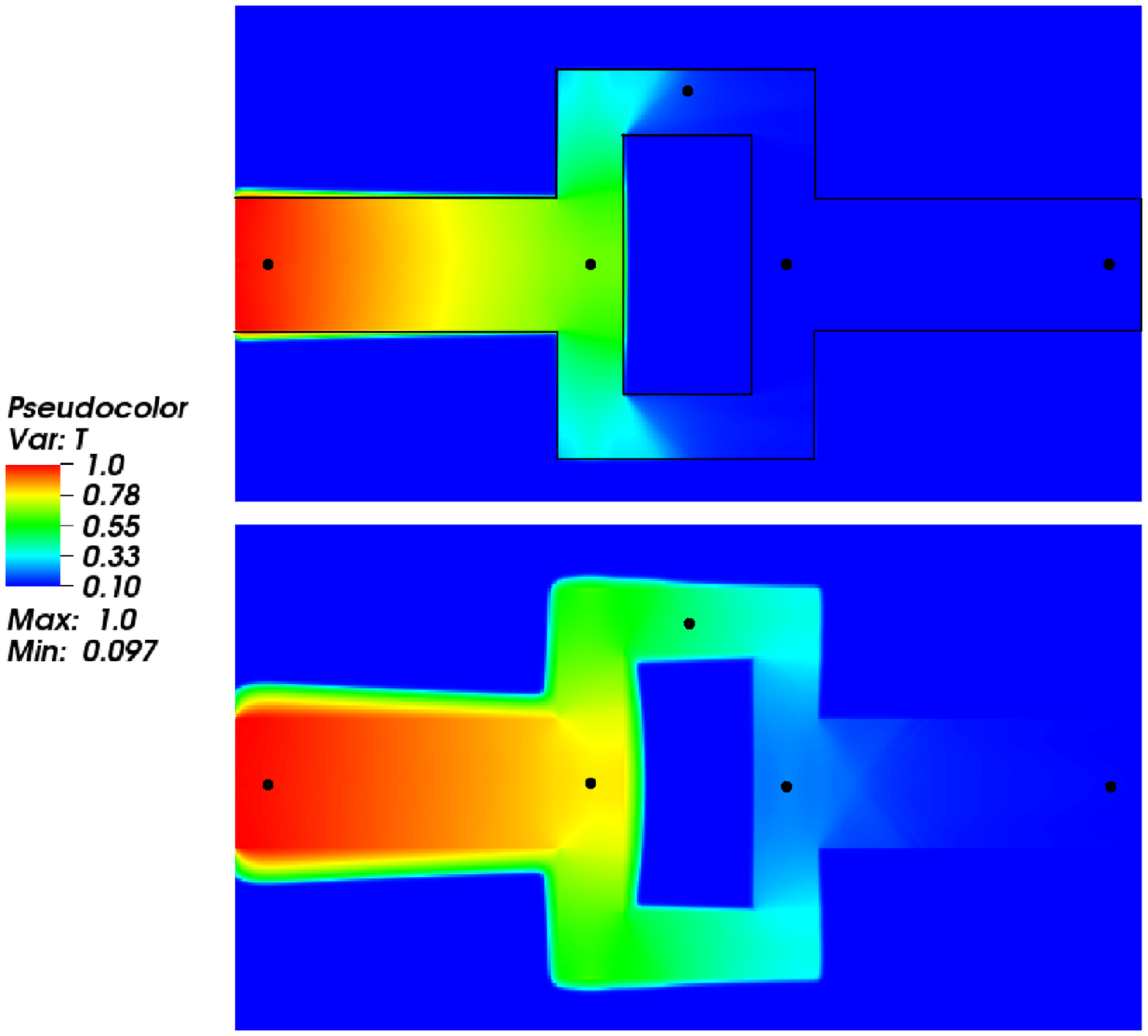}
\caption{Temperature of the pipe in the Tophat test at time 0.8 (top) 
and 9.4 (bottom) respectively. Temperature unit is $0.5$ keV. The five black
dots in each plot are the positions where the temperature probes are placed. The heating 
source with fixed temperature $0.5$ keV is located at the left hand side of the pipe. The black line 
in the top panel is the boundary of the pipe.}
\label{Tophatshape}
\end{figure}

The size of the simulation domain is $(0,7)\times(-2,2)$ with a
resolution of $512\times256$ grid points.  Dense, opaque material with
density $10$ g/cm$^3$ and opacity $\sigma_a=2000$ cm$^{-1}$ is located
in the following regions: $(3,4)\times(-1,1)$, $(0,2.5)\times(-2,-0.5)$,
$(0,2.5)\times(0.5,2)$, $(4.5,7)\times(-2,-0.5)$,  $(4.5,7)\times(0.5,2)$,
$(2.5,4.5)\times(-2,-1.5)$ and $(2.5,4.5)\times(1.5,2.5)$.  The pipe, with
density $0.01$ g/cm$^3$ and opacity $\sigma_a=0.2$ cm$^{-1}$, occupies
all other regions.  The structure of the pipe is shown by the black
line in the top panel of Figure \ref{Tophatshape}. The dimensionless
speed of light is $300$. Initially, the material has a temperature
$0.05$ keV everywhere, and the radiation and material temperature are
in equilibrium.  A heating source with a fixed temperature $0.5$ keV
is located on the left boundary for $-0.5<y<0.5$.   All other boundary
conditions are outflow.  We use the short-characteristic module to
calculate the VET.  An isotropic incoming specific intensity is applied
only along the left boundary in the region covering the heating source.
The incoming specific intensity is zero for all other boundaries.
During the evolution, velocities are always kept zero so the material
density does not change.  Thus, the modified Godunov step to update the
material quantities is not needed in this test, and we simply evolve the
material temperature through the following two equations \begin{eqnarray}
c_v\rho(T^{n+1}-T^{n})&=&-dt\Crat\sigma_a(a_rT^{n+1,4}-E_r^{n+1}),
\nonumber \\ E_r^{n+1}-E_r^{n}&=&dt\Crat\sigma_a(a_rT^{n+1,4}-E_r^{n+1}),
\end{eqnarray} where the heat capacity $c_v=10^{15}$ erg g$^{-1}$
keV$^{-1}$. Note that we only update material temperature in this step;
the radiation energy density is unchanged.   The radiation energy source
term added in this step is $dS_r(E)=c_v\rho(T^{n+1}-T^n)$.  The radiation
energy density and flux are then evolved using our first-order Backward
Euler update.  We call this algorithm method I.

\begin{figure}
\centering
\includegraphics[width=0.9\hsize]{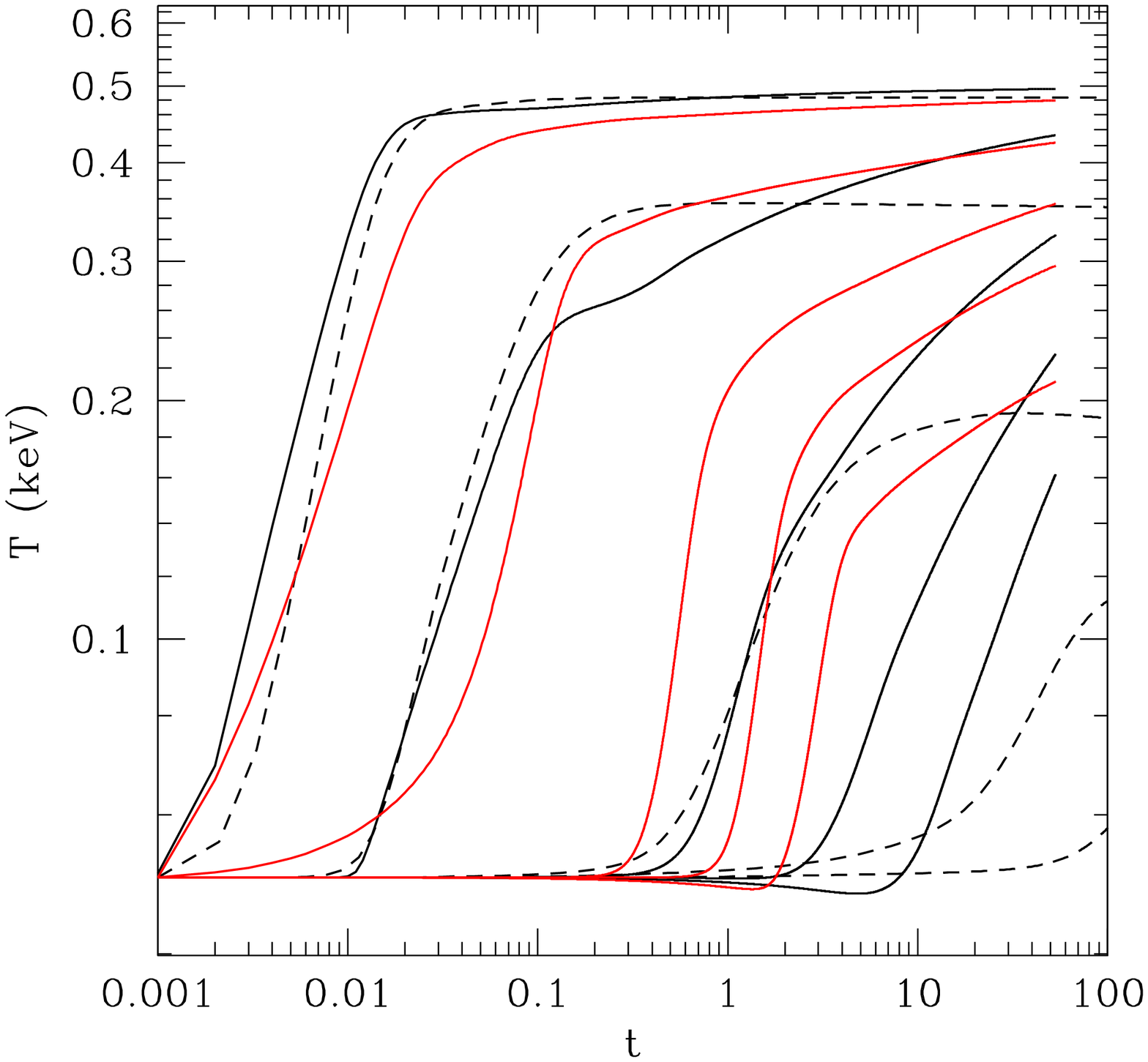}
\caption{Evolution of temperature for the five probes in the 
crooked pipe. From left to right, 
each line corresponds to one of the temperature probes labeled in Figure 
\ref{Tophatshape}. The solid line is the solution from method I with 
fixed time step $4\times10^{-5}$ to resolve the light crossing time. 
The dashed line is also the solution from method I with time step $10^{-3}$ to 
begin with, and increased by a factor of $1.1$ at each step until the time step
reaches $1$. The red line is the solution from method II with a fixed time 
step $10^{-3}$.}
\label{Tophat}
\end{figure}

First, we fix the time step to be $4\times10^{-5}$ to resolve the light
crossing time.  Snapshots of the temperature distribution at two different
times are shown in Figure \ref{Tophatshape}. Note that the shadows
around the corners of the pipe are captured properly.  Five probes are
placed at $(x=0.25, y=0)$, $(x=2.75, y=0)$, $(x=3.5, y=1.25)$, $(x=4.25,
y=0)$ and $(x=6.75,y=0)$ to monitor the change of the temperature in
the pipe.  The time evolution of the temperatures at the five points
are shown as solid black lines in Figure \ref{Tophat}. Compared with
Figure 9 of \cite{Gentile2001}, which shows the result calculated with a
Monte Carlo method, our first-order backward Euler method with VET
gives very similar results when the time step is small enough to resolve
the light crossing time.  For the fifth point, note that it cools off
slightly before being heated by the radiation wave, in agreement with
the behavior noted by \cite{Gentile2001}.  The most obvious difference
in our solution compared to \cite{Gentile2001} is that the temperature
of the first two points increases too quickly. This is likely due to the
fact that we neglect the light travel time in our short-characteristic
module when we calculate the VET. Thus the first two points are affected
by the heating source too early.  We also find the evolution of the last
three points to differ in detail from that shown by \cite{Gentile2001},
however for these points the difference between propagation in cylindrical
versus planar (Cartesian) geometry may be important.  In our solution,
there is no geometrical dilution of the flux as it propagates radially,
and therefore the heating rate per unit surface area on the walls of
the pipe is increased.  This can affect the detailed temporal evolution
at the third through fifth probes.  Most importantly, note that from
Figure \ref{Tophatshape} our first order implicit update can maintain
very sharp gradients in the temperature at the walls of the pipe.

As a second test of our method, we repeat the problem using a time
step of $10^{-3}$ initially, and increasing it by a factor of $1.1$
at each step until it reaches $1$. The result is shown as the dashed
line in Figure \ref{Tophat}.  With this larger time step, the error in
the temporal evolution of the probes is increased, although the basic
evolution is still correct.  This is consistent with our expectations.
If we are not interested following evolution on a light crossing or
thermalization time, we can use a large time step and still maintain a
stable solution with the correct asymptotic behavior.  Instead, if we
want to explore the evolution on these time scales, we must use a small
enough time step to resolve them.

A second method to evolve the material temperature with our algorithms
is to add the radiation energy source term based on the solution
of the time-independent transfer equation used
to calculate the VET.  Details of how to compute the material
heating rate from the solution of the transfer equation are given in
\cite{Davisetal2012}.  We call this algorithm method II.  The solution
computed using this method for the tophat test is shown as red lines in
Figure \ref{Tophat}.  The time step is fixed to be $10^{-3}$.  The result
does not change substantially when we decrease the time step further.
This method also gets a very similar solution as that shown in Figure
9 of \cite{Gentile2001}.  Once again, the material is heated up too
quickly everywhere but the location of the first probe.  Again, this
is likely due to the fact that the short characteristic module solves
the time-independent radiation transfer equation, so that we ignore the
delay due to the finite speed of propagation of radiation down the pipe.
The last three points also differ because of our use of a planar geometry.
Nevertheless, method II still captures the basic evolution correctly.

Other stringent test of our radiation transport module used to
calculate the VET (such as the beam test), are described in detail in
\cite{Davisetal2012}, and will not be repeated here.

\section{Tests of the Full Algorithm} \label{sec:tests}

In SS10, tests of various aspects of the one dimensional version of
the algorithm used here were presented, and we have verified that our
extension to multidimensions is able to pass all of these same tests.
However, more rigorous testing requires solutions to the full system of
equations of radiation hydrodynamics in multidimensions.  Unfortunately,
unlike the case of hydrodynamics or MHD, there are very few such
solutions available.  In the following subsections, we describe the
results of the suite of test problems we have found most useful.

\subsection{Linear Wave Convergence}
\label{sec:Linearwave}
 

The dispersion relation for linear wave solutions to the
radiation hydrodynamic equations in a uniform homogeneous background medium,
and assuming the Eddington approximation
(fixed Eddington factor ${\sf f}=1/3{\sf I}$), have been analyzed by a variety
of authors \citep[e.g.,][]{MihalasMihalas1984,Bogdanetal1996,JohnsonKlein2010}.
These authors solve the boundary value problem
in which a complex
wave number $k$ is found for each real frequency $\omega$.
\cite{JohnsonKlein2010} have used these
solutions to test the Lagrangian radiation hydrodynamic
code KULL by driving the waves with a time-dependent
boundary condition; similar tests were used earlier
by \cite{TurnerStone2001} to test the flux-limited diffusion module in 
the ZEUS code.

In contrast, \cite{Lowrieetal1999} give the dispersion relation for linear
waves (again in the Eddington approximation) as an initial value problem,
that is solving for the imaginary frequency $\omega$ at fixed real $k$.
Although their analysis focused on the properties of linear waves in a
moving fluid (they point out that in radiation hydrodynamics, linear wave
solutions are not Galilean invariant), their approach is ideally suited
for code tests.  It allows the propagation of linear waves to be followed
in a periodic domain, free from the complexity of the implementation
of driving boundary conditions in an Eulerian code, which may in and
of itself introduce spurious error into the solutions.  The dispersion
relation for linear waves in radiation MHD has also been considered by
\cite{BlaesSocrates2001} and \cite{BlaesSocrates2003}.  We use both the
hydrodynamic and MHD linear wave solutions for convergence testing in
$1D$, $2D$, and $3D$ in the following subsections.  We emphasize that
in order to keep the solutions to the dispersion relation analytic so that they 
can be used in the test problems,
the Eddington approximation must be adopted.  There may be regions of
parameter space where the Eddington approximation is not appropriate,
and the properties of linear waves in this regime should be computed
using a variable Eddington factor computed self-consistently with flow.
However such solutions are beyond the scope of this paper.

It is not possible to simply initialize this test problem using
previous solutions for linear waves given in the literature, because
there may be differences in the frame of reference, and in the order of
the source terms kept, in the fundamental system of dynamical equations
(our equation \ref{equations}). For this reason, we have rederived the
dispersion relation for linear waves in the system of equations solved
by our code.  Even with the Eddington approximation, these solutions are
non-trivial, requiring finding the roots of a fourth-order polynomial
for complex $k$ in hydrodynamics, and a tenth-order polynomial for $k$
in MHD.  Appendices B and C give the details of our solutions in the
case of hydrodynamics and MHD respectively, while in Tables 1 through 3
we give the complete eigenvector for linear waves for several parameter
values for use by others. 

\subsubsection{Linear Waves in Radiation Hydrodynamics}

We begin by studying the propagation of linear waves in one dimension with
hydrodynamics.  We consider a uniform, homogeneous medium with background
state $\rho=T=P=1$, adiabatic index $\gamma=5/3$, and zero velocity
initially.  The radiation flux is zero, and the radiation energy density
$E_r=1$ initially for all of the calculations.  We assume a constant
absorption opacity $\sigma_a$ (perturbations of the opacity will always
be second order, which can be neglected), and zero scattering opacity.
In order to adopt the Eddington approximation, we fix the Eddington
tensor to be diagonal with components $f_{ii} = 1/3$.  The dimensionless
speed of light $\Crat=10^4$.  We use a domain of size $L=1$ with 512 grid
zones, and periodic boundary conditions for all variables at each edge.
We initialize a wave solution (Appendix \ref{Radhydrowave}) by adding
an eigenmode (see Table 1) with a wavelength equal to the size of the
domain $L$, and an amplitude of $10^{-6}$.

We perform a series of calculations in which we study the effect 
of varying the radiation to gas pressure ratio by
varying the dimensionless pressure $\Prat$, while keeping the initial
conditions fixed as above.  We also study the effect of varying the
optical depth per wavelength $\tau_a = \sigma_a L$ by varying the
absorption opacity $\sigma_a$.  For each of these calculations, we measure the
phase velocity of the waves by determining how long it takes a fixed feature
in the wave (the density maximum) to propagate a distance equal to one
wavelength ($L=1$).  We also measure the damping rate of the wave by
determining the best fit amplitude to the entire wave profile after one
period.  We then compare these results to linear theory.

The properties of radiation modified linear acoustic waves vary
dramatically with optical depth and pressure ratio.  Two dimensionless
numbers can be used to define various regimes.  The first,
$\Prat\Crat\tau_a$, measure the importance of energy exchange between
the radiation field and the material, while the second, $\Prat\tau_a$,
measures the importance of momentum exchange.  When both $\Prat\Crat\tau_a
\ll 1$ and $\Prat\tau_a \ll 1$, energy exchange between the radiation
field and material is very weak, the momentum of the radiation field
can be neglected, so the waves are weakly damped and propagate at
nearly the adiabatic sound speed.  The damping rate {\em increases}
with optical depth until $\tau_a \sim 1$; thereafter the radiation and
material energy densities are strongly coupled, and the damping rate
decreases with increasing optical depth.  When $\Prat\Crat\tau_a \gg 1$
and $\Prat < 1$, the radiation and material energy densities are strongly
coupled but the radiation momentum is still negligible, therefore in this case
the damping rate {\em decreases} with optical depth until $\tau_a \sim
1$, and increases thereafter.  Finally, if $\Prat\Crat\tau_a \gg 1$ and
$\Prat > 1$, the momentum in the photons becomes important, and the damping
rate reaches a minimum when $\Prat\tau_a\sim 1$ and increase afterwards. 
To ensure our algorithms are accurate over a wide range of regimes,
it is important to perform tests that span these dimensionless parameter
values. These dimensionless quantities also clarify a potential 
shortcoming with numerical algorithm based on the reduced 
speed of light approximation \citep[e.g.,][]{GnedinAbel2001}. Any 
arbitrary reduction in the speed of light will reduce $\Crat$ and 
the above dimensionless numbers correspondingly, which will alter the 
damping rate and phase velocity of the radiation modified acoustic 
wave.

In addition to $\Prat$ and $\Crat$, the Boltzmann number $\text{Bo}$ is another dimensionless number 
which is useful to quantify the relative importance of radiative and 
material energy transport in a radiating flow \citep[e.g.,][]{MihalasMihalas1984};
it is defined to be the ratio between the material enthalpy flux and 
radiative flux from a free surface. It is related to our dimensionless numbers 
$\Prat$ and $\Crat$ as
\begin{eqnarray}
\text{Bo}=\frac{4\tilde{v}\gamma}{\mathbb{PC}(\gamma-1)},
\label{BoPC}
\end{eqnarray}
where $\tilde{v}\equiv v/a_0$, is the typical flow velocity in units of $a_0$. In the linear wave tests, 
the typical flow velocity $v$ can be replaced with the adiabatic sound speed. When $\text{Bo}$ is 
small, energy transport is dominated by radiation field and material energy 
transport is dominant when $\text{Bo}$ is large.

\begin{figure}[htp]
\centering
\includegraphics[width=0.8\hsize]{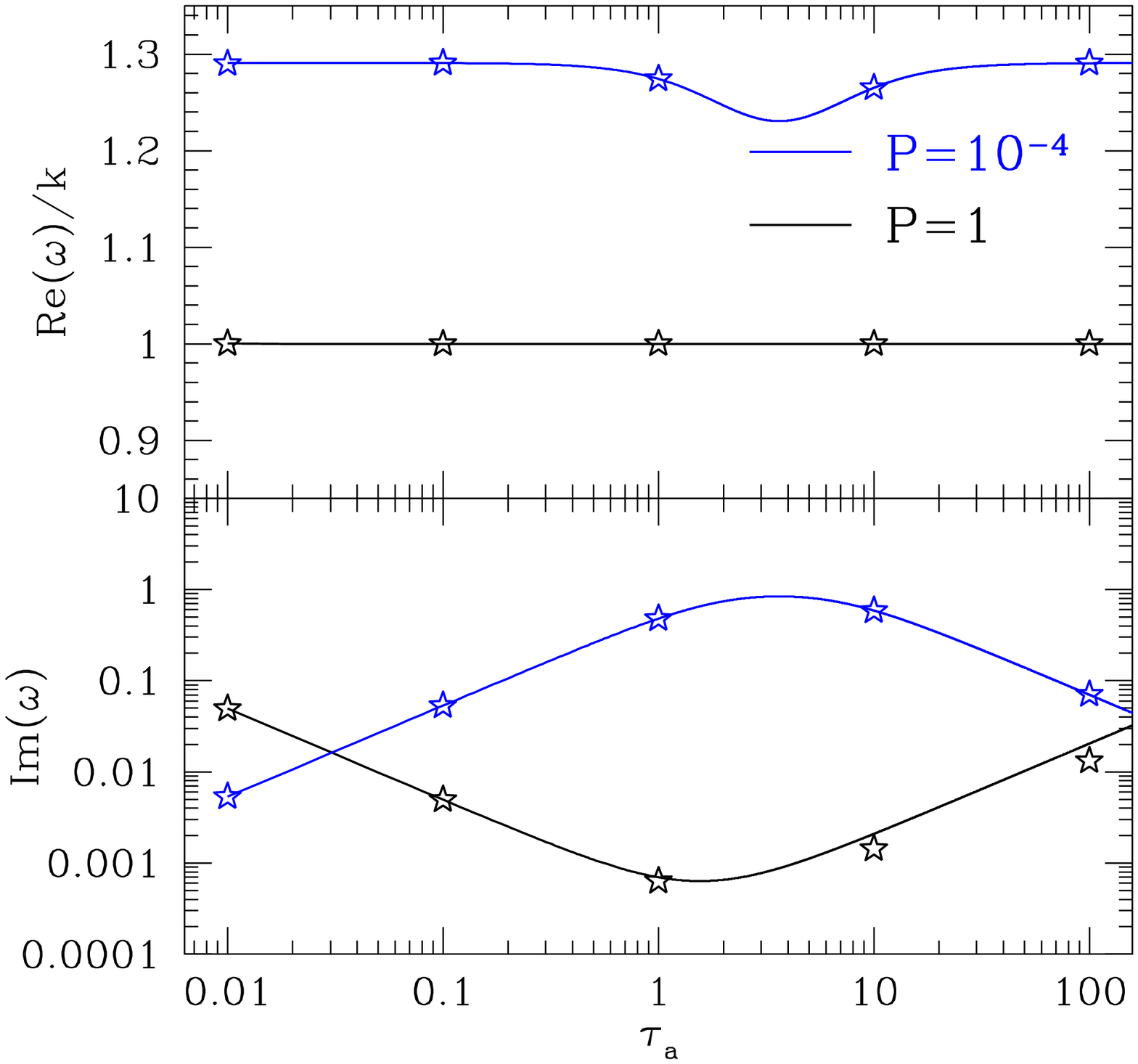}
\caption{Phase velocity scaled to the isothermal sound speed
(top panel) and damping rate (lower panel) of linear acoustic waves
versus the optical depth per wavelength for two
different radiation pressures.  The stars represent quantaties measured from
simulations, while the lines come from the solution to the linear dispersion
relation (equation \ref{dispersionEqn0}).  All simulations are performed in
1D with 512 grid points per wavelength.
}
\label{FigLinearwave1}
\end{figure}

Figure \ref{FigLinearwave1} compares the numerically measured phase
velocity and damping rate (shown as stars) for two different radiation
pressures, and over a wide range of optical depths $\tau_a$, in
comparison to the solution of the linear dispersion relation (equation
\ref{dispersionEqn0}) shown as solid lines.  The parameters are chosen
such that for the runs with $\Prat = 10^{-4}$, the dimensionless number
$\Prat\Crat\tau_a$ spans $10^{-2} \le \Prat\Crat\tau_a \le 10^{2}$, while
for the runs with $\Prat = 1$ both of these limits are a factor of $10^{4}$
larger.  Based on the discussion above, for $\Prat = 10^{-4}$ we expect the
damping rate to increase until $\tau_a \sim 1$ and decrease thereafter,
while for $\Prat = 1$ we expect the opposite.  The solid
lines from the analytic solution of the dispersion relation clearly show
these trends.  In addition, the numerically measured phase velocity 
agrees with the analytic results over the
entire range of optical depths for both values of the radiation pressure, 
while the damping rate is accurate in all cases except when $\Prat=1$ 
and $\tau_a\geq 1$. However, for these parameter values, the damping 
rate is small and the numerical measured damping rates are not converged 
at the resolution used for this plot. We have confirmed that if we increase 
the resolution, the numerical measured damping rate converges to the 
analytic values. At the same time, we have also found when $\tau_a\geq 100$, 
our algorithm no longer reproduces the correct damping rate. We discuss this further, 
and suggest a solution, below.
When $\Prat=10^{-4}$, Boltzmann number $\text{Bo}=10$ and it is $10^{-3}$ 
when $\Prat=1$. In the case $\Prat=100$, $\text{Bo}$ is only to $10^{-5}$.

\begin{figure}[htp]
\centering
\includegraphics[width=0.48\hsize]{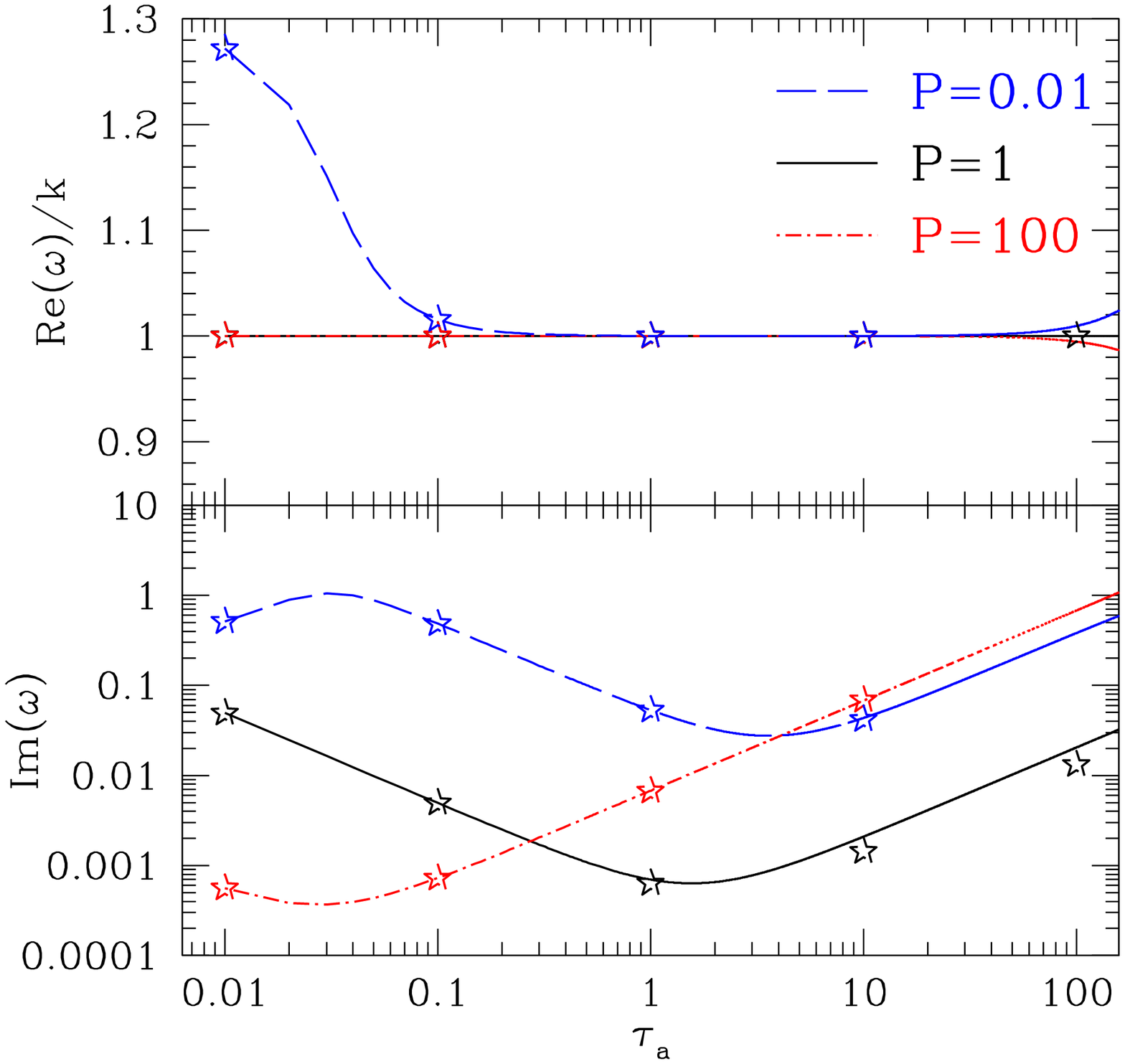}
\includegraphics[width=0.48\hsize]{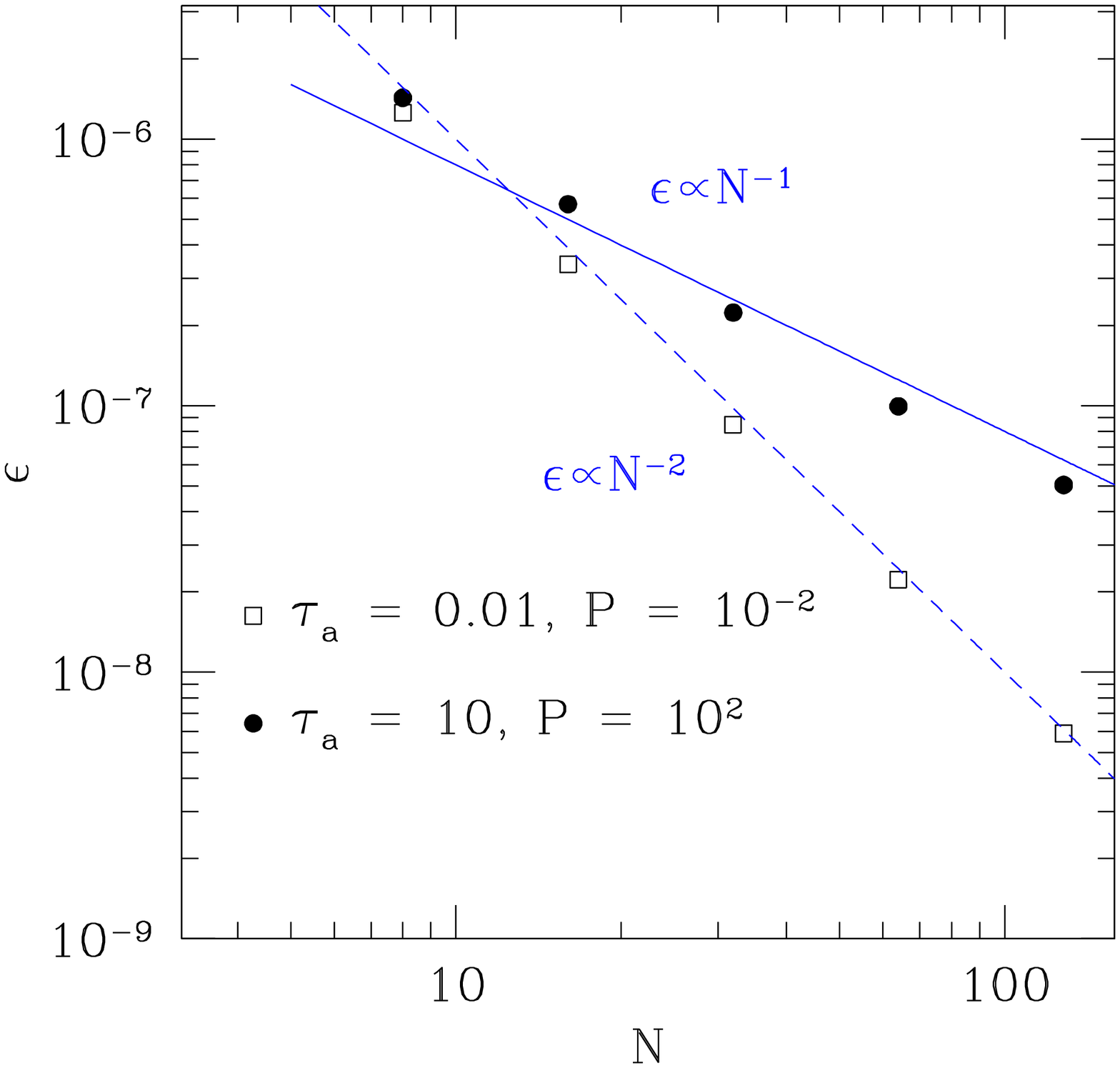}
\caption{{\em Left.} Phase velocity scaled to the isothermal sound speed
(top panel) and damping rate (lower panel) of linear acoustic waves
versus the optical depth per wavelength for three
different radiation pressures.  The stars represent quantities measured from 
simulations, while the lines come from the solution to the linear dispersion
relation (appendix \ref{Radhydrowave}).  All simulations are performed in 1D with 512 grid
points per wavelength.  Numerically measured values of both
quantaties agrees with linear theory when the absorption optical depth 
per wavelength is smaller than $\sim 100$.
{\em Right.} 
Convergence of L1 error with
resolution for two different sets of parameters.  Both simulations
are in full 3D.  The radiation dominated case converges at first order,
while the gas pressure dominated case converges at second order
as expected (see text).
}
\label{FigLinearwave2}
\end{figure}

In Figure
\ref{FigLinearwave2} we test linear wave propagation over a wide
range of radiation pressures.  Once again, the left panel compares the numerically
measured phase velocity and damping rate (shown as stars) for three
different radiation pressures, and over a wide range of optical depths
$\tau_a$, in comparison to the solution of the linear dispersion relation
(equation \ref{dispersionEqn0}) shown as solid lines.
Even though the properties of radiation
modified linear acoustic waves vary dramatically over this range of
optical depths and pressure ratios, there 
is good agreement between the numerical and analytic solutions, 
except when $\Prat=1$ and $\tau_a>1$. For these parameters, 
we also find at higher resolution, our algorithm converges to the analytic 
damping rate, suggesting that the discrepancy is dominated by numerical 
diffusion at this resolution.

Note that in Figure \ref{FigLinearwave2} we do not plot the numerical
solution for $\tau=100$ and $\Prat=100$ or $0.01$. We have found the
damping rate of linear waves stops converging for these parameter values.
Our tests reveal that the problem is associated with our implicit solution
of the radiation subsystem at a timestep determined by the adiabatic
sound speed in the material.  In this regime, an accurate solution
require a timestep which resolves the light crossing time across a cell.
Note we find this is only an issue for very large absorption opacity:
large scattering opacities do not couple the radiation and material energy
densities.  When the optical depth per wavelength due to the absorption
opacity becomes so large ($\Prat\tau_a\sim\Crat$) that the radiation
field and the material can be treated like a single fluid, we have found
it is more accurate to solve a single system of conservations laws for
the total (radiation plus material) energy and momentum, rather than
splitting the radiation subsystem from the material conservation laws.

As our next test, we measure the convergence rate of linear waves in 3D,
for two different values of the problem parameters.  We use a
computational domain of size $2L \times L \times L$, with the same
number of grid cells per unit length $L$ in each direction.  We initialize
the one dimensional solution inclined to the grid using the same
technique to compute the appropriate volume averages of the solution
as described in \citep[][]{Stoneetal2008}.  We propagate the wave for one
period, and then measure the L1 error in the solution from
\begin{equation}
\delta {\bf q} = \frac{1}{N} \sum_{i} \vert {\bf q}_{i} - {\bf q}_{i}^{0} \vert
\end{equation}
where ${\bf q}_{i}^{0}$ is the initial solution, and the sum is taken
over all grid points.
We repeat this calculation for a variety of different numerical
resolutions (grid cells per unit length $L$), and plot the change in
the L1 error with resolution.  The result is shown in the right panel of
Figure \ref{FigLinearwave2} for a radiation dominated, optically thick fluid
($\Prat=10^{2}$, $\tau_a=10$), denoted by the solid circles, and for a gas
pressure dominated, optically thin fluid ($\Prat=10^{-2}$, $\tau_a=10^{-2}$),
denoted by the open squares.  In the radiation dominated case the errors
converge close to first order, while in the gas pressure dominated case,
the errors converge close to second order.  This behavior is expected.
The errors in the former are dominated by the solution to the radiation
subsystem, which uses a first-order accurate backward Euler step for
stability.  Thus, we expect the overall rate of convergence should be
first order.  On the other hand, in the gas pressure dominated case, the
errors are dominated by the solution
to the material conservation laws, which uses a second-order accurate
modified Godunov step.  In this case, the overall rate of convergence should
be second order.  Of course, it might be possible to increase the
rate of convergence in the implicit solution of the radiation moment
equations by adopting higher order implicit differencing.  However, limiting
such methods to enforce monotonicity can be problematic (e.g., SS10).  We prefer
to adopt unconditionally stable first-order methods for the implicit
solver.

Linear wave solutions when only radiation energy source 
terms are added with radiation momentum source terms neglected are 
also given by \cite{Davisetal2012} and used to test our 
radiation transfer module (see Figure 9 of that paper). 
Because the radiation energy source term is added explicitly, 
the radiation diffusion mode needs to be resolved to make the code 
stable, which can limit the time step significantly when $\text{Bo}\lesssim 1$.  
Therefore the algorithm of \cite{Davisetal2012} is most suitable for the regime 
$\text{Bo}>0.01$. Because we include 
all the radiation momentum and energy source terms, 
the dispersion relations given in this paper differ 
from Figure 8 of \cite{Davisetal2012}. 
However, in regimes of parameter spaces which overlap, we do get very 
similar results, for example when $\Prat=10^{-4}$ 
and $\text{Bo}=10$ shown in Figure \ref{FigLinearwave1}.

\subsubsection{Linear Waves in radiation MHD}

Next, in order to test the accuracy of our MHD algorithms with radiation,
we consider the propagation of linear modes of radiation modified
magnetosonic waves.  We do not consider Alfv\'{e}n waves in this
subsection since they are incompressible and are affected less by radiation
than magnetosonic modes.  In the case of radiation modified MHD waves,
the transverse components of the velocity must be included even for
1D solutions.   However, as implemented our code does not include
the transverse velocities in 1D during the implicit solution of the
radiation moment equations.  Thus, all of the MHD tests presented here have
been performed in 2D using a grid of either $512 \times 64$, or
in some cases $1024 \times 64$ cells.

For these MHD tests, we use a uniform, homogeneous medium with the
hydrodynamic and radiation variables set to the identical values as
were used for the hydrodynamic test (see the previous subsection).
The strength of the magnetic field is $B_{0} = \sqrt{5/3}$ (which gives
a ratio of the Alfv\'{e}n to sound speed of 2).  The direction of the
magnetic field is 45 degrees to the $x-$axis, and it is confined to the
$x-y$ plane (so $B_{z}=0$).  We use a 2D domain of size $L \times L$ with
$L=1$ with periodic boundary conditions for all variables at each edge.
As before, we initialize a wave solution (Appendix \ref{RadMHDwave})
by adding an eigenmode (see Tables 2 or 3) propagating in the $x-$
direction with a wavelength equal to then size of the domain $L$,
and an amplitude of $10^{-6}$.  Once again, we perform a series of
calculations in which we study the effect of varying the radiation
to gas pressure ratio by varying the dimensionless pressure $\Prat$,
and the effect of varying the optical depth per wavelength $\tau_a =
\sigma_a L$ by varying the absorption opacity $\sigma_a$.

\begin{figure}[htp]
\centering
\includegraphics[width=0.48\hsize]{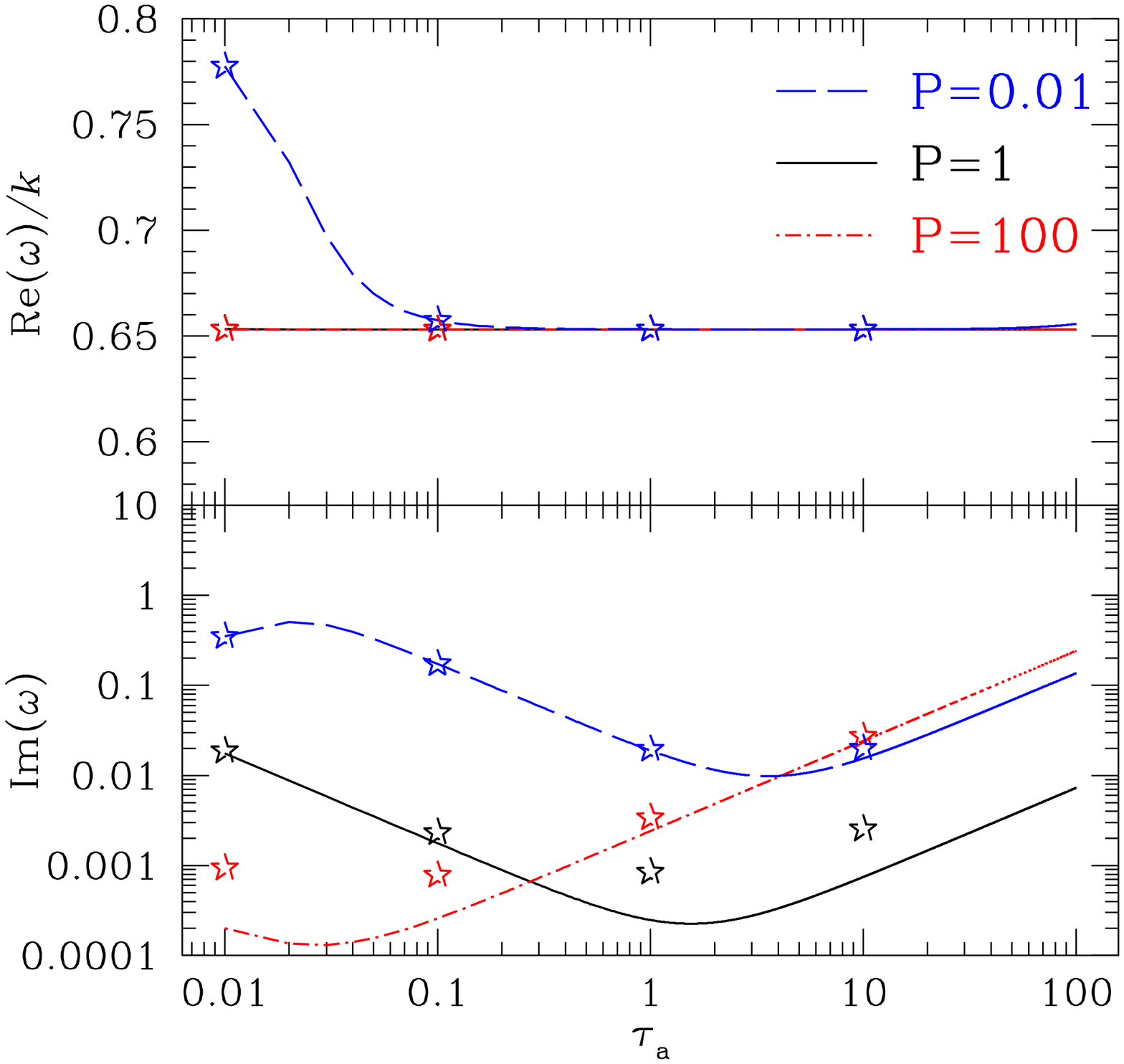}
\includegraphics[width=0.48\hsize]{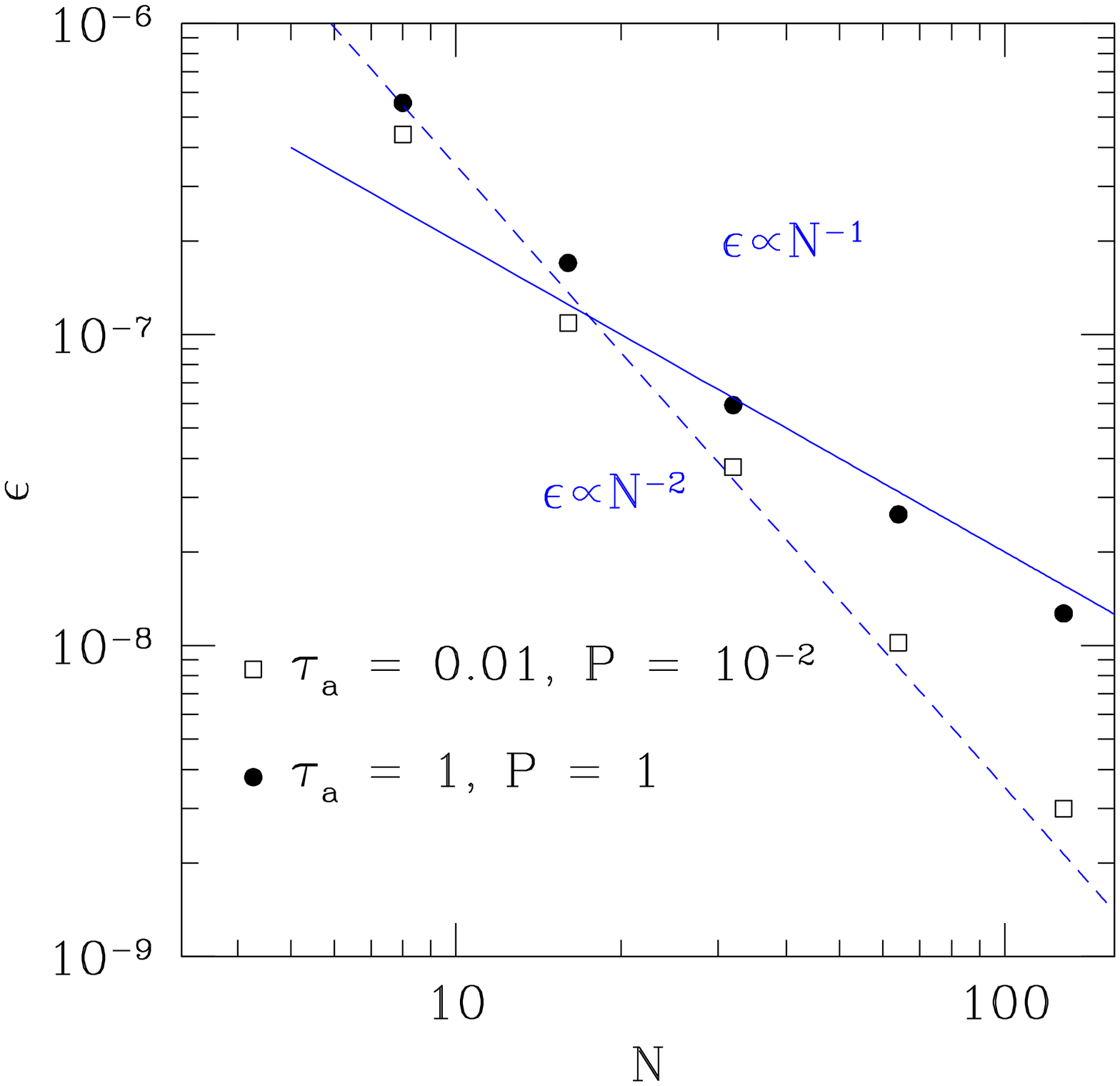}
\caption{Phase velocity scaled to the isothermal sound speed
(top panel) and damping rate (lower panel) of linear slow magnetosonic waves
versus the optical depth per wavelength for three
different radiation pressures.  The stars represent quantaties measured from
simulations, while the lines come from the solution to the linear dispersion
relation (appendix \ref{RadMHDwave}).  All simulations use an effective resolutions
of 512 grid points per wavelength
(except for the two runs with $P=100$ and
$\tau_a < 1$, which use 1024 grid points per wavelength).
Numerically measured values of both
quantaties agree with linear theory over the entire
range of parameter space, except for very low damping rates, where numerical
diffusion dominates.
{\em Right.}
Convergence of L1 error with
resolution for two different sets of parameters.  Both simulations
are in full 3D.  The radiation dominated case converges at first order,
while the gas pressure dominated case converges at second order,
as in the case of hydrodynmaics.}
\label{MHDwave_slow}
\end{figure}

\begin{figure}[htp]
\centering
\includegraphics[width=0.48\hsize]{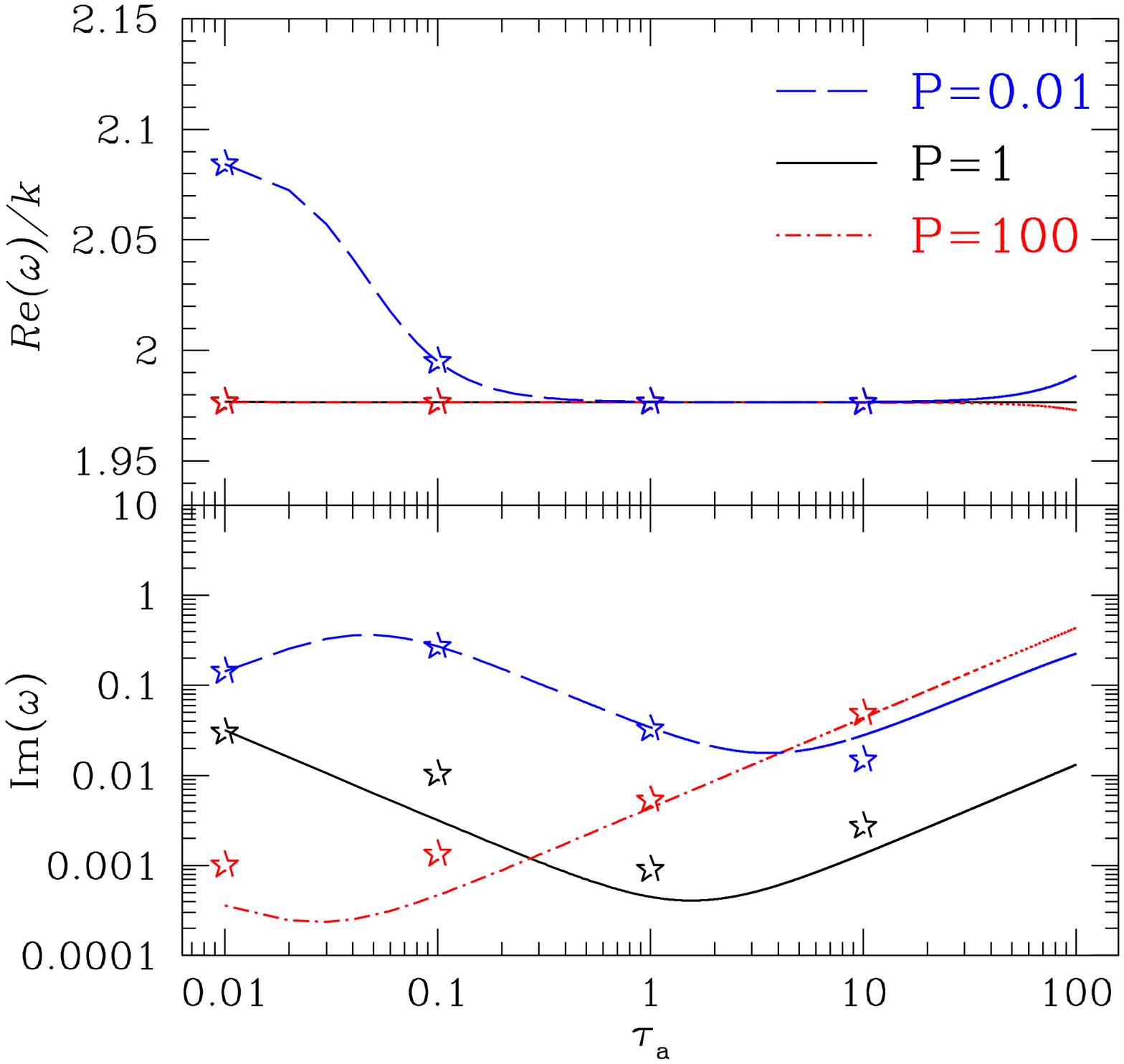}
\includegraphics[width=0.48\hsize]{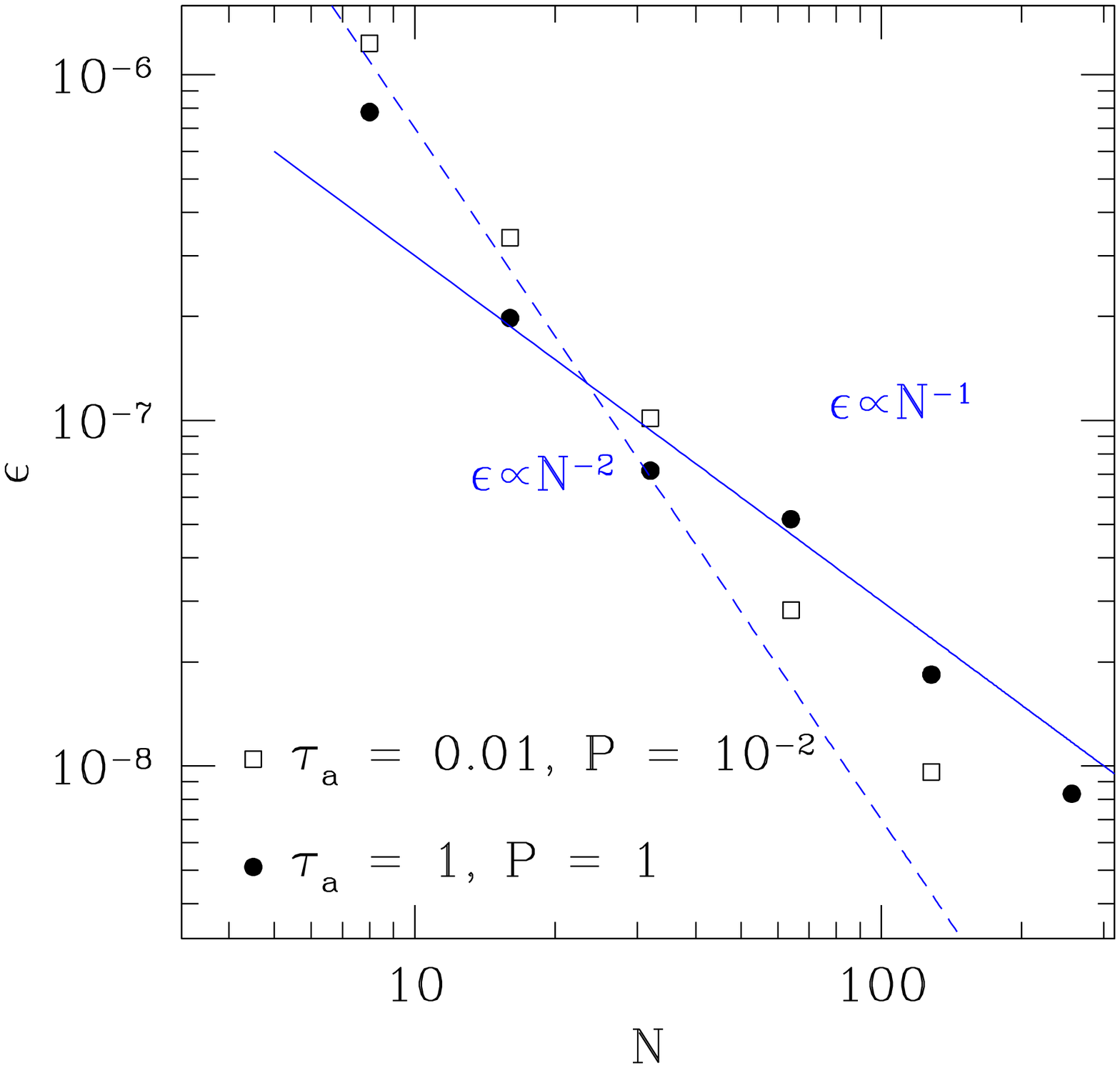}
\caption{Same as figure \ref{MHDwave_slow}, but for fast magnetosonic waves.}
\label{MHDwave_fast}
\end{figure}

The results of our tests for slow magnetosonic waves are shown
in Figure \ref{MHDwave_slow}, while the results for fast magnetosonic
wave are shown in Figure \ref{MHDwave_fast}.
The left panel in each figure compares the
numerically measured values of the phase velocity and damping rate
with linear theory.
Note that the solution
to the dispersion relation as a function of
optical depth and $\Prat$ for both magnetosonic modes is very
similar to that for radiation modified acoustic waves in hydrodynamics
(see Figure \ref{MHDwave_fast}).   Once again, we find good agreement
between our numerical values for the phase velocity versus linear theory 
over this parameter regime.  The damping rate also shows good agreement, 
except when $\Prat=100, \tau_a <1$ and $\Prat=1, \tau_a>0.1$. In these cases, 
the damping rate is so small that our numerical measured values are not 
converged at this resolution. Nevertheless, as Figure \ref{MHDwave_fast} 
demonstrates, our algorithm converges to the correct solution for $\Prat=1, \tau_a=1$, 
albeit at first order. We have confirmed our algorithm also converges for the other 
parameters shown in this figure as well.

The right panels of Figures
\ref{MHDwave_slow} and \ref{MHDwave_fast} show the convergence rate of the
L1 error with numerical resolution for the slow and fast magnetosonic waves
respectively in a fully 3D domain of size $2L \times L \times L$.
As before, we measure first-order convergence in a parameter regime where
the solution to the radiation moment equations dominates the error, and
second-order convergence in the opposite limit.

\subsection{Radiative shocks in the non-equilibrium diffusion limit}
\label{sec:RadShock}

Shock tubes have long been used as a test of hydrodynamic and MHD
codes in the nonlinear regime, since the structure of a planar shock
can be computed analytically in these cases.  However, in radiation
hydrodynamics and MHD, shock structure is much more complicated, and
is difficult to compute analytically.
At low Mach numbers $\mathcal{M}$, radiation can diffuse upstream of the shock,
heating the gas and forming a smooth precursor in the temperature.
As $\mathcal{M}$ increases, the gas temperature near the shock front
can begin to exceed the downstream value, forming a Zel'dovich spike
\citep[e.g.,][]{ZeldovichRaizer1967,MihalasMihalas1984}.  This spike is
followed by a relaxation region where the gas temperature
cools to its far downstream value.
When a Zel'dovich spike is formed, if the
downstream gas temperature (after the relaxation region) is larger
than the value immediately upstream of the shock (at the end of the
precursor region), the shock is called subcritical.  On the other hand,
if the downstream gas temperature after the relaxation region is the
same as the upstream gas temperature, the shock is termed supercritical.
Subcritical shocks are formed at lower  $\mathcal{M}$ than supercritical
shocks.  Finally, as $\mathcal{M}$ is increased further, the downstream
radiation pressure can exceed the gas pressure, the Zel'dovich spike
disappears, and the solution becomes everywhere smooth again.

Although many authors have presented numerical solutions to radiating
shocks as a test of their algorithms, the lack of analytic solutions
inhibits quantitive testing.
Recently \cite{LowrieEdwards2008} have studied the structure of
radiation modified shocks in the non-equilibrium diffusion limit as a
function of Mach number.
They give a very clear explanation of how these structures can be
computed by combining smooth solutions to ordinary differential equations
with discontinuous jumps determined by the Rankine-Hugoniot relations when
needed.  These solutions not only provide interesting insight into
the structure of radiating shocks, but also provide an quantative
test of time-dependent radiation hydrodynamic codes.

To test our algorithms, we follow the
semi-analytic method described in \cite{LowrieEdwards2008} (hereafter LE) 
to calculate
the shock structure at a variety of Mach numbers, using
a non-dimensional pre-shock solution of $\rho_{0}=T_{0}=E_{r,0}=1$ and
$v_{0}=\mathcal{M}$ in the upstream state.  We then initialize
this solution on a 1D grid by volume averaging each conserved variable
to our numerical mesh.   The LE solutions are given in the co-moving
frame, thus we transform the co-moving radiation flux and energy
density to the Eulerian frame, using  equation \ref{Comoving}.
In order to make our solutions match those presented in LE, we
use a dimensionless speed of light $\Crat=1.732\times10^3$ and
pressure $\Prat=10^{-4}$ and absorption and scattering opacities
of $\sigma_a=\sigma_t=577.4$.
The gas temperature $T$ is calculated as $T=P/(0.6\rho)$ and
$\gamma=5/3$.  The Eddington approximation is used so $P_r=E_r/3$.  We use
a computational domain of size $L$, where $L$ is large enough to capture the
upstream precursor and downstream radiative relaxation regions (the size
of these regions are given by the LE solution itself), with a grid of
$1024$ cells in all cases (except $\mathcal{M}=50$ where we we use $2048$
cells).  Input boundary conditions at the upstream values are used on
the left, and outflow boundary conditions are used on the right, with
the shock propagating in the negative $x$-direction.  We then let the code
evolve this solution for several flow crossing times, $t_f = L/\mathcal{M}$.
Ideally, the solution should remain stationary
on the mesh.  A small shift in the position of the shock front can be
expected due to truncation error in the averaging of the initial
solution to the hydrodynamic grid.  A serious error in the algorithm
or its implementation would be revealed if the code cannot hold the
input solution.

\begin{figure}[htp]
\includegraphics[width=0.98\hsize]{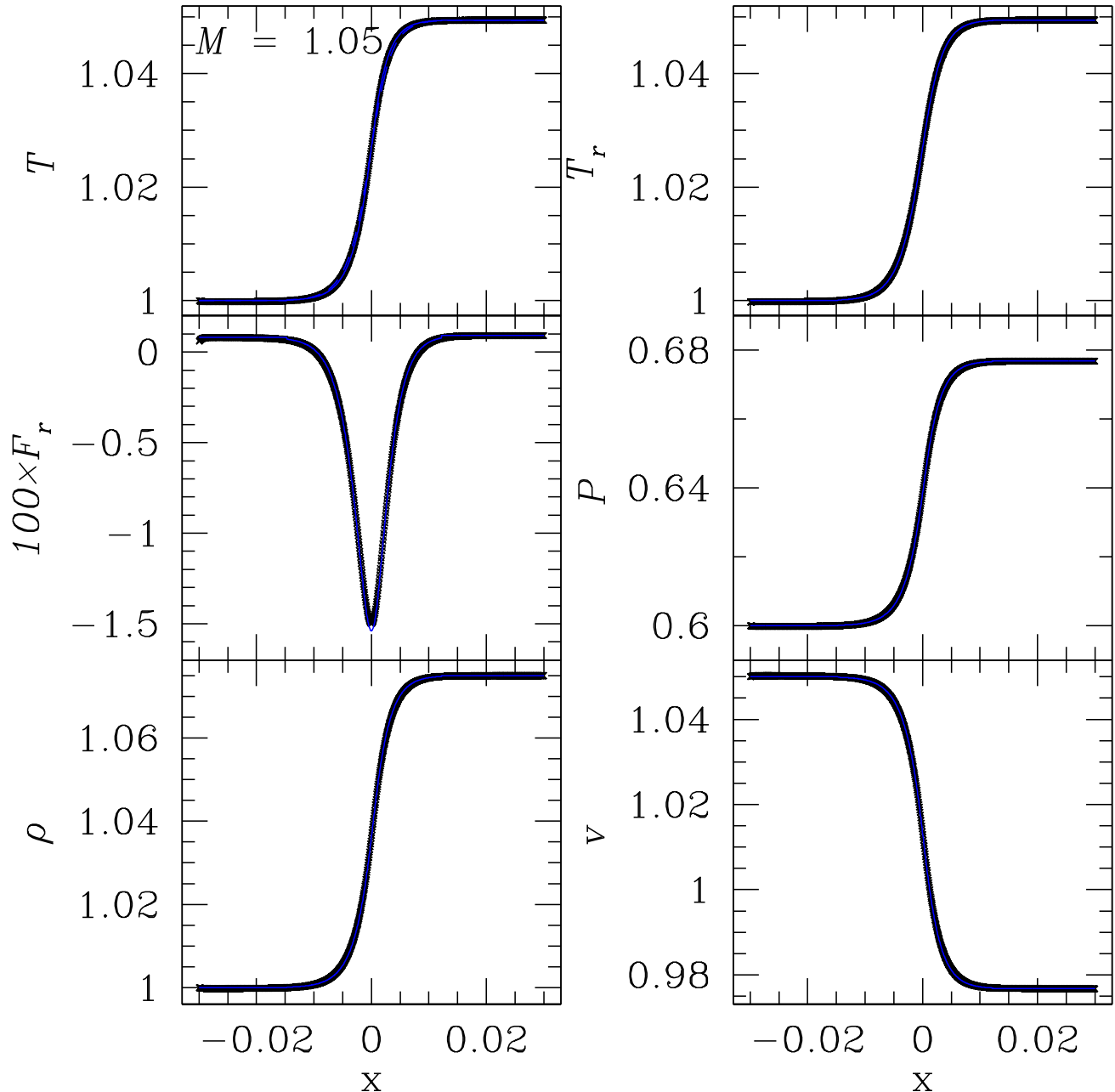}
\caption{Strucure of a radiation modified shock for Mach number
$\mathcal{M}=1.05$.
Blue lines are initial conditions, while black dots are numerical results 
after a few flow crossing times. 
}
\label{RadShock1}
\end{figure}

\begin{figure}[htp]
\includegraphics[width=0.98\hsize]{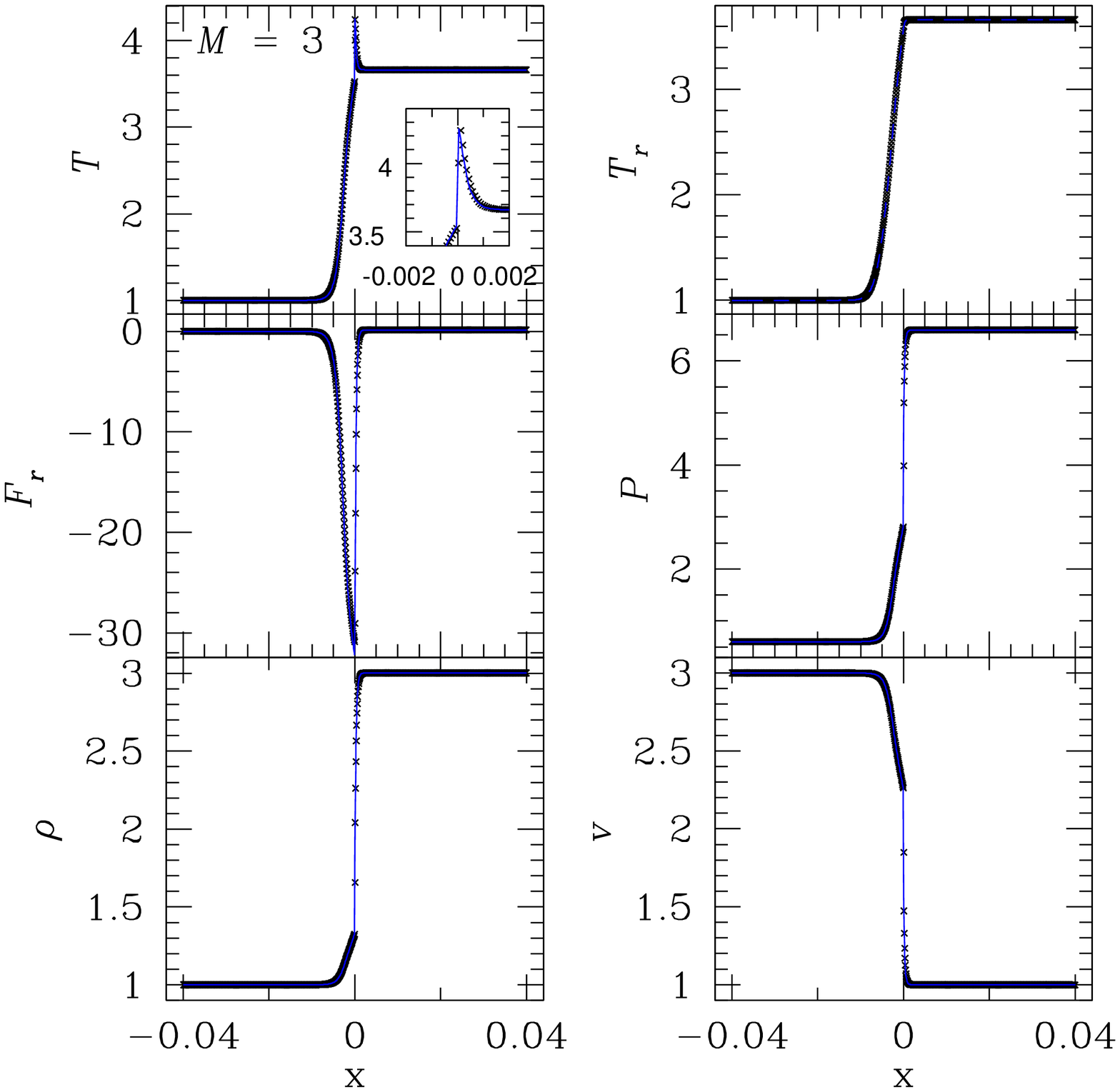}
\caption[]{Same as figure \ref{RadShock1}, but for $\mathcal{M}=3$.  At this
Mach number, a subcritical shock is formed.
Inset in the gas temperature panel shows details of the Zel'dovich spike.}
\label{RadShock2}
\end{figure}

\begin{figure}[htp]
\includegraphics[width=0.98\hsize]{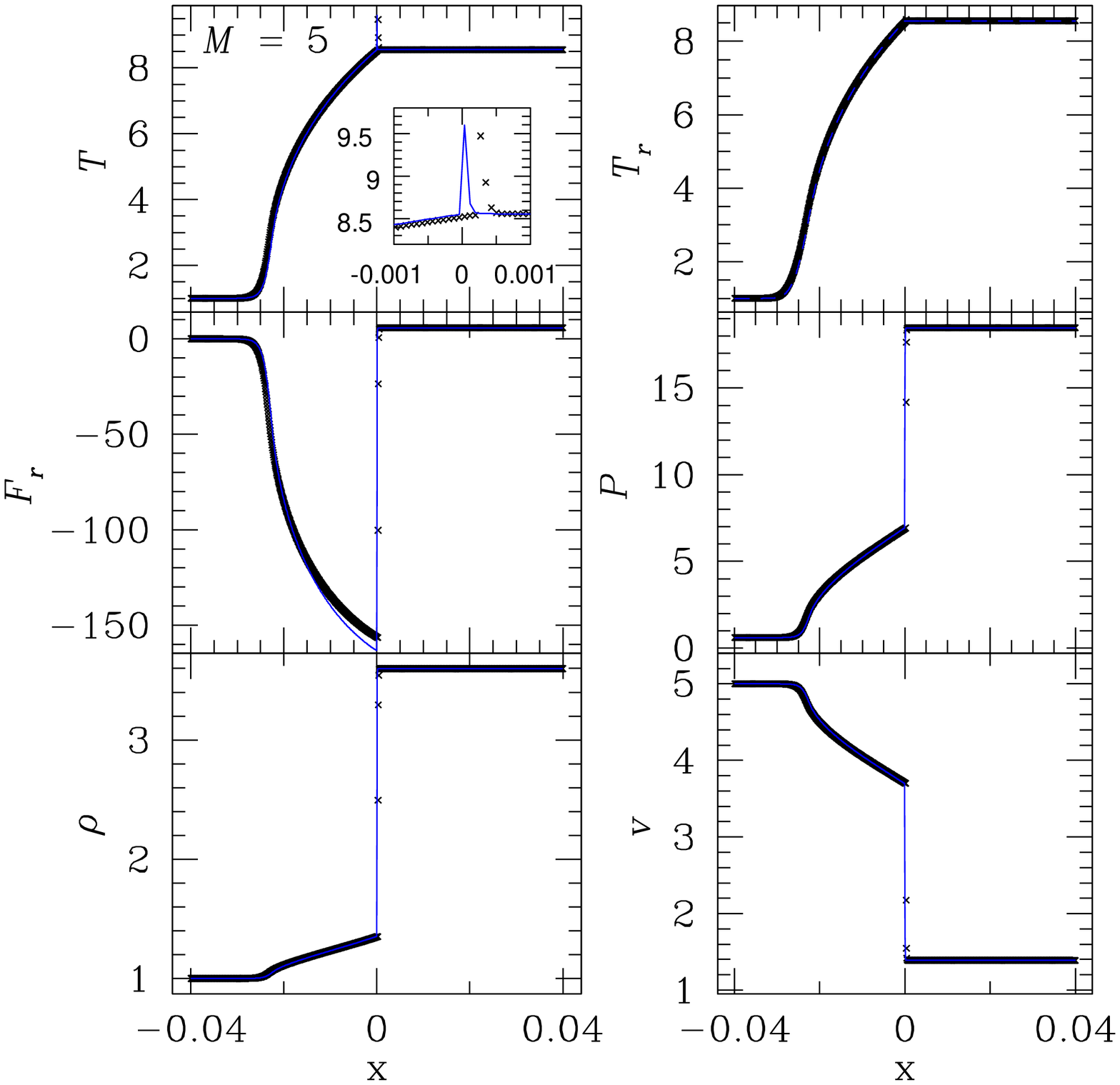}
\caption[]{Same as figure \ref{RadShock1}, but for $\mathcal{M}=5$.
At this Mach number, a supercritical shock is formed.
Inset in the gas temperature panel shows details of the Zel'dovich spike.}
\label{RadShock3}
\end{figure}

\begin{figure}[htp]
\includegraphics[width=0.98\hsize]{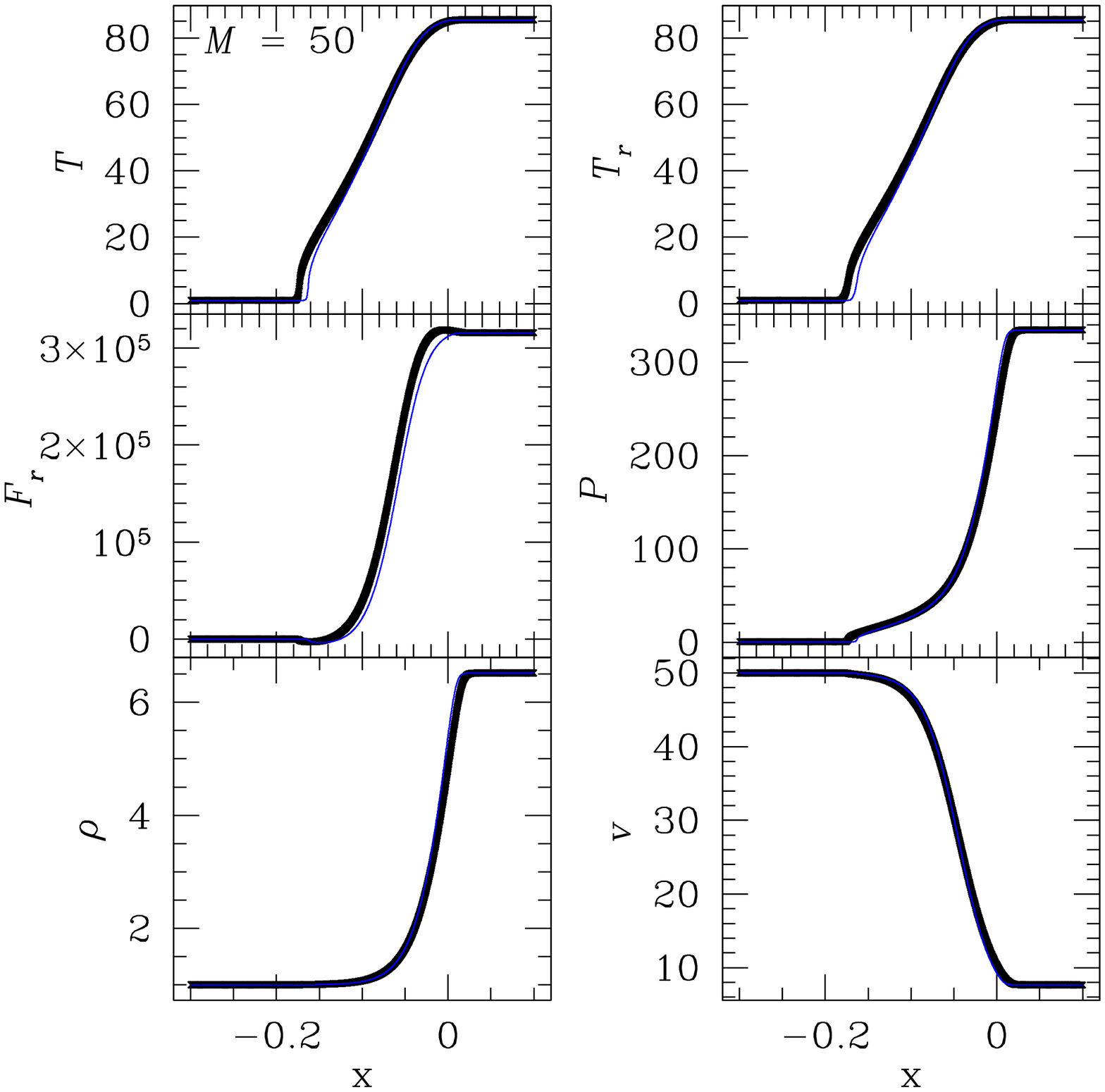}
\caption[]{Same as figure \ref{RadShock1}, but for $\mathcal{M}=50$.
At this Mach number, the downstream flow is radiation pressure dominated.}
\label{RadShock4}
\end{figure}

We begin by presenting our numerical solution
for $\mathcal{M}=1.05$ in Figure
\ref{RadShock1}.  In this case there is no discontinuous jump in
any variable, but rather only a smooth precursor.  Next, Figure
\ref{RadShock2} shows the solution for $\mathcal{M}=3$.  Now a
discontinous shock jump has appeared, along with a Zel'dovich spike.
For these parameters, this shock is subcritical.  The inset
panel in the plot of gas temperature shows a blow-up of the region
near the spike.  The numerical solution shows agreement with
the semi-analytic solution of LE.   Next, Figure \ref{RadShock3} shows the
solution for $\mathcal{M}=5$.  Now there is a discontinuous jump in some
variables (such as density and velocity), but the gas temperature is
continuous apart from the Zel'dovich spike.  This solution corresponds
to a supercritical shock.  Again, the blow-up of the gas temperature
near the spike region reveals agreement between the numerical
and semi-analytic solutions.  The position of the spike has moved a
few cell widths, but has quickly relaxed to a
steady structure at the correct shock speed.
Finally, Figure \ref{RadShock4} shows the solution
for $\mathcal{M}=50$.  In this case, the radiation pressure is about $10$
times gas pressure in the downstream flow.  Now all variables are smooth,
with no jumps.

\begin{figure}[htp]
\includegraphics[width=0.98\hsize]{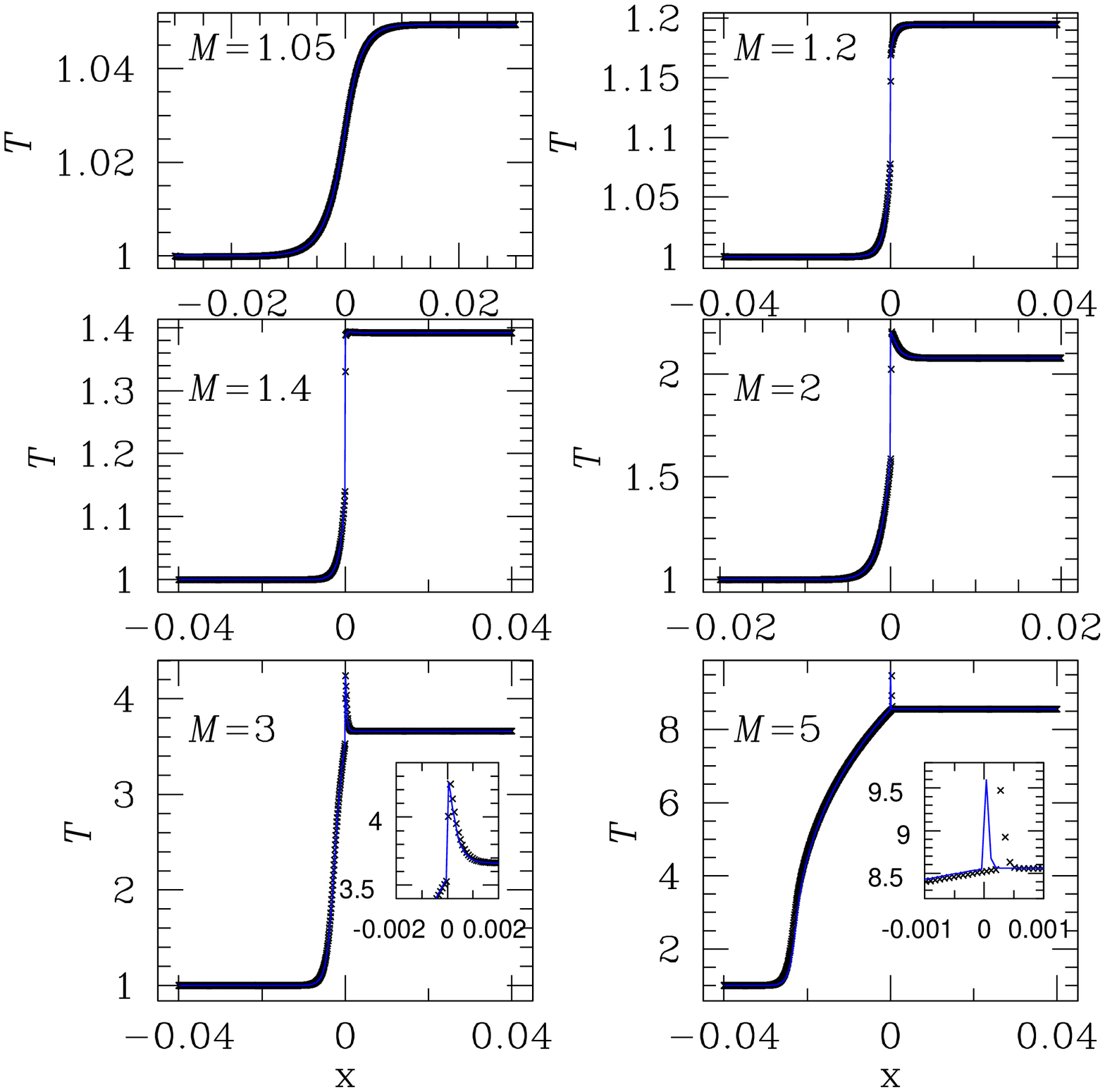}
\caption{Gas temperature profiles with different Mach number for radiation
modified shocks. Blue 
lines are the initial conditions, while black dots are the numerical solution
after a few flow crossing times.   Note the subtle changes in the
shock structure between $\mathcal{M}=1$ and 2, which are clearly captured
in the numerical solutions. }
\label{RadShockT}
\end{figure}

LE have shown that the gas temperature profile near the shock front
is very sensitive to small changes in $\mathcal{M}$ when $\mathcal{M}$
is small.  To explore whether our algorithms can capture these subtle
changes, Figure \ref{RadShockT} plots the gas temperature for six
Mach numbers between $\mathcal{M}=1.05$ and 5.  The values are chosen
to correspond to figures 4 through 11 of LE, where these changes are
noted and discussed.  This region covers the formation of a Zel'dovich
spike, and the increase in the amplitude of the spike with increasing
$\mathcal{M}$.   It is clear from the figure, in which the numerical
solution is compared to the corresponding solution from LE, that our
algorithm captures each phase accurately.

Figure \ref{RadShockT} shows that our algorithm captures the transition
from subcritical to supercritical shocks accurately at low $\mathcal{M}$,
while figure \ref{RadShock4} shows the solution with $\mathcal{M}=50$
(in the stongly radiation pressure dominated regime) is also captured accurately.
However, we have found solutions at $\mathcal{M} \approx 30$
are inaccurate at the resolutions we use.  As the Mach number increases
beyond $\mathcal{M}=5$, the width if the Zel'dovich spike compared
to the width of the precursor region gets smaller, until the spike
disappears in the radiation pressure dominated case near $\mathcal{M}=50$.
We find that unless the grid resolution is smaller than the thickness
of the Zel'dovich spike, our algorithm will not hold a steady solution.
For example, at $\mathcal{M}=30$ this would require a resolution of about $10^6$
cells for a uniform grid.  Clearly, in this case either static or adaptive
mesh refinement would be useful to resolve the spike.  With enough resolution
to resolve the spike, our algorithms provide an accurate solution in this regime.

\subsection{Radiative shocks with a variable Eddington tensor}

Calculating the structure of radiation modified shocks without invoking
the Eddington approximation or other simplifying assumptions is extremely
challenging.  \cite{Sincelletal1999} have used a time dependent radiation
hydrodynamics code
to compute the structure of radiation modified shocks.  We have found that
comparison of our solutions in this regime to the results reported by
\cite{Sincelletal1999} is a useful code test.

In many ways, the numerical algorithm used by \cite{Sincelletal1999}
is similar to that described here.  They solve the radiation moment
equations using a variable Eddington factor computed from a formal
solution of the transfer equation, assuming LTE and gray opacities.
However, the underlying hydrodynamic solver they use is quite different from
that adopted in this work.  \cite{Sincelletal1999} use methods based
on artificial viscosity for shock capturing, and solve the internal
rather than total gas energy equation.  The Godunov methods used here
use Riemann solvers rather than artificial viscosity for shock capturing,
and are based on the total gas energy equation.

Since the Mach number $\mathcal{M}$ is not known independently of the
shock structure, it is not possible to compute the solution in the frame
of the shock.  Instead, we initialize a flow that generates a shock,
and allow the shock to propagate a large enough distance to settle into
a steady structure.  To generate the shock, we collide two symmetric
flows at the center of the domain, producing two identical shocks that
propagate symmetrically away from the center.  This avoids having to
implement reflecting boundary conditions in the radiative transfer
solver that is used to compute the VET \citep[][]{Davisetal2012}
necessary if the shock is generated by reflection of the flow off a wall.
The initial conditions for the flow are chosen to match those used by
\cite{Sincelletal1999} (see also SS10) for the subcritical shock solution
discussed in \S2.1 of their paper.
We use a computational domain that spans $-2 \le x \le 2$.  Initially,
the density is one, the gas pressure is $P=0.923\rho T$, and the
gas temperature is $T=1+7.5 |x|/2$.  For $x<0$, the flow velocity is
$v=20$, while for $x>0$ it is $v=-20$.  The dimensionless pressure
and speed of light are $\Prat=1.08\times10^{-10}$ and $\Crat=10^6$,
and $\gamma=5/3$.
The radiation temperature is the same as gas temperature, the radiation
flux is $\bF_r=-(\partial E_r/\partial x)/(3\sigma_a)$ (the Eddington
approximation is assumed in the initial condition only because the density 
is uniform initially), where the
absorption opacity $\sigma_a=10$.  The VET is initialized with only
non-zero diagonal components of $1/3$; thereafter the VET is computed
self-consistently with the flow using short characteristics.

\begin{figure}[htp]
\centering
\includegraphics[width=0.98\hsize]{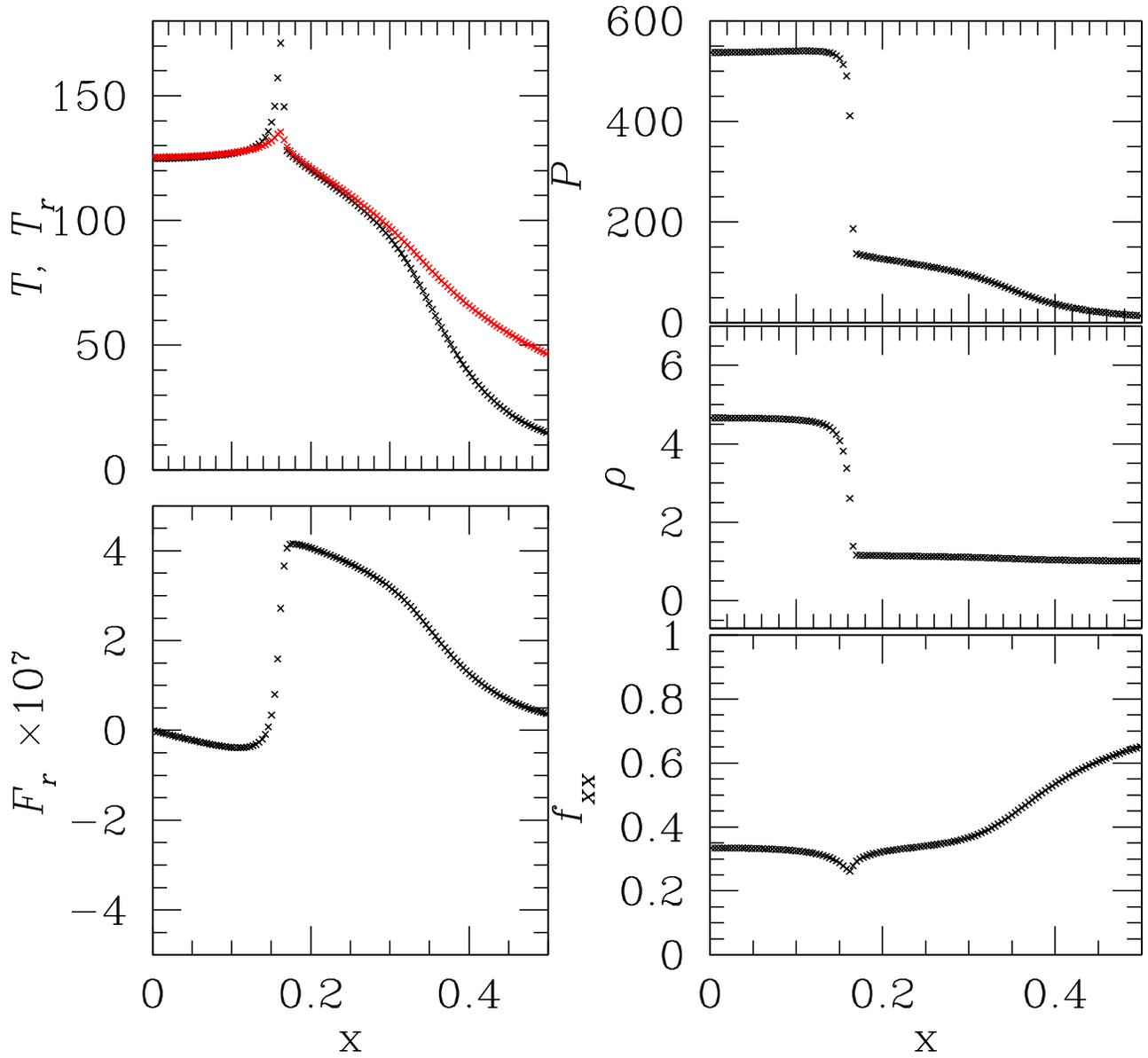}
\caption{Structure of a subcritical shock computed with a VET at time $t=0.03$
in one dimension with $1024$ cells. 
In the upper left panel, the red line is the radiation temperature $T_r$ while the black line is the gas temperature $T$.   Note that $x-x$ component of the
Eddington tensor (lower right panel) can be smaller than $1/3$ near the shock
front.  }
\label{ShockEdd}
\end{figure}

Profiles of various quantities at $t=0.03$
are shown in Figure \ref{ShockEdd} in the region $0 \le x \le 0.5$.
(We have checked, as another test of our code, that the solution is
exactly symmetric with respect to $x=0$.)
Profiles of each variable show the characteristic structure of a
subcritical shock, including a strong radiative precursor and Zel'dovich
spike.   Perhaps of greatest interest is the profile of the $x-x$
component of the VET shown in the lower right panel.
Upstream of the shock $f_{xx}$ is much bigger than 1/3, as expected.
Downstream of the shock, it is close to $1/3$.  However, in the vicinity
of the shock itself, $f_{xx}$ is smaller than $1/3$, in agreement
with the results of
\cite{Sincelletal1999} (see their figure 8).
The reason for the behavior
is discussed in \S2.3 of \cite{Sincelletal1999}.  Most of the upstream radiation
is generated within one optical depth of the shock front. Rays
parallel to the shock front go through a longer path of source than rays
perpendicular to the shock front.  Thus, the intensity along rays parallel
to the shock front is larger than those perpendicular, leading to a
value of $f_{xx}$ less than
$1/3$.  Note that since the empirical relations for most flux limiters \citep[e.g.,][]{LevermorePomraning1981}
bound the Eddington factor between $1/3$ and one along the direction of 
radiation energy density gradient, FLD cannot get the right answer in this case.

\begin{figure}[htp]
\centering
\includegraphics[width=0.98\hsize]{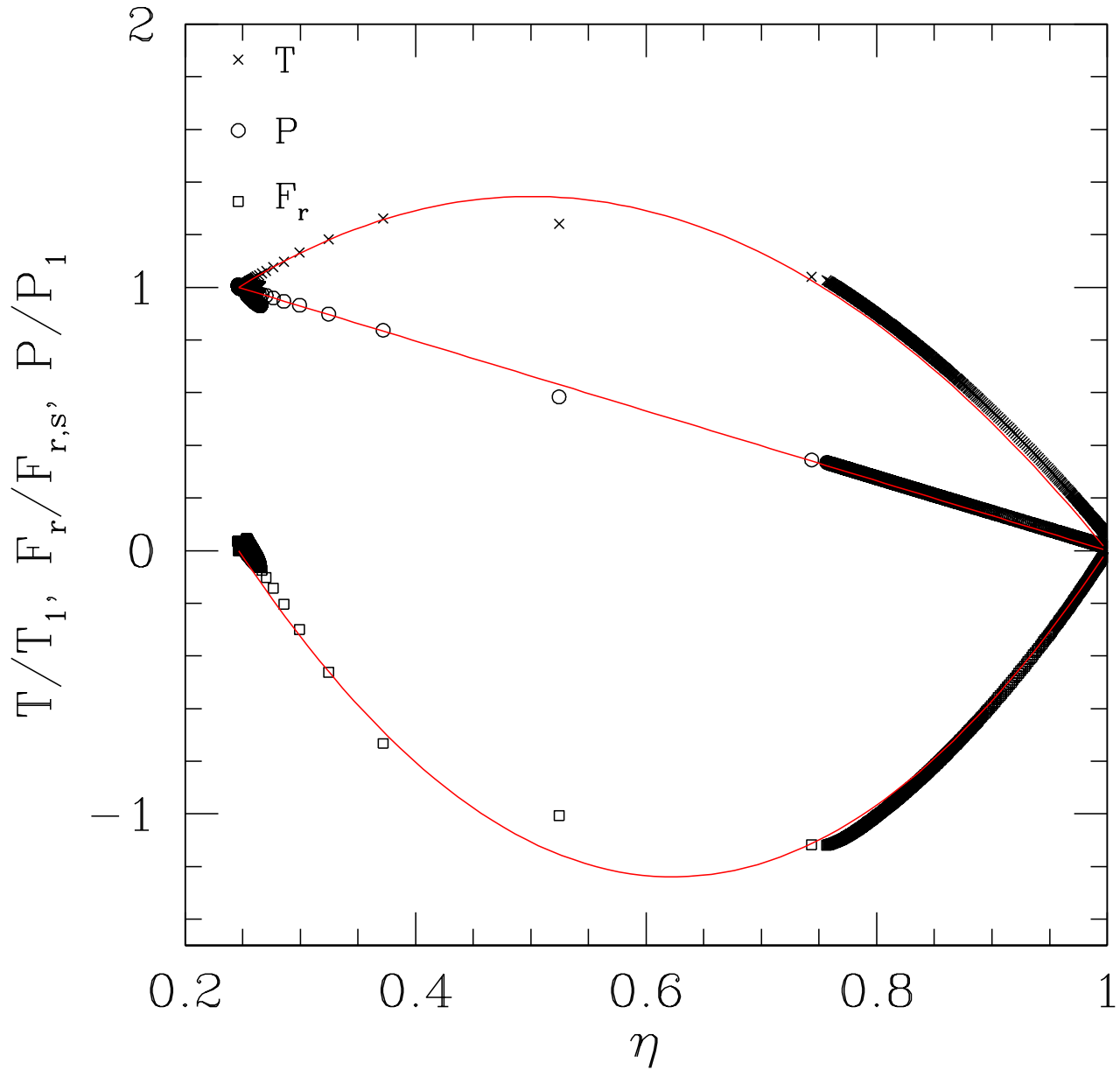}
\caption{Gas temperature, radiation flux, and gas pressure as a function
of inverse compression ratio $\eta\equiv \rho_0/\rho$, where $\rho_0$
is upstream density, for the subcritical radiative shock shown in Figure
\ref{ShockEdd} at time $0.3$. 
Temperature and pressure are scaled by the downstream temperature 
$T_1$ and pressure $P_1$ at $x=0$, while the radiative flux is scaled by the   
flux $F_{r,s}$ at the shock front. Note $F_r$ is the co-moving flux.
The red lines are 
analytic solutions given by \cite{ZeldovichRaizer1967}
(see also \citealt{Sincelletal1999}).}
\label{shockcompression}
\end{figure}

We find the most quantitative comparison to the previously published
results of \cite{Sincelletal1999} is to plot the gas temperature,
pressure, and radiative flux against the inverse compression ratio
$\eta\equiv \rho_0/\rho$, where $\rho_0$ is upstream density.
Figure \ref{shockcompression} shows our result at $t=0.3$, for direct
comparison to figures 2 and 3 in \cite{Sincelletal1999}.  An approximate
analytic expression for these quantities by \cite{ZeldovichRaizer1967}
(and also given in \S1.3 of \cite{Sincelletal1999}) is shown as solid
lines.  The numerical solution of \cite{Sincelletal1999}
is in fact in poor agreement with the analytic predictions, and these
authors argue this is due to approximations made in the latter.  It is
clear from Figure \ref{shockcompression} that our solutions agree very
well with the approximate analytic solution, and therefore do not agree
with the results of \cite{Sincelletal1999} for these
quantities.  Recall these authors use a
numerical method based on artificial viscosity for shock capturing, and
the internal rather than total energy equation.  The plot measures the
internal structure of the shock, which could be quite sensitive to both
the form of viscosity used, and the degree to which energy is conserved.
Our solution may be in better agreement with the analytic expectations
since our algorithm uses a Riemann solver rather than artificial viscosity
to capture shocks, and is based on the total energy equation.

\subsection{Photon Bubble Instability}

Photon bubble instability is an overstable mode present in
radiation supported, magnetized atmospheres.  It was first noticed
in numerical simulations of neutron star polar cap accretion flows
\citep{KleinArons1989}, and was subsequently confirmed in a linear
stability analysis for neutron star atmospheres \citep{Arons1992}.
More recently, \cite{Gammie1998} and \cite{BlaesSocrates2003} recognized a
similar instability occurs in accretion disk atmospheres, and analyzed the
properties in the linear regime.  Numerical simulations that investigate
the non-linear regime were performed by \cite{Turneretal2005} using
the FLD module in the ZEUS code \citep{TurnerStone2001}.   Since both the
linear and non-linear regimes of this instability have been studied in
detail in the literature, and since fundamentally it relies on radiation
modified MHD waves, it provides is an excellent test of our radiation
MHD algorithms in multidimensions.  In this section, we repeat one set
of parameters reported by \cite{Turneretal2005} and compare our results
with solutions from ZEUS.   In order to make direct the comparisons,
we adopt the Eddington approximation, so that the diagonal components
of the VET are fixed to be $1/3$.

The initial conditions consist of a stratified atmosphere in mechanical
and thermal equilibrium, in which a uniform radiation flux balances
gravity.  The structure of the atmosphere is computed from the
solution of the hydrostatic equilibrium equations \begin{eqnarray}
\frac{dP}{dy}&=&-\rho g+\sigma F_r,\ \ P=\rho T,\nonumber\\
\frac{1}{3}\frac{dE_r}{dy}&=&-\sigma F_r,\ \ E_r=T^4.  \end{eqnarray}
with a constant gravitational acceleration $g=1$ in the $-\hat{y}$
direction, and an opacity of $\sigma=137.11\rho+0.15\rho^2T^{-3.5}$
(which includes both electron scattering opacity and Kramer's opacity).
The co-moving vertical radiation flux $\hat{F}_{r,0}=2.52\times10^{-4}$.  These equations
are integrated in a domain of size $(0,2.75)\times(26.11, 28.86)$ in our dimensionless units, starting from the midplane $y=27.485$ where the dimensionless
pressure, temperature, density  and radiation energy density are all
set to $1$.  
In this test, the dimensionless parameter $\Prat=762$
(so the ratio between radiation pressure and gas pressure in the midplane
is $254$), and the dimensionless speed of light is $\Crat=7557.74$.
The initial conditions include a uniform, horizontal magnetic field in the
$+\hat{x}$ direction with magnetic pressure $10\%$ of mid-plane radiation
pressure.  Our simulations use a resolution of $256\times256$ grid points.
The instability is seeded with random perturbations of the density with
amplitude $10^{-3}$ at the midplane.  The initial condition for this
simulation is similar (but not identical, the amplitude of perturbation is $10^{-8}$ in 
\citealt{Turneretal2005}) 
to the initial condition used
for Figure 9 of \cite{Turneretal2005}.  This model is appropriate for
the surface layers of an accretion disk at 20 Schwarzschild radii from
a $10^8\msun$ black hole.  In this case, the middle of the simulation domain
corresponds to a location one scale height above the disk midplane, and
the simulation domain extends from $0.95H$ to $1.05H$.  The total optical
across the domain in the vertical direction is 364,
thus the Eddington approximation should be an
excellent description.  In our dimensionless units, one orbital period
is $6.28$ time units.

\begin{figure}[htp]
\centering
\vspace{-4cm}
\includegraphics[width=1.0\hsize]{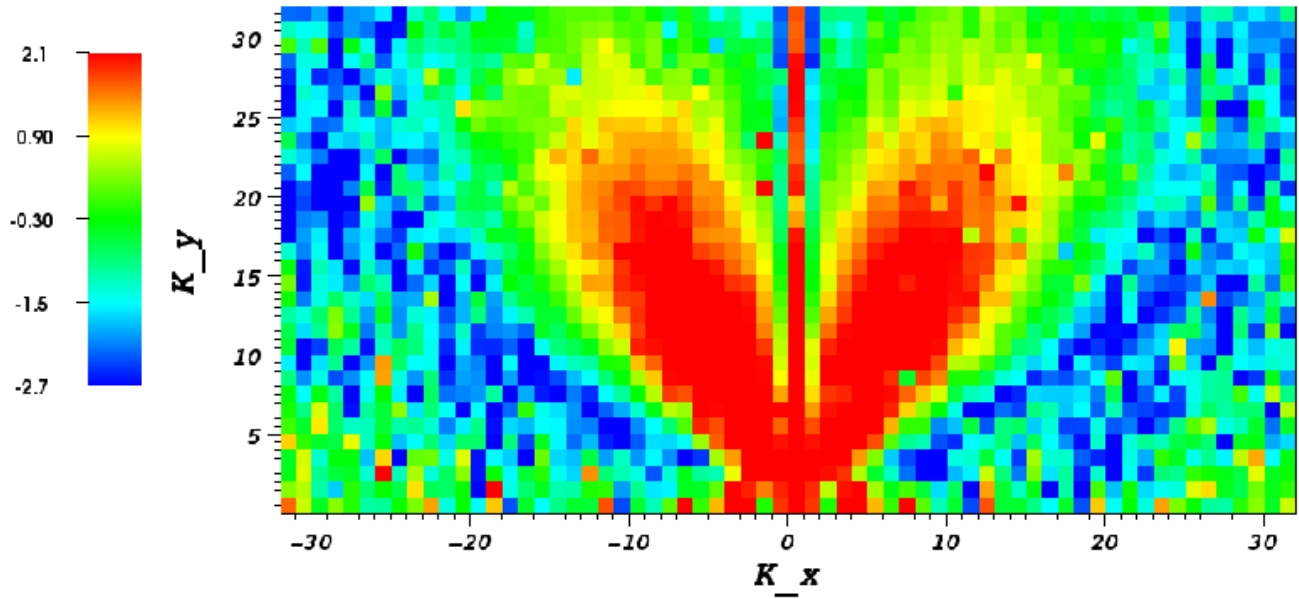}
\caption{Growth rate of photon bubble instability as a function of 
horizontal and vertical wavenumber. Unit for the wave number
is one over cell size. Similar pattern is shown in Figure 9 of \cite{Turneretal2005}. }
\label{PBGrowth}
\end{figure}

To measure the linear growth rate of the photon bubble instability as a
function of the vertical and horizontal wavenumbers $k_x$ and $k_y$, we
follow the method used by \cite{Turneretal2005}.  We Fourier transform
the horizontal velocity $v_x$ at 50 snapshots between times of $t=0.8$
and 2.8 equally spaced at a time interval of 0.04.  We assume exponential
growth at each wavenumber independently, and therefore calculate the
slope of the logarithm of the power with time between each neighboring
time, and average the resulting rates from all the snapshots together.
This approach only approximates the true growth rates in a stratified
atmosphere.  The result is shown in in Figure \ref{PBGrowth}, which
can be compared directly to Figure 9 of \cite{Turneretal2005} (which
also shows solution of linear analysis from \cite{BlaesSocrates2003}).
There is significant noise in the measurement of the growth rates,
especially at small $k$.  However, clearly the correct trends predicted
by the linear analysis are reproduced, with the fastest growth occurring
for modes with $k_x = k_y$. The noise in Figure 9 of \cite{Turneretal2005} 
is much smaller because they can use an initial amplitude as small 
as $10^{-8}$. This cannot be achieved in Athena because the pressure 
gradient and gravity cannot be balanced to roundoff error (without special 
modification to the reconstruction algorithm) and there is always 
noise in the background state. 

\begin{figure}[htp]
\centering
\vspace{0.8cm}
\includegraphics[width=0.49\hsize]{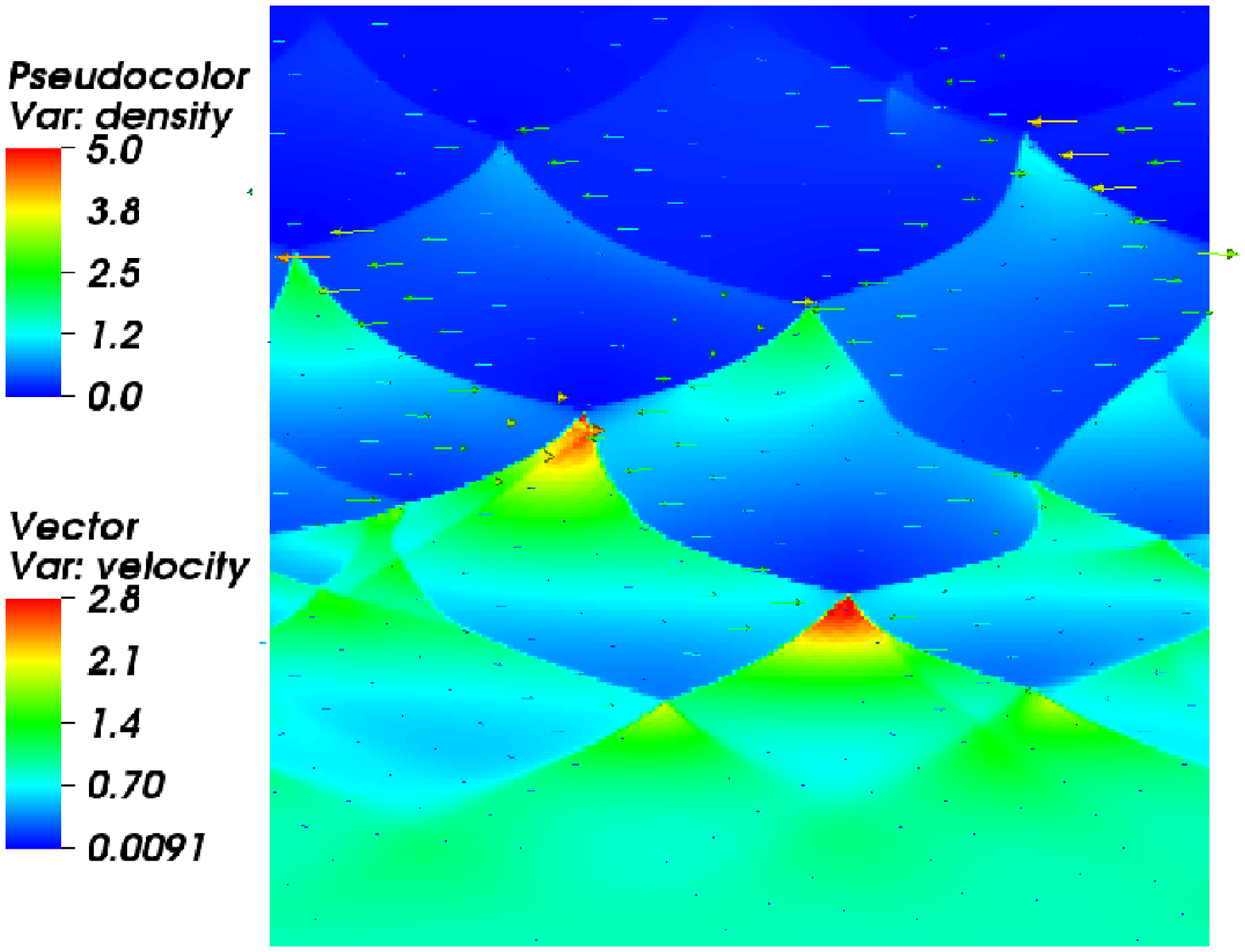}
\includegraphics[width=0.49\hsize]{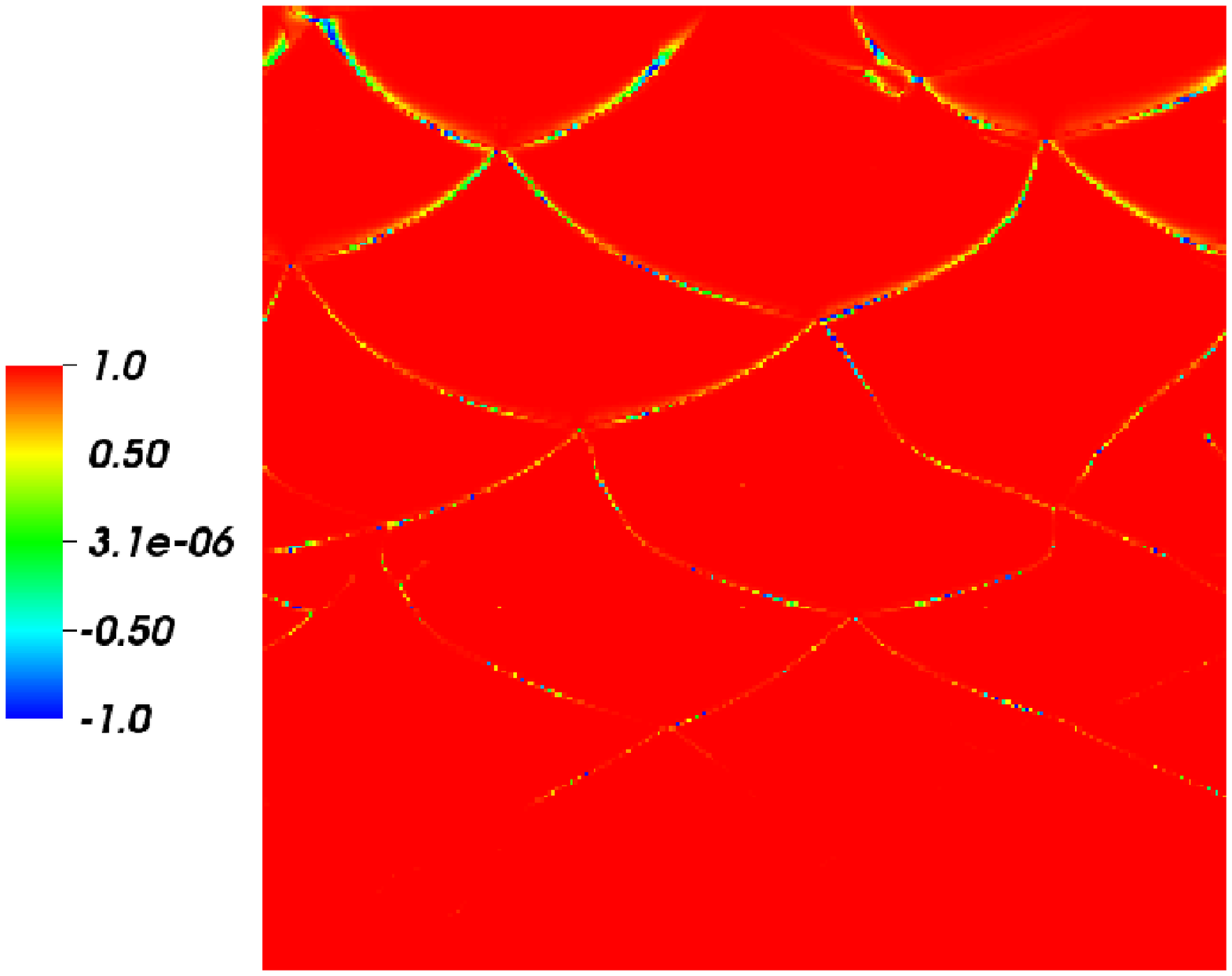}
\caption{Non-linear outcome of Photon Bubble instability at time 
$9.83$, corresponding to $1.56$ orbital periods. The left panel shows 
the density and velocity vectors.  The right panel plots
$\cos\theta$, where $\theta$ is the angle between the perturbed co-moving
radiation flux and the gradient of the perturbed co-moving radiation 
energy density.  Deviations of $\cos\theta$ from one near the shock fronts
indicate the difference between the direction of the flux in our simulations
versus the value assumed in FLD.}
\label{PBdensity}
\end{figure}

In the non-linear regime, \cite{Turneretal2005} found the instability
produces shock trains that propagate from the bottom to the top of
the atmosphere.  With a purely horizontal background field, there is no
preference for shocks propagating in either direction, resulting in a
symmetric pattern of shocks in the atmosphere.   The left panel of Figure
\ref{PBdensity} shows the density and velocity distribution at time $9.83$
when the shock train pattern becomes relatively strong.  The velocity
vectors are nearly horizontal direction because flow is confined to be
parallel to the strong horizontal magnetic field.  This figure can be
compared with the first panel of Figure 19 of \cite{Turneretal2005}:
our results are very similar.

In the FLD approximation used by \cite{Turneretal2005}, the co-moving
radiation flux is always assumed to be along the direction of the
gradient of the co-moving radiation energy density.   Although we adopt the
Eddington approximation in our simulation, we make no assumption about the
direction of the flux.  Thus, it is possible for us to test whether the
radiation flux is indeed everywhere parallel to the gradient of $E_r$ in
this highly nonlinear flow.  Figure \ref{PBdensity} shows the distribution
of $\cos\theta$ at $t=9.83$, where the angle $\theta$ is the angle between
the perturbed co-moving radiation flux in our simulation $\bF_{r,0} -
\hat{F}_{r,0}$ and the direction of the flux predicted by FLD $-\bfnabla
E_{r,0}/(3\sigma_t)$.  If the FLD approximation is valid, $\cos\theta$
will be one everywhere.  We find that near the shock fronts, where there
is a large density gradient, this assumption clearly fails.  Since the
optical depth across the domain is so large, the Eddington approximation
should apply, and the flux-limiter used by \cite{Turneretal2005} should
have a numerical value very close to the value of $1/3$ adopted here.
Thus, the discrepancy in the direction of the radiation flux is not due
to the {\em value} of the Eddington factor, but rather because in FLD the
radiation flux is always along the gradient of radiation energy density.
 
\subsection{Shadow Test}
\label{sec:ShadowTest}

Since the direction of the flux is not known independently of the energy
density, it is well-known that methods based on the diffusion approximation
cannot represent shadows.  This is often demonstrated through the irradiation
of very optically thick structures with a beamed radiation field 
\citep[e.g.,][]{HayesNorman2003,Gonzalezetal2007}.
In this section, we demonstrate our methods capture shadows accurately,
and moreover we present the dynamical evolution of a cloud ablated by
an intense radiation field. VET is crucial to to get the solution correctly, which 
is calculated from the transfer module using short characteristic. The
tests described in \cite{Davisetal2012} (see Figure 6 of that paper) 
show the accuracy of the radiation transfer solver 
for two radiation beams. 
Here we show that when coupled to the VET method, our solver captures 
shadows correctly.

Our test is performed in a rectangular box of size
$(-0.5,0.5)\times(-0.3,0.3)$ cm.  Initially the background medium has
density $\rho_0=1$ $g/cm^3$ and temperature $T_0=290$ $K$.  An over-dense
clump is located in an elliptical region $r\equiv x^2/a^2+y^2/b^2\le
1$, with $a=0.1$ $cm$ and $b=0.06$ $cm$.  The density inside this
region is $\rho(x,y)=\rho_0+(\rho_1-\rho_0)/[1.0+\exp(10(r-1))]$
with $\rho_1=10\rho_0$.  The clump is in pressure equilibrium
with its surroundings, so the interior is much colder than the
ambient medium.  The initial radiation temperature is the same as
the gas temperature everywhere.  Pure absorption opacity is used with
$\sigma=(T/T_0)^{-3.5}(\rho/\rho_0)^2$ $cm^{-1}$.  The radiation flux
$\bF_r$ is zero everywhere initially.  Outflow boundary conditions are
used on both $y$ boundaries, and on the right $x$ boundary.  A constant
radiation field with temperature $T_r=6T_0$ is input through the left
$x$ boundary at angles of $\pm14$ degrees with respect to the $x$ axis.
This radiation field is also input along the top $y$ boundary at $-14$
degrees, and along the bottom $y$ boundary at $+14$ degrees, in order
to mimic the radiation field from an infinite plane along the left $x$
boundary at $x=-0.5$.  At the left $x$ boundary, the gas temperature and
density are fixed to $T_0$ and $\rho_0$ respectively.  A resolution of
$512\times256$ cells is used for this test.  The dimensionless speed of
light $\Crat=1.9\times10^5$, and the parameter $\Prat=2.2\times10^{-15}$.

In the ambient medium, the photon mean free path is the length of simulation
domain, so that the radiation can propagate freely across the box.  However,
inside the clump the photon mean free path is only $3.2\times10^{-6}$ of 
the length of simulation box due to the high density and low temperature. 
Thus, radiation coming from the left boundary cannot penetrate the clump, and
shadows are cast towards the right boundary.  By using an input radiation field
at two angles, both umbra and penumbra are formed.  This makes the test
more difficult, because ad-hoc closures that capture only one direction for
the flux will not represent both the umbra and penumbra correctly.

\begin{figure}[htp]
\centering
\includegraphics[width=1.0\hsize]{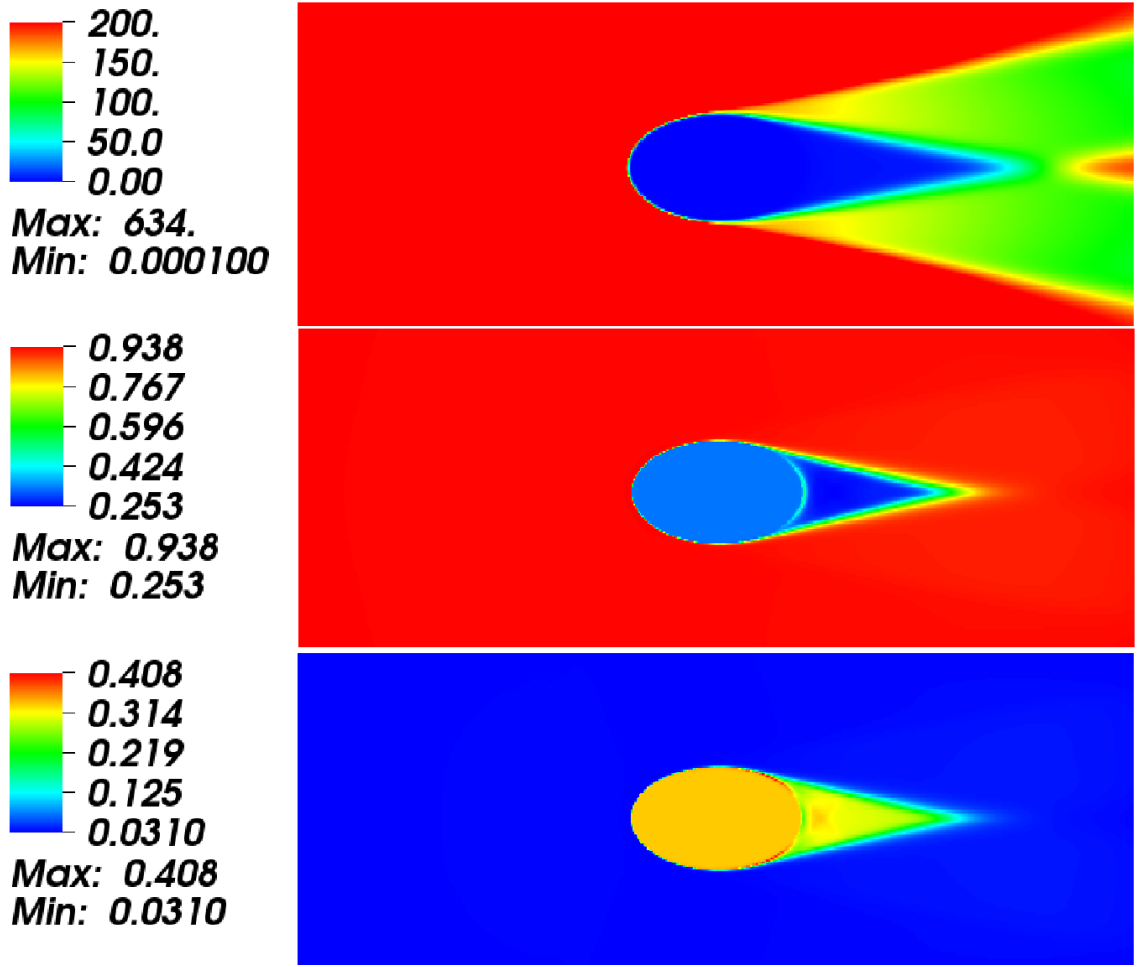}
\caption{Shadows created by the irradiation of an optically thick clump by
two beams at $\pm 14$ degrees with respect to the $x$ axis.
From top to bottom, the three panels show the radiation energy density
$E_r$, and the $xx$ and $yy$ components 
of Eddington tensor $f_{xx}$ and $f_{yy}$ respectively. 
Inside the clump, the Eddington tensor is close to $1/3$ due to the large 
opacity, while $E_r$ retains its initial value. 
Umbra and penumbra are clearly formed behind the clump.}
\label{shadow}
\end{figure}

In Figure \ref{shadow} , we show the radiation energy density,
and $x-x$ and $y-y$ components of the VET after one timestep at
$\delta t=1.0e^{-3}$.  This timestep corresponds to the CFL stability
condition set by the adiabatic sound speed in the ambient gas and the
grid resolution.  The ratio of this timestep to the light crossing time
of the domain is about 500, thus after one timestep both radiation beams
have crossed the domain and a steady-state shadow structure is formed. 

\begin{figure}[htp]
\centering
\includegraphics[width=1.0\hsize]{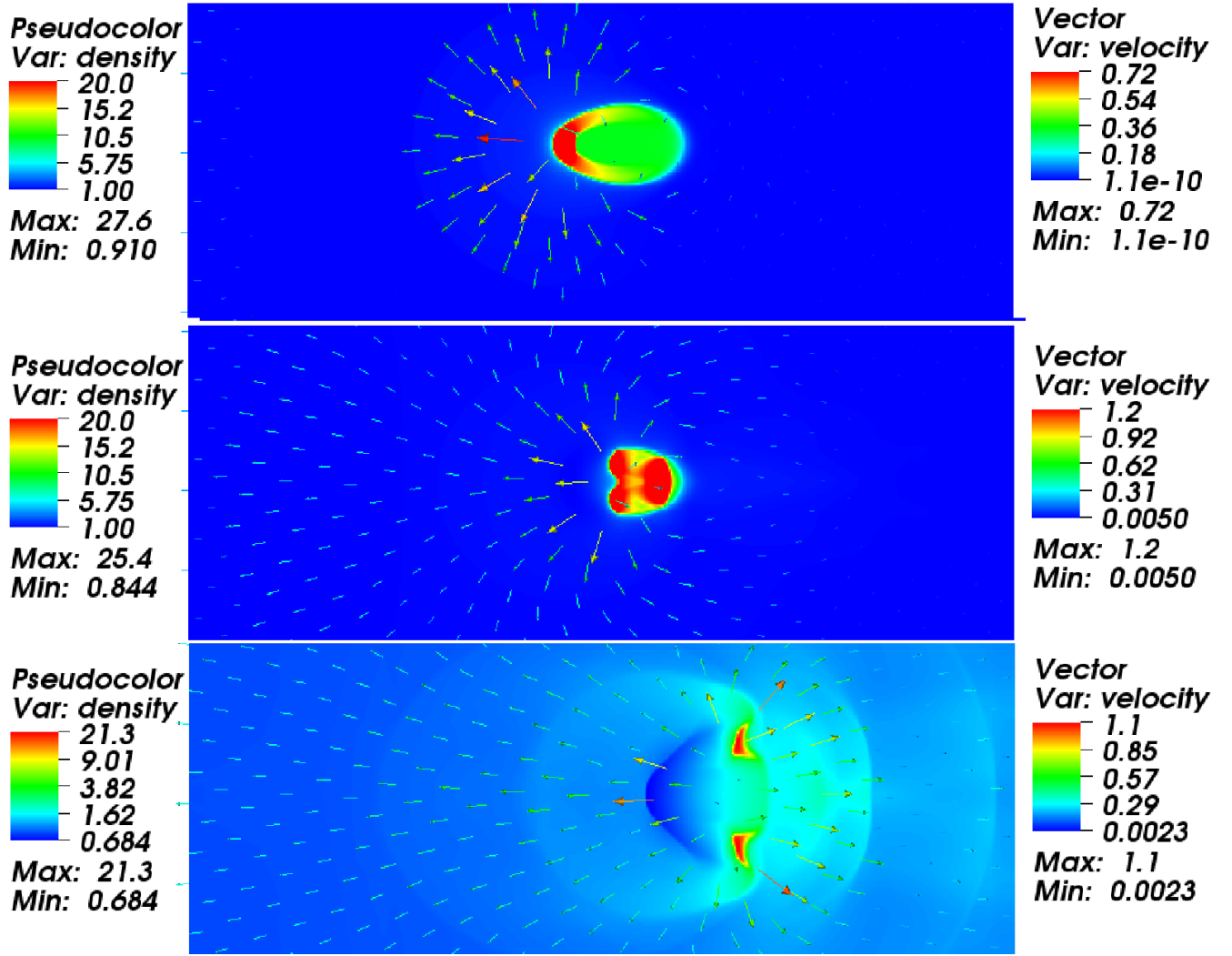}
\caption{Ablation of an optically thick clump by two radiation beams
at $\pm 14$ degrees with respect to the $x$ axis.  From top to bottom, 
the density (color) and velocity (vectors) are shown times
$t=4.9\times10^{-7},\ 1.3\times10^{-6} $ and  
$2.6\times10^{-6} s$ respectively.}
\label{shadow_Pr}
\end{figure}

Since the radiation temperature is low in this first test, the clump
undergoes negligible dynamical evolution.  To demonstrate the dynamics
of the ablation of a clump,
we have repeated this test with an increase the 
background temperature $T_0$ by a factor of $357$ 
such that the dimensionless $\Prat=10^{-7}$. Then the radiation pressure 
of the applied radiation field at the left $x$ boundary is about $10^{-3}$ 
of the gas pressure.   This non-negligible radiation field pushes and compresses 
the clump from the left and the shape of the clump is changed. A shock 
is driven into the clump during this process. The clump 
is also heated up by the radiation field and outflow from the surface of the 
clump is formed on the left side.  The density distribution and velocity field 
at three different times are shown in Figure \ref{shadow_Pr}. The adiabatic 
sound speed in the ambient medium is $1.3$ in our units and $0.39$ inside the 
clump. In the first panel of Figure \ref{shadow_Pr}, the density on the left side 
of the clump is increased by a factor of $2$ due to the radiation pressure. Outflow is generated 
from the surface of the clump as shown by the velocity vectors. The flow is 
subsonic in the ambient medium and supersonic with respect to the sound speed inside the clump. 
In the second 
panel, we clearly see a shock driven to the clump  with Mach number up to $3$. 
In the third panel, the clump is completely destroyed by the radiation field and 
several shock fronts are formed. The ablation of clouds by strong radiation 
field is in fact of great interest in the interstellar medium (ISM) \citep[e.g.,][]{BertoldiMcKee1990,Bally1995} and near 
the central region of quasars \citep[e.g.,][]{Mathews1983}. Further explore of this flow is warranted, but 
beyond the scope of this paper.

\section{Discussion}
\label{sec:discussFLD}

\subsection{Comparison with methods based on the diffusion approximation}
\label{sec:CompareDiffusion}

A central ingredient of the algorithm presented in this paper
is that it uses a VET to close the radiation moment equations, rather
than relying on the
diffusion approximation.  It is therefore worthwhile to discuss
specific examples where there will be differences between these approaches.

The first difference is that by solving the moment equations for the
radiation flux, we are able to solve for the {\em direction} of
the flux self-consistently with the flow, rather than assuming it
lies in the direction of $\nabla E_{r}$.  Both the
photon bubble instability (see Figure \ref{PBdensity}) and the shadow test 
(see Figure \ref{shadow}) demonstrate the importance of keeping the direction of the flux.

A second difference is that the value of the flux limiter is
always an approximation (of unknown reliability) to the true value
of the diagonal components of the VET.  In fact, the planar subcritical shock
test (see Figure \ref{ShockEdd}) shows that near the shock front,
the VET can be {\em less} than $1/3$ along the    
direction of the radiation energy density gradient, which clearly
cannot be represented by the formulae used for most 
limiters \citep[e.g.,][]{LevermorePomraning1981}. Based on the 
density distribution from a snapshot of simulations for black hole accretion 
disk done by \cite{Hiroseetal2006}, \cite{Davisetal2012} 
compares the Eddington tensor used in the flux-limited diffusion approximation 
and the VET calculated based on our transfer module. They also show 
significant difference (see Figure 7 of that paper). In the photosphere, the 
Eddington tensor given by the diffusion approximation can be much larger than 
the values given by the VET.

\begin{figure}[htp]
\centering
\includegraphics[width=1.0\hsize]{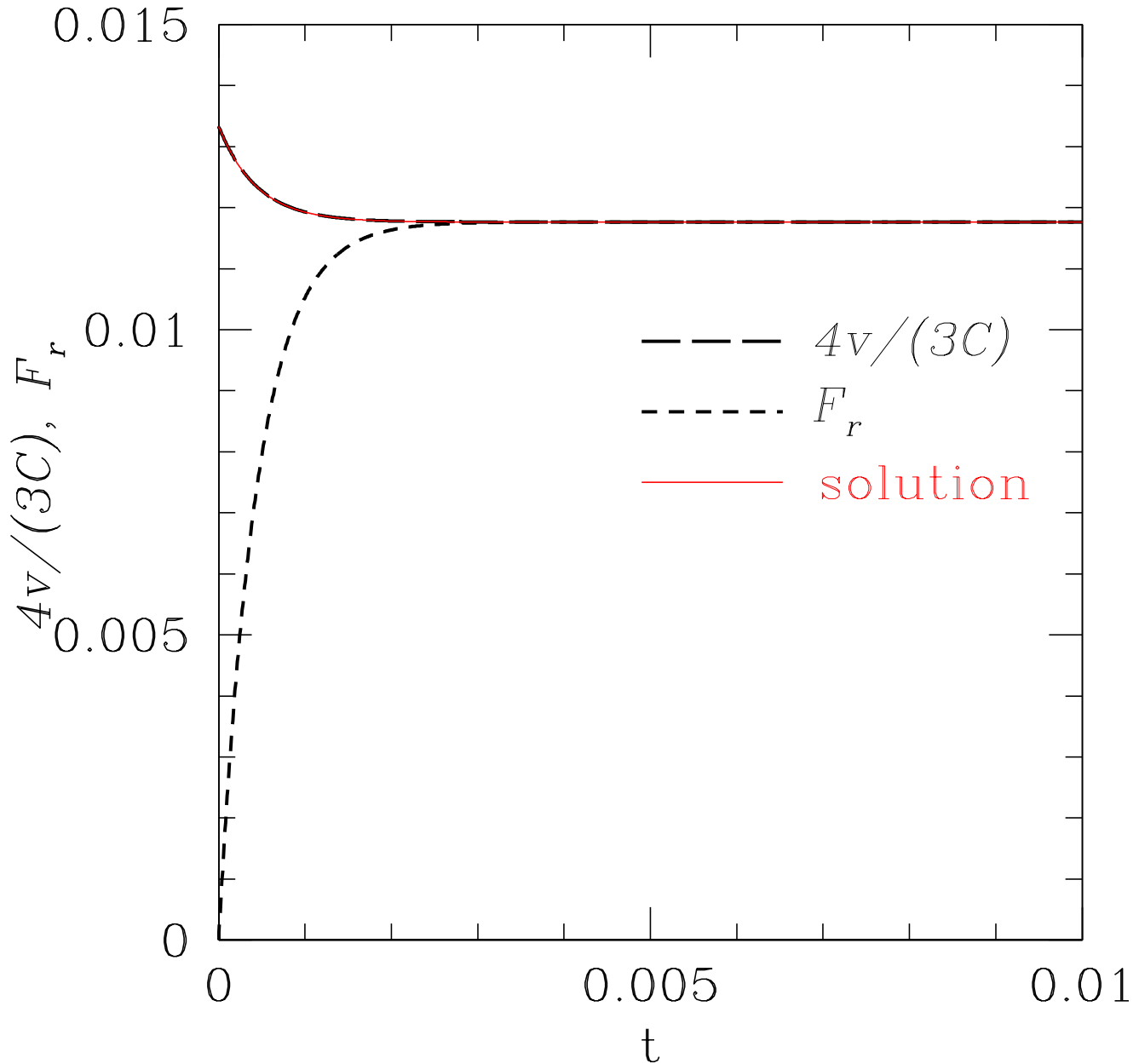}
\caption{Momentum exchange between radiation and material due to radiation drag. 
Red line is the analytic solution given by equation \ref{Radwork:v},
while the black lines are our numerical solutions for normalized velocity and 
radiation flux. The solution reaches equilibrium when the 
co-moving radiation flux is zero.
The velocity changes by $12\%$ in this example. }
\label{Radwork}
\end{figure}

A third difference is that we do not drop the time-derivative
of the flux $\partial F_{r}/\partial t$ in the moment equations for the
flux. This term is important when the inertia
of the radiation field is important, and dropping this term can lead
to incorrect behavior in this regime.  For example, consider
a uniform medium with density $\rho=1$, absorption opacity $\sigma_a=10$
and scattering opacity $\sigma_s=10$ in thermal equilibrium ($E_r=T^4=1$
in our dimensionless units) with zero fixed fram radiation flux, and
an initially uniform velocity along the $x$-direction $v_0=1$.  Adopt
values for the dimensionless speed of light $\Crat=100$ and parameter
$\Prat=1000$, and use the Eddington approximation.  In the diffusion
approximation, because the radiation energy density is always uniform,
the co-moving radiation flux will be always be zero and the system will
always be in thermal equilibrium.  However, this is not true if the
$\partial \bF_r/\partial t$ term is kept.  Since the gas has a non-zero
velocity with respect to the radiation field, the co-moving radiation flux
is {\em not} zero, and there will be a drag force from the radiation field
on the fluid.
(a consequence of the fact that solutions to the equations of radiation
hydrodynamic are not Galilean invariant).  Because of the drag with the
radiation field, momentum will be exchanged 
and the gas will be heated.  In this example, this process can be
described by 
\begin{eqnarray}
\frac{\partial( \rho\bv)}{\partial t} &=&\mathbb{P}\sigma_t\left(\bF_r-\frac{\bv E_r+\bv\cdot\bP _r}{\mathbb{C}}\right),\  \nonumber \\
\frac{\partial \bF_r}{\partial t}&=&-\mathbb{C}\sigma_t\left(\bF_r-\frac{\bv E_r+\bv\cdot\bP _r}{\mathbb{C}}\right).
\end{eqnarray}
In solutions to these equations, $\rho$ and $E_r$ are unchanged while the 
change of velocity with time $t$ can be described by
\begin{eqnarray}
v(t)=v_0\left(1-\frac{\Crat}{\Crat+4\Prat/(3\Crat)}\right)\exp\left[-\Prat\sigma_t\left(\frac{\Crat}{\Prat}+\frac{4}{3\Crat}\right)t\right]+\frac{v_0\Crat}{\Crat+4\Prat/(3\Crat)}.
\label{Radwork:v}
\end{eqnarray}
We compare this analytic solution for this problem, and a numerical
solution generated by our algorithms, in Figure \ref{Radwork}.
Our solutions captures this momentum exchange process accurately, which
is not possible with flux-limited diffusion.  The importance of this
process is characterized by the parameter $\Prat/\Crat^2$, which measures
the importance of momentum carried by the radiation field compared with
the material momentum.

A fourth difference between our VET algorithm and FLD is that
we keep the non-diagonal components of the VET.
 In the FLD approximation,
the radiation pressure is always a diagonal tensor.  Thus, 
radiation viscosity
\citep[e.g.,][]{MihalasMihalas1984,Castor2004} 
cannot be captured in this approximation.
However, with the VET approach, the off-diagonal components are computed
self-consistently with the flow.  Thus, the effects of
radiation viscosity are captured naturally in simulations of accretion flows.

The importance of the differences between VET and FLD depend on the
particular application of interest.  For example, in the inner regions of
accretion disks around supermassive black holes, we expect the momentum
carried by the radiation field to be a significant fraction of material
momentum ($\Prat/\Crat^{2}$ is not negligible), and photon viscosity may be significant
\citep[e.g.,][]{AgolKrolik1998}.  Thus, adopting algorithms based on VET is important for our
applications.

\subsection{Performance and Scaling}
\label{sec:performance}

For solving the application problems of interest, our algorithms must
not only be accurate, but also efficient.  In this subsection, we
report preliminary performance and scaling of the method on parallel
systems.  We continue to work on optimization of the algorithms, and
it is likely substantial improvements in performance are still possible.

For reference, it is useful to compare to the performance of Athena in
ideal MHD without radiation.  On a single core of a $2.5$ GHz Intel Xeon
processor, Athena updates roughly $1 \times 10^{5}$ grid cells per second 
for 3D MHD calculations with a $32^3$ grid.
Weak scaling is very good, with an efficiency (defined
as the performance per core in a parallel calculation, compared to the
performance of a serial calculation) of 0.9 on up to $10^{5}$ processors.

With radiation hydrodynamics and MHD, there are two potential bottlenecks.
The first is the implicit solution of the linear radiation subsystem,
which requires inverting one matrix per timestep (this is already an
advantage over methods that require implicit solution of nonlinear systems
which generally require many matrix inversions per timestep).  To invert
the matrix, we have explored the use of algorithms implemented in both
the LIS \citep[e.g.,][]{Nishida2010} and {\em Hypre} \citep[e.g.,][]{Falgoutetal2006} 
libraries, as well as our
own implementation of the multigrid \citep[e.g.,][]{Pressetal1992} method.  Both libraries
provide a variety of iterative algorithms and preconditioners for solving
sparse matrices, and can be used on both serial and parallel systems.
The convergence rate of the matrix solver is limited by several things,
such as the opacity and error tolerance.

To check the serial performance of each of these methods, we
use the linear hydrodynamic wave problem as described in Section
\ref{sec:Linearwave} on a 3D domain and a $32^3$ grid.  With the LIS
library on a single $2.5$ GHz Intel Xeon processor with tolerance level
$10^{-8}$, we achieve roughly $1.7 \times 10^4$ cell updates per cpu second
in optical thick regimes with the ILU pre-conditioner.  The {\em Hypre}
library is somewhat faster, achieving about $3 \times 10^4$ cell updates
per cpu with the GMRES solver and the BoomerAMG preconditioner. 
Because the sparsity pattern of the matrix we invert is not common (see
Appendix A), we must use the general matrix solvers in these libraries,
which have very poor cache performance.  By writing our own multigrid
algorithm for inverting the matrix, we can take advantage of the
specific structure of our matrix to substantially improve performance.
With multigrid, we can achieve $4.3 \times 10^4$ cell updates per cpu
for this test, and the convergence rate is independent of whether the
linear waves are optically thick or thin.  With both the LIS and {\em
Hypre} libraries, performance is about $2\sim 5$ times slower in the optically
thin regime.  Currently, the multigrid solver is our workhorse algorithm.

A second bottleneck to performance in simulations with radiation is
the formal solution of the transfer equation.  The cost of this step
depends on the number of rays required to resolve the angular dependence
of the radiation field, and whether the LTE approximation can be used
to compute the source function (see \citealt{Davisetal2012}).  In our
tests, we have found sufficient accuracy is obtained with roughly 24
angles in 3D, although this number may be highly application dependent.
If absorption opacity is dominant and the LTE approximation can be used,
this step consumes a negligible fraction of the overall execution time.
Moreover, since the cost to compute the VET scales linearly with the
number of angles, using many more angles (for example, 80 instead of 24)
does not affect the overall cost per timestep significantly.  For example,
using 80 angles increases the cost per timestep by less than factor of
two compared to 24 angles.

However, if scattering opacity is dominant and the non-LTE integrator
must be used to compute the VET, then this step can dominate the cost.
With 24 angles in a non-LTE calculation, the overall performance of the
code can be decreased by a factor of up to $5$.  Moreover, using more
angles decreases the convergence rate of the iteration scheme used
for non-LTE problems, thus roughly tripling the number of angles (from 24 to
80) will increase the cost per timestep by more than a factor of three.
There are several simple steps that can be used to increase performance
in this regime, however.  For example, in many cases it is possible to
compute the VET only every several MHD time steps, since the geometry of
the radiation field (and therefore the VET) changes only slowly compared
to the flow structure.

It is interesting to compare the performance of the VET algorithm to FLD.
Since we have not implemented FLD directly in Athena, we cannot perform a 
head-to-head comparison.  However, we note that because the implicit
differencing most often used with FLD \citep[e.g.,][]{TurnerStone2001,Zhangetal2011} involves nonlinear terms, Newton-Raphson iterations are required with
one matrix solver per iteration.  Depending on the convergence rate of the
iterations, this could require several to many matrix solvers per timestep.
Thus, \cite{TurnerStone2001} reported that in the optically thin limit
the FLD module in ZEUS made the code run $10\times$ slower, whereas we find
the VET algorithm (including the short-characteristics RT solver)
is only $2-3\times$ slower than the standard MHD integrators
in Athena.  ZEUS itself is several times faster than Athena, so the VET may
not in fact be faster than FLD, but it is at least comparable in performance.  
Using FLD because it is faster does not seem to be supported by these comparisons.

The parallel performance of the code is also limited by the matrix
solver we use.  Generally, in parallel calculations, the iterative linear
system solver requires more steps to converge compared to the serial case.
Using the linear hydrodynamic wave problem as described above, with $32^3$
grids on each processor, we have performed a weak scaling test for both
the LIS and {\em Hypre} libraries, and our own multigrid solver.  With 8
cores, the efficiency of the two libraries are about the same: $0.65$, and on
256 cores this drops to $0.25$.  With our multigrid solver, the efficiency
on 8 cores is $0.76$, and drops to $0.56$ on 256 cores.  While this is
substantially lower than the efficiency of the MHD integrator alone,
it is still sufficient to enable large 3D applications.  However,
clearly further effort to improve performance is warranted.

\section{Summary} \label{sec:summary}

We have described a Godunov algorithm for multidimensional radiation
hydrodynamics and MHD,
designed to study the dynamics of radiation dominated
accretion flows around compact objects, and radiation driven winds
from stars, disks, and galaxies.  The algorithm has been implemented
in the Athena MHD code \citep[][]{Stoneetal2008}, and tested 
over a wide range of parameter space.

Compared with previous algorithms for radiation hydrodynamics, there
are several features of our approach that bear special mention.

{\em Use of a VET rather than FLD.}  Most multidimensional algorithms
for radiation hydrodynamics adopt the FLD approximation.  However, this
means that the direction of the radiation flux is lost, that the value
of the flux is determined by an {\em ad-hoc} limiter, that the inertia of
the radiation field is lost 
and that the off-diagonal
components of the radiation pressure are assumed to be zero.  We have
shown through a variety of test problems that using methods based on
a VET overcomes these limitations.

{\em Formal solution of the transfer equation via short characteristics.}
To compute the VET, we discretize the specific intensity over many angles,
and solve the transfer equation using short characteristics.  While this
has been deemed prohibitively expensive for 3D applications, advances in
algorithms and hardware have overcome this difficulty.  We find we can
solve the transfer equation in full 3D using 24 angles at a cost that
is no more than a single timestep of the MHD algorithm.  We have also
explored the use of Monte Carlo methods to compute the VET, and have
found for our applications, short characteristics is substantially faster
\citep[][]{Davisetal2012}.

{\em Use of Godunov methods for the underlying MHD solver.}  
The VET method has been implemented previously in the ZEUS
\citep[e.g.,][]{Stoneetal1992,HayesNorman2003}, and TITAN  \citep[][]{GehmeyrMihalas1994}  codes,
however these methods used an operator split method with an artificial
viscosity for shock capturing.  Instead, we have adopted a higher-order
Godunov method with a dimensionally unsplit integrator.  Comparison of
the structure of radiation modified shocks show the advantages of
this approach.
Since the addition of stiff source terms to Godunov methods
is problematic, we extend the modified Godunov method of SS10 to
multi-dimensions.  We show this method results in accurate and
stable evolution.

To test our algorithms, we have introduced a variety of quantitative
test problems.  We show the results of error convergence tests for 
linear radiation modified acoustic waves in both hydrodynamics and MHD
over a wide range of
parameters, and give examples of eigenmodes that can be used by others for
testing.  We also use the structure of radiating shocks in non-equilibrium
diffusion computed using the methods outlined by \cite{LowrieEdwards2008}.  Our algorithm
reproduces the subtle changes in shock structure that occur as the Mach
number is increased,
even including very high Mach numbers where radiation pressure dominates
the postshock flow.   We also have performed multidimensional tests
based on the photon bubble instability and shadowing.

There are still some limitations to our algorithms, which could be
improved with future development.  For example, we assume LTE, and our
methods are limited to the grey approximation.  Extension to multiple
frequency bands requires analytic expressions for the interaction between
different frequency bands in order to calculate the propagation operator
used to update the stiff source terms.  This is an important direction 
for future work. Our algorithm cannot propagate
linear waves very well when absorption opacity is very large, 
in particular when $\tau_a>\Crat/\Prat$. Note that the algorithm is 
still capable of capturing flows in this regime. However, the damping 
rate and phase velocities of radiation modified acoustic waves will not be 
accurate. This can be improved by solving the total conservative form of the
equations in this regime. If absorption opacity is so large that radiation temperature is 
locked to the gas temperature, 
the $\sigma_a(T^4-E_r)$ term can be dropped in the radiation energy source term. Then damping rate and phase velocity of the radiation modified acoustic waves can be captured accurately in this regime. 
Since we solve the radiation transfer 
equations in the mixed-frame, our algorithm cannot be used 
for problems when spectral lines dominate the transport.
The current algorithm is 
also not appropriate for problems when the radiation field 
changes rapidly (on a light-crossing time) because the VET
is calculated using
the time-independent transfer equation, although the error introduced by
this assumption is acceptable if the
time step is comparable to the light-crossing time of each cell,
as we demonstrated
in Section \ref{sec:Tophat}.

Not all the applications of interest requiring including the effects
of radiation pressure.  In this regime, it is possible to use a much
simpler algorithm which only adds the radiation source terms to the
energy equations, computed from a formal solution of the transfer equation.
This method is described in a companion
paper \citep[][]{Davisetal2012}, and is currently being used to study 
black hole accretion disks in the gas pressure dominated regime.
We use the same module for computing the VET in this work.

Currently we are using the radiation MHD version of Athena to
study several applications, including the nonlinear regime of the
Rayleigh-Taylor instability in a radiation supported atmosphere, 
radiation feedback in AGN, and the
saturation of the magneto-rotational instability (MRI) in radiation dominated
disks (Jiang et al. in preparation).  Eventually the code will be publicly available through
the Athena Trac site.

\section*{Acknowledgements}
We thank Aristotle Socrates and Jeremy Goodman for helpful discussions on
the properties of linear waves in radiation hydrodynamics and MHD. Y.-F. J. 
thanks Francesco Miniati for helpful discussions on the basic numerical algorithm 
and Julian Krolik for the suggestion of multi-grid solver. 
We thank Bryan Johnson, Richard Klein and
Louis Howell for suggestions for tests.  We also thank the anonymous referee 
for helpful comments that improved the paper. This work was supported by the
NASA ATP program through grant NNX11AF49G, and by computational resources
provided by the Princeton Institute for Computational Science and Engineering.

\begin{appendix}
\section{Matrix for the 3D radiation subsystem} 
\label{3Dmatrix}
  
Implicit backward Euler differencing of the radiation subsystem
(equation \ref{3DImplicit} and \ref{3DImplicitEr}) 
results in a set of linear equations for the radiation energy density and
flux in each cell.  For every cell with indices $(i,j,k)$, we have the following 
four equations  
  
\begin{eqnarray}
\label{Matrix3DEr} 
& &\theta_0 E_r^{n+1}(i,j,k-1) + \theta_1 F_{r3}^{n+1}(i,j,k-1)\nonumber\\
&+&\theta_2E_r^{n+1}(i,j-1,k)+\theta_3F_{r2}^{n+1}(i,j-1,k)\nonumber\\
&+&\theta_4E_r^{n+1}(i-1,j,k)+\theta_5F_{r1}^{n+1}(i-1,j,k)\nonumber\\
&+&\theta_6E_r^{n+1}(i,j,k)+\theta_7F_{r1}^{n+1}(i,j,k)+\theta_8F_{r2}^{n+1}(i,j,k)+\theta_9F_{r3}^{n+1}(i,j,k)\nonumber\\
&+&\theta_{10}E_r^{n+1}(i+1,j,k)+\theta_{11}F_{r1}^{n+1}(i+1,j,k)\\ \nonumber
&+&\theta_{12}E_r^{n+1}(i,j+1,k)+\theta_{13}F_{r2}^{n+1}(i,j+1,k)\\ \nonumber
&+&\theta_{14}E_r^{n+1}(i,j,k+1)+\theta_{15}F_{r3}^{n+1}(i,j,k+1)\\ \nonumber
&=&E_r^n(i,j,k)-(1-\mathcal{R})dS_r(E)/\Prat+\mathcal{R}d_2\sigma_a \tilde{T}_{i,j,k}^{4},
\end{eqnarray}
\begin{eqnarray}
\label{Matrix3DFr1}
& &\phi_0 E_r^{n+1}(i,j,k-1) + \phi_1 F_{r1}^{n+1}(i,j,k-1)\\ \nonumber
&+&\phi_2 E_r^{n+1}(i,j-1,k) + \phi_3 F_{r1}^{n+1}(i,j-1,k) \\ \nonumber
&+&\phi_4E_r^{n+1}(i-1,j,k)+\phi_5F_{r1}^{n+1}(i-1,j,k)\\ \nonumber
&+&\phi_6E_r^{n+1}(i,j,k) +\phi_7F_{r1}^{n+1}(i,j,k) \\ \nonumber
&+&\phi_8E_r^{n+1}(i+1,j,k)+\phi_9F_{r1}^{n+1}(i+1,j,k)\\ \nonumber
&+&\phi_{10}E_r^{n+1}(i,j+1,k)+\phi_{11}F_{r1}^{n+1}(i,j+1,k)\\ \nonumber
&+&\phi_{12}E_r^{n+1}(i,j,k+1)+\phi_{13}F_{r1}^{n+1}(i,j,k+1)\\ \nonumber
&=&F_{r1}^n(i,j,k)+d_2\sigma_a v_x \tilde{T}_{i,j,k}^{4}/\mathbb{C},
\end{eqnarray}
\begin{eqnarray}
\label{Matrix3DFr2}
& &\psi_0 E_r^{n+1}(i,j,k-1) + \psi_1 F_{r2}^{n+1}(i,j,k-1), \\ \nonumber
&+&\psi_2 E_r^{n+1}(i,j-1,k) + \psi_3 F_{r2}^{n+1}(i,j-1,k)\\ \nonumber
&+&\psi_4E_r^{n+1}(i-1,j,k)+\psi_5F_{r2}^{n+1}(i-1,j,k)\\ \nonumber
&+&\psi_6E_r^{n+1}(i,j,k) +\psi_7F_{r2}^{n+1}(i,j,k)\\ \nonumber
&+&\psi_8E_r^{n+1}(i+1,j,k)+\psi_9F_{r2}^{n+1}(i+1,j,k)\\ \nonumber
&+&\psi_{10}E_r^{n+1}(i,j+1,k)+\psi_{11}F_{r2}^{n+1}(i,j+1,k)\\ \nonumber
&+&\psi_{12}E_r^{n+1}(i,j,k+1)+\psi_{13}F_{r2}^{n+1}(i,j,k+1)\\ \nonumber
&=&F_{r2}^n(i,j,k)+d_2\sigma_a v_y \tilde{T}_{i,j,k}^{4}/\mathbb{C},
\end{eqnarray}
\begin{eqnarray}
\label{Matrix3DFr3}
& &\varphi_0 E_r^{n+1}(i,j,k-1) + \varphi_1 F_{r3}^{n+1}(i,j,k-1), \\ \nonumber
&+&\varphi_2 E_r^{n+1}(i,j-1,k) + \varphi_3 F_{r3}^{n+1}(i,j-1,k)\\ \nonumber
&+&\varphi_4E_r^{n+1}(i-1,j,k)+\varphi_5F_{r3}^{n+1}(i-1,j,k)\\ \nonumber
&+&\varphi_6E_r^{n+1}(i,j,k) +\varphi_7F_{r3}^{n+1}(i,j,k)\\ \nonumber
&+&\varphi_8E_r^{n+1}(i+1,j,k)+\varphi_9F_{r3}^{n+1}(i+1,j,k)\\ \nonumber
&+&\varphi_{10}E_r^{n+1}(i,j+1,k)+\varphi_{11}F_{r3}^{n+1}(i,j+1,k)\\ \nonumber
&+&\varphi_{12}E_r^{n+1}(i,j,k+1)+\varphi_{13}F_{r3}^{n+1}(i,j,k+1)\\ \nonumber
&=&F_{r3}^n(i,j,k)+d_2\sigma_a v_z \tilde{T}_{i,j,k}^{4}/\mathbb{C}.
\end{eqnarray}

Here $v_x$, $v_y$ and $v_z$ are the velocities along the $x-$, $y-$
and $z-$ directions respectively, 
and $d_2\equiv \Delta t\mathbb{C}$. The temperature $T^r_{i,j,k}$ is the value 
we estimate in Section \ref{sec:TErterm}.  The velocities and opacity are
at time step $n$ and are kept constant during the solution of the radiation
moment equations.

Let the components of the symmetric $3\times 3$ VET $\bfr$ be labelled as
\begin{equation}
\bfr
=
\left(
\begin{array}{ccc}
  f_{11} 	& f_{12}  	&	f_{13}	\\
  f_{21} 	& f_{22}  	&	f_{23	}	\\
  f_{31}	& f_{32}	&	f_{33}	\\
\end{array}
\right).
\end{equation}
We also define the following quantities
 \begin{eqnarray}
d_{1x}&\equiv&\Delta t\mathbb{C}/(2\Delta x), \ d_{1y}\equiv\Delta t\mathbb{C}/(2\Delta y),\ d_{1z}\equiv\Delta t\mathbb{C}/(2\Delta z), \\ \nonumber
Ci0&\equiv&\frac{f_{11}^{1/2}(i,j,k)-f_{11}^{1/2}(i-1,j,k)}{f_{11}^{1/2}(i,j,k)+f_{11}^{1/2}(i-1,j,k)}, \ Ci1\equiv\frac{f_{11}^{1/2}(i+1,j,k)-f_{11}^{1/2}(i,j,k)}{f_{11}^{1/2}(i+1,j,k)+f_{11}^{1/2}(i,j,k)}, \\ \nonumber
Cj0&\equiv&\frac{f_{22}^{1/2}(i,j,k)-f_{22}^{1/2}(i,j-1,k)}{f_{22}^{1/2}(i,j,k)+f_{22}^{1/2}(i,j-1,k)}, \ Cj1\equiv\frac{f_{22}^{1/2}(i,j+1,k)-f_{22}^{1/2}(i,j,k)}{f_{22}^{1/2}(i,j+1,k)+f_{22}^{1/2}(i,j,k)}, \\ \nonumber
Ck0&\equiv&\frac{f_{33}^{1/2}(i,j,k)-f_{33}^{1/2}(i,j,k-1)}{f_{33}^{1/2}(i,j,k)+f_{33}^{1/2}(i,j,k-1)}, \ Ck1\equiv\frac{f_{33}^{1/2}(i,j,k+1)-f_{33}^{1/2}(i,j,k)}{f_{33}^{1/2}(i,j,k+1)+f_{33}^{1/2}(i,j,k)}.
\end{eqnarray}

Then the coefficients $\theta_0-\theta_{15}$, $\phi_0-\phi_{13}$, $\phi_0-\phi_{13}$ and 
$\varphi_0-\varphi_{13}$ in equations \ref{Matrix3DEr} $-$ \ref{Matrix3DFr3}
are  
\begin{eqnarray}
\theta_0&=&-d_{1z}(1+Ck0)f_{33}^{1/2}(i,j,k-1), \\ \nonumber
\theta_1&=&-d_{1z}(1+Ck0), \\ \nonumber
\theta_2&=&-d_{1y}(1+Cj0)f_{22}^{1/2}(i,j-1,k), \\ \nonumber
\theta_3&=&-d_{1y}(1+Cj0),\\ \nonumber
\theta_4&=&-d_{1x}(1+Ci0)f_{11}^{1/2}(i-1,j,k),\\ \nonumber
\theta_5&=&-d_{1x}(1+Ci0),\\ \nonumber
\theta_6&=&1+d_{1x}(1+Ci1)f_{11}^{1/2}(i,j,k)+d_{1x}(1-Ci0)f_{11}^{1/2}(i,j,k)+d_{1y}(1+Cj1)f_{22}^{1/2}(i,j,k) \\ \nonumber
&+&d_{1y}(1-Cj0)f_{22}^{1/2}(i,j,k) +d_{1z}(1+Ck1)f_{33}^{1/2}(i,j,k)+d_{1z}(1-Ck0)f_{33}^{1/2}(i,j,k) \\ \nonumber
&+&\mathcal{R}d_2(\sigma_a-\sigma_s)v_x\left[v_x(1+f_{11}(i,j,k))+v_yf_{12}(i,j,k)+v_zf_{13}(i,j,k)\right]/\mathbb{C}^2 \\ \nonumber 
&+&\mathcal{R}d_2(\sigma_a-\sigma_s)v_y\left[v_y(1+f_{22}(i,j,k))+v_xf_{12}(i,j,k)+v_zf_{23}(i,j,k)\right]/\mathbb{C}^2 \\ \nonumber
&+&\mathcal{R}d_2(\sigma_a-\sigma_s)v_z\left[v_z(1+f_{33}(i,j,k))+v_xf_{13}(i,j,k)+v_yf_{23}(i,j,k)\right]/\mathbb{C}^2 \\ \nonumber
&+&\mathcal{R}d_2\sigma_a, \\ \nonumber
\theta_7&=&d_{1x}(1+Ci1)-d_{1x}(1-Ci0)-\mathcal{R}d_2v_x(\sigma_a-\sigma_s)/\mathbb{C}, \\ \nonumber
\theta_8&=&d_{1y}(1+Cj1)-d_{1y}(1-Cj0)-\mathcal{R}d_2v_y(\sigma_a-\sigma_s)/\mathbb{C}, \\ \nonumber
\theta_9&=&d_{1z}(1+Ck1)-d_{1z}(1-Ck0)-\mathcal{R}d_2v_z(\sigma_a-\sigma_s)/\mathbb{C}, \\ \nonumber
\theta_{10}&=&-d_{1x}(1-Ci1)f_{11}^{1/2}(i+1,j,k), \\ \nonumber
\theta_{11}&=&d_{1x}(1-Ci1), \\ \nonumber
\theta_{12}&=&-d_{1y}(1-Cj1)f_{22}^{1/2}(i,j+1,k),  \\ \nonumber
\theta_{13}&=&d_{1y}(1-Cj1), \\ \nonumber
\theta_{14}&=&-d_{1z}(1-Ck1)f_{33}^{1/2}(i,j,k+1),  \\ \nonumber
\theta_{15}&=&d_{1z}(1-Ck1);
\end{eqnarray}

\begin{eqnarray}
\phi_0&=&-d_{1z}(1+Ck0)f_{31}(i,j,k-1), \\ \nonumber
\phi_1&=&-d_{1z}(1+Ck0)f_{33}^{1/2}(i,j,k-1),\\ \nonumber
\phi_2&=&-d_{1y}(1+Cj0)f_{21}(i,j-1,k), \\ \nonumber
\phi_3&=&-d_{1y}(1+Cj0)f_{22}^{1/2}(i,j-1,k),\\ \nonumber
\phi_4&=&-d_{1x}(1+Ci0)f_{11}(i-1,j,k),\\ \nonumber
\phi_5&=&-d_{1x}(1+Ci0)f_{11}^{1/2}(i-1,j,k),\\ \nonumber
\phi_6&=&d_{1x}(1+Ci1)f_{11}(i,j,k)-d_{1x}(1-Ci0)f_{11}(i,j,k)+d_{1y}(1+Cj1)f_{21}(i,j,k) \\ \nonumber
&-&d_{1y}(1-Cj0)f_{21}(i,j,k)+d_{1z}(1+Ck1)f_{31}(i,j,k)-d_{1z}(1-Ck0)f_{31}(i,j,k) \\ \nonumber
&-&d_2\sigma_t\left[v_x(1+f_{11}(i,j,k))+v_yf_{21}(i,j,k)+v_zf_{31}(i,j,k)\right]/\mathbb{C} +
d_2\sigma_a v_x/\mathbb{C}, \\ \nonumber
\phi_7&=&1+d_{1x}(1+Ci1)f_{11}^{1/2}(i,j,k)+d_{1x}(1-Ci0)f_{11}^{1/2}(i,j,k)\\ \nonumber
&+&d_{1y}(1+Cj1)f_{22}^{1/2}(i,j,k)+d_{1y}(1-Cj0)f_{22}^{1/2}(i,j,k)
+d_{1z}(1+Ck1)f_{33}^{1/2}(i,j,k)\\ \nonumber
&+&d_{1z}(1-Ck0)f_{33}^{1/2}(i,j,k)
+d_2\sigma_t , \\ \nonumber
\phi_8&=&d_{1x}(1-Ci1)f_{11}(i+1,j,k), \\ \nonumber
\phi_{9}&=&-d_{1x}(1-Ci1)f_{11}^{1/2}(i+1,j,k), \\ \nonumber
\phi_{10}&=&d_{1y}(1-Cj1)f_{21}(i,j+1,k),  \\ \nonumber
\phi_{11}&=&-d_{1y}(1-Cj1)f_{22}^{1/2}(i,j+1,k), \\ \nonumber
\phi_{12}&=&d_{1z}(1-Ck1)f_{31}(i,j,k+1),  \\ \nonumber
\phi_{13}&=&-d_{1z}(1-Ck1)f_{33}^{1/2}(i,j,k+1);
\end{eqnarray}

\begin{eqnarray}
\psi_0&=&-d_{1z}(1+Ck0)f_{32}(i,j,k-1), \\ \nonumber
\psi_1&=&-d_{1z}(1+Ck0)f_{33}^{1/2}(i,j,k-1), \\ \nonumber
\psi_2&=&-d_{1y}(1+Cj0)f_{22}(i,j-1,k), \\ \nonumber
\psi_3&=&-d_{1y}(1+Cj0)f_{22}^{1/2}(i,j-1,k), \\ \nonumber
\psi_4&=&-d_{1x}(1+Ci0)f_{21}(i-1,j,k),\\ \nonumber
\psi_5&=&-d_{1x}(1+Ci0)f_{11}^{1/2}(i-1,j,k),\\ \nonumber
\psi_6&=&d_{1x}(1+Ci1)f_{21}(i,j,k)-d_{1x}(1-Ci0)f_{21}(i,j,k)+d_{1y}(1+Cj1)f_{22}(i,j,k) \\ \nonumber
&-&d_{1y}(1-Cj0)f_{22}(i,j,k)+d_{1z}(1+Ck1)f_{32}(i,j,k)-d_{1z}(1-Ck0)f_{32}(i,j,k) \\ \nonumber
&-&d_2\sigma_t\left[v_y(1+f_{22}(i,j,k))+v_xf_{12}(i,j,k)+v_zf_{32}(i,j,k)\right]/\mathbb{C} +
d_2\sigma_a v_y/\mathbb{C}, \\ \nonumber
\psi_7&=&1+d_{1x}(1+Ci1)f_{11}^{1/2}(i,j,k)+d_{1x}(1-Ci0)f_{11}^{1/2}(i,j,k)\\ \nonumber
&+&d_{1y}(1+Cj1)f_{22}^{1/2}(i,j,k)+d_{1y}(1-Cj0)f_{22}^{1/2}(i,j,k) \\ \nonumber 
&+&d_{1z}(1+Ck1)f_{33}^{1/2}(i,j,k)+d_{1z}(1-Ck0)f_{33}^{1/2}(i,j,k)+d_2\sigma_t , \\ \nonumber
\psi_8&=&d_{1x}(1-Ci1)f_{12}(i+1,j,k), \\ \nonumber
\psi_9&=&-d_{1x}(1-Ci1)f_{11}^{1/2}(i+1,j,k), \\ \nonumber
\psi_{10}&=&d_{1y}(1-Cj1)f_{22}(i,j+1,k),  \\ \nonumber
\psi_{11}&=&-d_{1y}(1-Cj1)f_{22}^{1/2}(i,j+1,k), \\ \nonumber
\psi_{12}&=&d_{1z}(1-Ck1)f_{32}(i,j,k+1),  \\ \nonumber
\psi_{13}&=&-d_{1z}(1-Ck1)f_{33}^{1/2}(i,j,k+1);
\end{eqnarray}

\begin{eqnarray}
\varphi_0&=&-d_{1z}(1+Ck0)f_{33}(i,j,k-1), \\ \nonumber
\varphi_1&=&-d_{1z}(1+Ck0)f_{33}^{1/2}(i,j,k-1), \\ \nonumber
\varphi_2&=&-d_{1y}(1+Cj0)f_{23}(i,j-1,k), \\ \nonumber
\varphi_3&=&-d_{1y}(1+Cj0)f_{22}^{1/2}(i,j-1,k), \\ \nonumber
\varphi_4&=&-d_{1x}(1+Ci0)f_{31}(i-1,j,k),\\ \nonumber
\varphi_5&=&-d_{1x}(1+Ci0)f_{11}^{1/2}(i-1,j,k),\\ \nonumber
\varphi_6&=&d_{1x}(1+Ci1)f_{31}(i,j,k)-d_{1x}(1-Ci0)f_{31}(i,j,k)+d_{1y}(1+Cj1)f_{32}(i,j,k) \\ \nonumber
&-&d_{1y}(1-Cj0)f_{32}(i,j,k)+d_{1z}(1+Ck1)f_{33}(i,j,k)-d_{1z}(1-Ck0)f_{33}(i,j,k) \\ \nonumber
&-&d_2\sigma_t\left[v_z(1+f_{33}(i,j,k))+v_xf_{31}(i,j,k)+v_yf_{32}(i,j,k)\right]/\mathbb{C} +
d_2\sigma_a v_z/\mathbb{C}, \\ \nonumber
\varphi_7&=&1+d_{1x}(1+Ci1)f_{11}^{1/2}(i,j,k)+d_{1x}(1-Ci0)f_{11}^{1/2}(i,j,k)\\ \nonumber
&+&d_{1y}(1+Cj1)f_{22}^{1/2}(i,j,k)+d_{1y}(1-Cj0)f_{22}^{1/2}(i,j,k) \\ \nonumber 
&+&d_{1z}(1+Ck1)f_{33}^{1/2}(i,j,k)+d_{1z}(1-Ck0)f_{33}^{1/2}(i,j,k)+d_2\sigma_t , \\ \nonumber
\varphi_8&=&d_{1x}(1-Ci1)f_{31}(i+1,j,k), \\ \nonumber
\varphi_9&=&-d_{1x}(1-Ci1)f_{11}^{1/2}(i+1,j,k), \\ \nonumber
\varphi_{10}&=&d_{1y}(1-Cj1)f_{32}(i,j+1,k),  \\ \nonumber
\varphi_{11}&=&-d_{1y}(1-Cj1)f_{22}^{1/2}(i,j+1,k), \\ \nonumber
\varphi_{12}&=&d_{1z}(1-Ck1)f_{33}(i,j,k+1),  \\ \nonumber
\varphi_{13}&=&-d_{1z}(1-Ck1)f_{33}^{1/2}(i,j,k+1).
\end{eqnarray}

The above equations can be written in matrix form, with the vector
of unknowns containing elements $(E_{r}, F_{r1}, F_{r2}, F_{r3})^{n+1}_{i,j,k}$
with index $i$ varying fastest.  The result is a large sparse banded matrix
that must be inverted (see section \ref{sec:performance}).  Note that some
of the elements of the matrix depend on the boundary conditions.  We have
implemented periodic, reflecting, inflow and outflow conditions.

\section{Dispersion relation for linear waves in radiation hydrodynamics}
\label{Radhydrowave}

In this appendix, we derive the dispersion relation for linear
waves in radiation hydrodynamics used for the tests in section
\ref{sec:Linearwave}. 
The opacity $\sigma_a=\sigma_t=\sigma$ is a
constant.  The background state is chosen to be $\rho_0=1,P_0=1,E_{r,0}=1, T_0=1$, 
which is uniform with zero velocity and radiation flux, and is in mechanical
and thermal equilibrium.

Following the analysis done by \cite{JohnsonKlein2010},
for each perturbed quantity, we look for wave solutions of the form
$\delta \tilde{f}=\delta f\times\mathrm{e}^{\mathrm{i}(\omega t-kx)}$,
where $k$ is the wave number, $\omega$ the angular frequency and
$\delta f$ the amplitude. Note that we only need the real part of
$\delta \tilde{f}$.  The amplitude $\delta f$ can be
a complex number, the imaginary part of which represents the relative
phase shift between different quantities. Then the dispersion relation between $\omega$ and $k$ has the form
\begin{equation}
 c_4 k^4+c_2 k^2 + c_0=0,
\label{dispersionEqn0}
\end{equation}
where the coefficients are
\begin{eqnarray}
 c_4&=&\mathrm{i}\frac{\gamma}{\gamma-1}\frac{\omega \mathbb{C}^2}{3\mathbb{P}\sigma}+\frac{4\mathbb{C}^3}{3},\nonumber\\
 c_2&=&-\mathrm{i}\frac{\mathbb{C}^2+3\gamma}{3\mathbb{P}\sigma(\gamma-1)}\omega^3-\left(\frac{4}{3}\mathbb{C}^3
 +\frac{2\mathbb{C}\gamma}{\mathbb{P}(\gamma-1)}+4\mathbb{C}+\frac{4\mathbb{C}}{9(\gamma-1)}\right)\omega^2\nonumber\\
 &+&\mathrm{i}\left(\frac{\sigma\mathbb{C}^2\gamma}{\mathbb{P}(\gamma-1)}+\frac{20}{3}\mathbb{C}^2\sigma+\frac{16}{9}\mathbb{C}^2\mathbb{P}\sigma\right)\omega,\nonumber\\
 c_0&=&\mathrm{i}\frac{\omega^5}{\mathbb{P}\sigma(\gamma-1)}+\left(\frac{2\mathbb{C}}{\mathbb{P}(\gamma-1)}+\frac{4}{3\mathbb{C}(\gamma-1)}+4\mathbb{C}\right)\omega^4\nonumber\\
 &-&\mathrm{i}\left(4\mathbb{C}^2\sigma+\frac{4\sigma}{3(\gamma-1)}+\frac{\mathbb{C}^2\sigma}{\mathbb{P}(\gamma-1)}
 +\frac{16\mathbb{P}\sigma}{3}\right)\omega^3.
\label{dispersionEqn}
\end{eqnarray}
 
\begin{figure}
\hspace{-1cm}
\includegraphics[width=1.00\hsize]{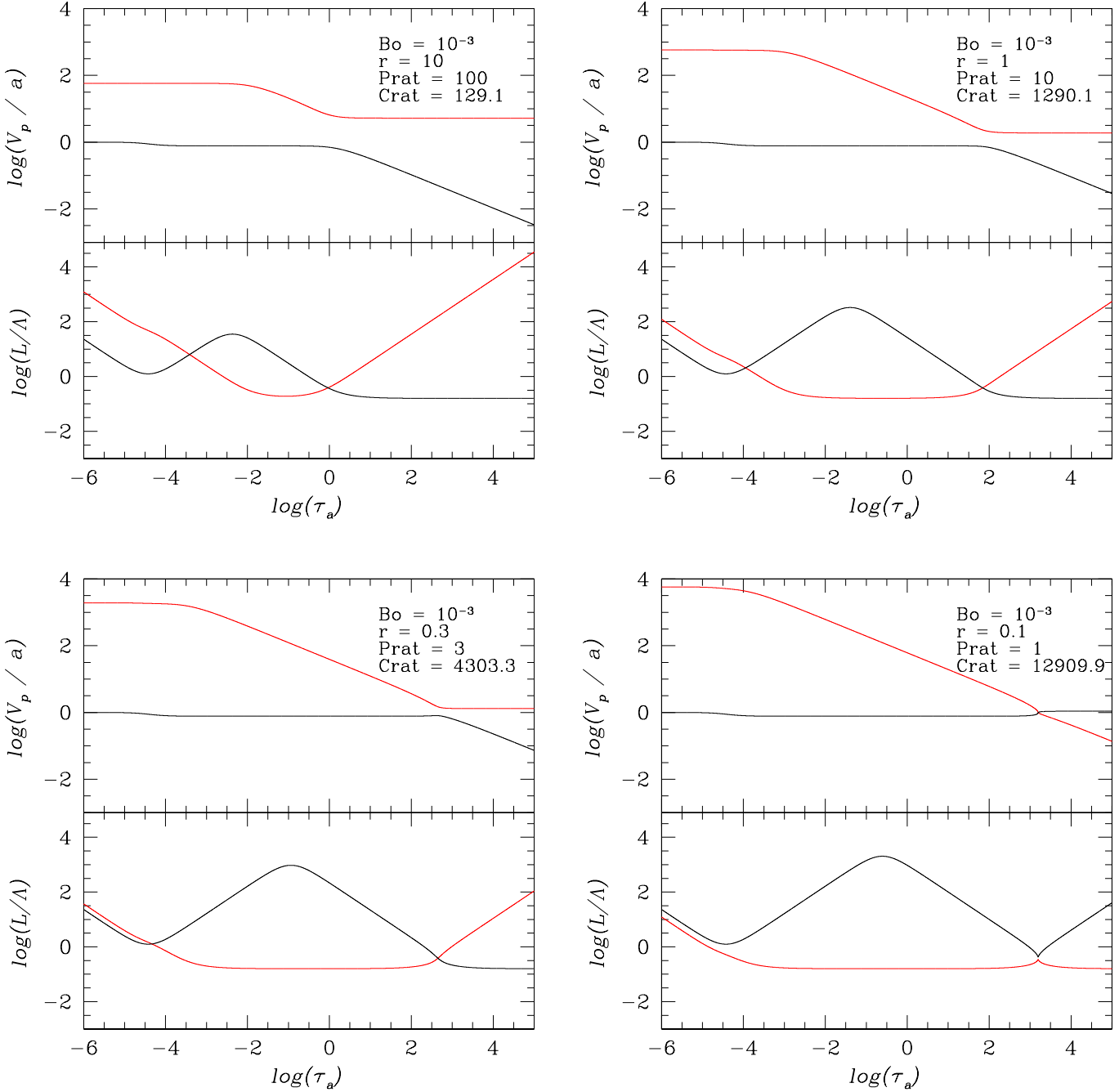}
\caption{Solutions to the dispersion relation for linear waves in
radiation hydrodynamics.  The red lines are
radiation diffusion modes, while the black lines are radiation modified acoustic
waves. Both the phase velocity $V_p$ and damping rate
per wavelength $L/\Lambda$ are shown.  These solutions can be compared to
Figures 101.4 and 101.5 of \cite{MihalasMihalas1984}.
}
\label{Linearwave_test}
\end{figure}
 
The dispersion relation is fourth order in the wave number
$k$, and fifth order in the angular frequency $\omega$.  It is identical
to the dispersion relation given in equation 18 of
\cite{JohnsonKlein2010}; although they use co-moving radiation energy
density and radiation flux for their analysis, there is no difference for
the linearized equations.   The asymptotic behavior of the dispersion relation
for different parameters $\sigma$ and $\Prat$ are discussed in detail by
\cite{Bogdanetal1996} and \cite{JohnsonKlein2010} and will not be repeated
here. The analysis is done for a static background medium. Dispersion
relation for a moving background medium is given by \cite{Lowrieetal1999}.
In particular, they point out that the radiation hydrodynamic system is
not Galilean invariant.  The damping rate of some radiation dominated modes
depends on the flow velocity, because the co-moving
radiation flux seen by the gas depends on the flow velocity.

In order to check that the dispersion relation we get is consistent
with previous results, we reproduce Figures 101.4 and 101.5 of
\cite{MihalasMihalas1984} in Figure \ref{Linearwave_test}. Those
solutions are for boundary value problems, which means we assume a
real number of angular frequency $\omega$ and calculate a complex wave
number $k$. The dimensionless number $\text{Bo}$ used in
\cite{MihalasMihalas1984} is related to our dimensionless number $\Prat$
and $\Crat$ by equation (\ref{BoPC}) and another dimensionless number $r$ 
is determined by the following equation
\begin{eqnarray} 
r=\frac{(\gamma-1)\Prat}{4\gamma}.
\label{rPC}
\end{eqnarray}
The Boltzmann number $\text{Bo}$ is a
dimensionless number which measures the importance of energy flux carried
by the radiation field.  When $\text{Bo}\rightarrow \infty$, there is no
energy exchange between radiation field and material. Equation \ref{BoPC}
shows that the combination of our dimensionless numbers $1/(\Prat\Crat)$
has the same physical meaning.   $\Prat$, or equivalent $r$, is a dimensionless number which
measures the importance of momentum exchange between the radiation field
and material, as discussed in section \ref{sec:Linearwave}.

The dispersion relation given above must be solved using numerical root
finding methods.  For convenience, we give
four numerical examples of eigenmodes that span the parameter space defined
by $\Prat$ and $\sigma$ which can be used by others to test their own codes
In all cases, $\Crat=10^4$ and $k=2\pi$.
\begin{table}
\caption{Examples of solutions to the dispersion relation (\ref{dispersionEqn0})}
\begin{center}
\begin{tabular}{c|cccc}
\hline\hline
$\Prat$		& 0.01	&0.01	\\
$\sigma$		&0.01	&10		\\
$\delta\rho$	&$10^{-3}$	&$10^{-3}$	\\
$\delta v	$	&$1.27177\times10^{-3}+8.15409\times10^{-5} i$	&$1.00012\times10^{-3}+6.91295\times10^{-6} i$	\\
$\delta P 	$	&$1.61075\times10^{-3}+2.07402\times10^{-4} i$	&$1.00019\times10^{-3}+1.36663\times10^{-5} i$	\\
$\delta E_r$	&$1.79137\times10^{-8}+8.56498\times10^{-9} i$	&$6.64619\times10^{-7}+4.83820\times10^{-5}i$	\\
$\delta F_r$	&$-1.32035\times10^{-6}+3.88814\times10^{-6} i$	&$-9.99976\times10^{-6}+1.40748\times10^{-7}i$	\\
$\omega$		&$7.99077+0.512336 i$	&$6.28393+4.34354\times10^{-2} i$	\\
\hline
$\Prat$		&	$100$	&	$100$	\\
$\sigma$		&	$0.01$	&	$10$		\\
$\delta\rho$	&	$10^{-3}$	&	$10^{-3}$	\\
$\delta v	$	&$1.00000\times10^{-3}+8.93099\times10^{-8} i$	&$9.99947\times10^{-4}+1.07703\times10^{-5}i$	\\
$\delta P 	$	&$1.00000\times10^{-3}+1.57240\times10^{-7} i$	&$9.99998\times10^{-4}+1.60499\times10^{-7}i$	\\
$\delta E_r$	&$-3.55851\times10^{-13}+6.41394\times10^{-10}i$&$-6.60096\times10^{-9}+6.41367\times10^{-7}i$	\\
$\delta F_r$	&$-1.00000\times10^{-9}+2.10690\times10^{-13}i$		&$-1.00130\times10^{-9}+5.35966\times10^{-11}i$			\\
$\omega$		&$6.28319+5.61151\times10^{-4}i$	&$6.28285+6.76716\times10^{-2}i$		\\
\hline
$\Prat$		&	$100$	&	$1$	\\
$\sigma$		&	$0.1$	&	$10$		\\
$\delta\rho$	&	$10^{-3}$	&	$10^{-3}$	\\
$\delta v	$	&$1.00000\times10^{-3}+1.15555\times10^{-7} i$	&$1.00000\times10^{-3}+3.32795\times10^{-7}i$	\\
$\delta P 	$	&$1.00000\times10^{-3}+1.73114\times10^{-8} i$	&$1.00000\times10^{-3}+2.94229\times10^{-7}i$	\\
$\delta E_r$	&$-1.0093\times10^{-12}+6.41394\times10^{-9}i$&$3.419000\times10^{-10}+1.11408\times10^{-6}i$	\\
$\delta F_r$	&$-1.00000\times10^{-9}+5.51807\times10^{-13}i$		&$-1.00000\times10^{-7}+1.22263\times10^{-10}i$			\\
$\omega$		&$6.28319+7.26052\times10^{-4}i$	&$6.28319+2.09101\times10^{-3}i$		\\
\hline\hline
\end{tabular}
\end{center}
\label{EigenModes}
\end{table}

\section{Dispersion relation for linear waves in radiation MHD}
\label{RadMHDwave}

In this appendix, we derive the dispersion relation for linear
waves in radiation MHD used for the tests in section
\ref{sec:Linearwave}.  As in Appendix \ref{Radhydrowave}, we assume
a uniform, static background state, with a uniform
magnetic field $\bb_0=(B_{x,0}, B_{y,0}, B_{z,0})$.  We assume the
the wave propagation direction is along $x$ axis but the background
magnetic field $\bb_0$ can be along any direction.
In general, $v_x$, $v_y$ and $v_z$ are all non-zero.
The analysis proceeds as outlined in section \ref{Radhydrowave}
and will not be repeated here.
In terms of the
perturbed primitive variables $\delta \bW=(\delta\rho,\delta v_x,\delta
v_y,\delta v_z,\delta P,\delta E_r,\delta F_{r,x},\delta F_{r,y},
\delta F_{r,z}, \delta B_y, \delta B_z)^T$, the linearized equations can
be written as $\mathbf{A}\delta \bW=0$, where the coefficient matrix
$\mathbf{A}$ is given in equation \ref{matrix_MHDwave}.  The dispersion
relation for radiation MHD waves is given by $det(\mathbf{A})=0$.  This
gives an eleventh order polynomial for $\omega$ and a tenth order
polynomial for $k$.  The solution can be simplified by assuming 2D solutions,
in which $\bb_0$, the wave vector, 
and all the perturbed quantities are on the same plane.  In this case, the
dispersion relation will be reduced to an eighth order polynomial for
$\omega$ and a seventh order polynomial for $k$.  The dispersion relation
for radiation MHD waves including gravity in the short wavelength
limit is discussed in detail by \cite{BlaesSocrates2003}.

\begin{sidewaystable}
\centering
\begin{equation}
\scriptsize
\left[
\begin{array}{cccccccccccc}
\omega 	&   -k\rho_0	&	0	&0	&0	&0	&0	&0	&0	&0	&0	   \\
0		&\mathrm{i}\omega\rho_0+\frac{4}{3}\mathbb{P}\sigma\frac{E_{r,0}}{\mathbb{C}} &0	&0	& -\mathrm{i}k	&0	&-\mathbb{P}\sigma	&0	&0	&-\mathrm{i}kB_{y,0}	&-\mathrm{i}kB_{z,0}	\\
0	&0	& \mathrm{i}\omega\rho_0+\frac{4}{3}\mathbb{P}\sigma\frac{E_{r,0}}{\mathbb{C}}	&0	&0	&0	&0	&-\mathbb{P}\sigma	&0	&\mathrm{i}kB_{x,0}	&0	\\
0	&0	&0	& \mathrm{i}\omega\rho_0+\frac{4}{3}\mathbb{P}\sigma\frac{E_{r,0}}{\mathbb{C}}	&0	&0	&0	&0	&-\mathbb{P}\sigma	&0	&\mathrm{i}kB_{x,0}	\\
\frac{4T_0^4}{\rho_0}\mathbb{P}\mathbb{C}\sigma	&\mathrm{i}k\left(\frac{\gamma}{\gamma-1}P_0+B_0^2-B_{x,0}^2\right)&-\mathrm{i}kB_{x,0}B_{y,0}&-\mathrm{i}kB_{x,0}B_{z,0}&-\left(\frac{4T_0^4}{P_0}\mathbb{P}\mathbb{C}\sigma+\frac{\mathrm{i}\omega}{\gamma-1}\right)&\mathbb{P}\mathbb{C}\sigma&0	&0	&0	&-\mathrm{i}\omega B_{y,0}	&-\mathrm{i}\omega B_{z,0}	\\
\frac{4T_0^4}{\rho_0}\mathbb{C}\sigma	&0	&0	&0	&-\frac{4T_0^4}{P_0}\mathbb{C}\sigma		&\mathrm{i}\omega+\mathbb{C}\sigma	&-\mathrm{i}k\mathbb{C}	&0	&0	&0	&	0	\\
0	&\frac{4}{3}\sigma E_{r,0}	&0	&0	&0	&\mathrm{i}k\frac{\mathbb{C}}{3}	&-(\mathrm{i}\omega+\mathbb{C}\sigma)	&0	&0	&0	&0	\\
0	&0	&\frac{4}{3}\sigma E_{r,0}	&0	&0	&0	&0	&-(\mathrm{i}\omega+\mathbb{C}\sigma)	&0	&0	&0	\\
0	&0	&	0&\frac{4}{3}\sigma E_{r,0}	&0	&0	&0	&0	&-(\mathrm{i}\omega+\mathbb{C}\sigma)	&0	&0	\\
0	&kB_{y,0}	&-kB_{x,0}&0	&0	&0	&0	&0	&0	&-\omega&	0\\
0	&kB_{z,0}	&0	&-kB_{x,0}&0	&0	&0	&0	&0	&0	&-\omega\\
\end{array}
\right].
\label{matrix_MHDwave}
\end{equation}
\end{sidewaystable}

For convenience, we give
four numerical examples of eigenmodes for both slow and fast magnetosonic
modes that span the parameter space defined
by $\Prat$ and $\sigma$ which can be used by others to test their own codes.
In all cases, $\Crat=10^4$ and $k=2\pi$.
\begin{table}
\caption{Examples of solutions for radiation modified slow mode}
\begin{center}
\begin{tabular}{c|cccc}
\hline\hline
$\Prat$		& 0.01	&0.01	\\
$\sigma$		&0.01	&10		\\
$\delta\rho$	&$10^{-3}$	&$10^{-3}$	\\
$\delta v_x$	&$7.77601\times10^{-4}+5.56463\times10^{-5} i$	&$6.53191\times10^{-4}+2.46133\times10^{-6} i$	\\
$\delta v_y$	&$1.20178\times10^{-3}+1.84722\times10^{-4} i$	&$8.77920\times10^{-4}+5.58924\times10^{-6} i$	\\
$\delta P 	$	&$1.52579\times10^{-3}+2.97056\times10^{-4} i$	&$1.00009\times10^{-3}+8.92617\times10^{-6} i$	\\
$\delta B_y$	&$-7.15903\times10^{-4}-1.63064\times10^{-4} i$	&$-4.44181\times10^{-4}-4.50308\times10^{-6} i$	\\
$\delta E_r$	&$1.54496\times10^{-8}+1.02737\times10^{-8} i$	&$3.069869\times10^{-7}+3.16009\times10^{-5}i$	\\
$\delta F_{r,x}$	&$-1.89110\times10^{-6}+3.34728\times10^{-6} i$	&$-6.53138\times10^{-6}+6.48914\times10^{-8}i$	\\
$\delta F_{r,y}$	&$1.61623\times10^{-7}+1.67917\times10^{-8} i$	&$1.17056\times10^{-7}+7.40429\times10^{-10}i$	\\
$\omega$		&$4.88581+0.349636 i$	&$4.10401+1.54776\times10^{-2} i$	\\
\hline
$\Prat$		&	$100$	&	$100$	\\
$\sigma$		&	$0.01$	&	$10$		\\
$\delta\rho$	&	$10^{-3}$	&	$10^{-3}$	\\
$\delta v_x$	&$6.53158\times10^{-4}+3.18349\times10^{-8} i$	&$6.53158\times10^{-4}+3.83780\times10^{-6}i$	\\
$\delta v_y$	&$8.77864\times10^{-4}+7.22107\times10^{-8} i$	&$8.77819\times10^{-4}+8.70693\times10^{-6} i$	\\
$\delta P 	$	&$1.00000\times10^{-3}+1.02703\times10^{-7} i$	&$9.99999\times10^{-4}+1.04836\times10^{-7}i$	\\
$\delta B_y$	&$-4.44142\times10^{-4}-5.81571\times10^{-8} i$	&$-4.44093\times10^{-4}-7.01467\times10^{-6} i$	\\
$\delta E_r$	&$-1.47792\times10^{-13}+4.18932\times10^{-10}i$&$-2.33054\times10^{-9}+4.18933\times10^{-7}i$	\\
$\delta F_{r,x}$	&$-6.53158\times10^{-10}+9.61494\times10^{-14} i$&$-6.53472\times10^{-10}+2.36270\times10^{-11}i$	\\
$\delta F_{r,y}$	&$1.16852\times10^{-7}-4.78590\times10^{-9} i$	&$1.17043\times10^{-7}+1.15612\times10^{-9}i$	\\
$\omega$		&$4.10391+2.000251\times10^{-4}i$	&$4.10391+2.41136\times10^{-2}i$		\\
\hline
$\Prat$		&	$100$	&	$1$	\\
$\sigma$		&	$0.1$	&	$10$		\\
$\delta\rho$	&	$10^{-3}$	&	$10^{-3}$	\\
$\delta v_x$	&$6.53158\times10^{-4}+4.11763\times10^{-8} i$	&$6.53158\times10^{-4}+1.18584\times10^{-7}i$	\\
$\delta v_y$	&$8.77864\times10^{-4}+9.34194\times10^{-8} i$	&$8.77865\times10^{-4}+2.69043\times10^{-7} i$	\\
$\delta P 	$	&$1.00000\times10^{-3}+1.13071\times10^{-8} i$	&$1.00000\times10^{-3}+1.92178\times10^{-7}i$	\\
$\delta B_y$	&$-4.44142\times10^{-4}-7.52612\times10^{-8} i$	&$-4.44142\times10^{-4}-2.16754\times10^{-7} i$	\\
$\delta E_r$	&$-3.78493\times10^{-13}+4.18932\times10^{-9}i$&$1.71921\times10^{-10}+7.27673\times10^{-7}i$	\\
$\delta F_{r,x}$	&$-6.53158\times10^{-10}+2.43530\times10^{-13} i$	&$-6.53158\times10^{-8}+5.44986\times10^{-11}i$	\\
$\delta F_{r,y}$	&$1.17047\times10^{-7}-4.67894\times10^{-10} i$	&$1.17049\times10^{-7}+3.10689\times10^{-11}i$	\\
$\omega$		&$4.10391+2.58718\times10^{-4}i$	&$4.10391+7.45083\times10^{-4}i$		\\
\hline\hline
\end{tabular}
\end{center}
\end{table}

\begin{table}
\caption{Examples of solutions for radiation modified fast mode.}
\begin{center}
\begin{tabular}{c|cccc}
\hline\hline
$\Prat$		& 0.01	&0.01	\\
$\sigma$		&0.01	&10		\\
$\delta\rho$	&$10^{-3}$	&$10^{-3}$	\\
$\delta v_x$	&$2.08441\times10^{-3}+2.26355\times10^{-5} i$	&$1.97671\times10^{-3}+4.44472\times10^{-6}i$	\\
$\delta v_y$	&$-1.29633\times10^{-3}+3.15959\times10^{-5} i$	&$-1.47025\times10^{-3}+8.22391\times10^{-6} i$	\\
$\delta P 	$	&$1.64144\times10^{-3}+1.30879\times10^{-4} i$	&$1.00104\times10^{-3}+2.69749\times10^{-5} i$	\\
$\delta B_y$	&$2.09358\times10^{-3}-2.82848\times10^{-5} i$	&$2.25121\times10^{-3}-7.53016\times10^{-6} i$	\\
$\delta E_r$	&$1.89341\times10^{-8}+7.85298\times10^{-9} i$	&$3.68258\times10^{-6}+9.54978\times10^{-5}i$	\\
$\delta F_{r,x}$	&$-8.33183\times10^{-7}+4.08350\times10^{-6} i$	&$-1.97374\times10^{-5}+7.74322\times10^{-7}i$	\\
$\delta F_{r,y}$	&$-1.69620\times10^{-7}+2.64650\times10^{-8} i$	&$-1.96034\times10^{-7}+1.12087\times10^{-9}i$	\\
$\omega$		&$13.0967+0.142223 i$	&$12.4200+2.79270\times10^{-2} i$	\\
\hline
$\Prat$		&	$100$	&	$100$	\\
$\sigma$		&	$0.01$	&	$10$		\\
$\delta\rho$	&	$10^{-3}$	&	$10^{-3}$	\\
$\delta v_x$	&$1.97654\times10^{-3}+5.75439\times10^{-8} i$	&$1.97651\times10^{-3}+6.93247\times10^{-6}i$	\\
$\delta v_y$	&$-1.47061\times10^{-3}+1.06106\times10^{-7} i$	&$-1.47055\times10^{-3}+1.28333\times10^{-5} i$	\\
$\delta P 	$	&$1.00000\times10^{-3}+3.10791\times10^{-7} i$	&$9.99999\times10^{-4}+3.17242\times10^{-7}i$	\\
$\delta B_y$	&$2.25153\times10^{-3}-9.72688\times10^{-8} i$	&$2.25147\times10^{-3}-1.17512\times10^{-5} i$	\\
$\delta E_r$	&$-1.20333\times10^{-12}+1.26774\times10^{-9}i$&$-3.24677\times10^{-9}+1.26773\times10^{-6}i$	\\
$\delta F_{r,x}$	&$-1.97654\times10^{-9}+1.11447\times10^{-12} i$	&$-1.97803\times10^{-9}+2.44573\times10^{-10}i$	\\
$\delta F_{r,y}$	&$-1.93101\times10^{-7}+2.39955\times10^{-8} i$	&$-1.96073\times10^{-7}+1.73546\times10^{-9}i$	\\
$\omega$		&$12.4190+3.61559\times10^{-4}i$	&$12.4188+4.3558\times10^{-2}i$		\\
\hline
$\Prat$		&	$100$	&	$1$	\\
$\sigma$		&	$0.1$	&	$10$		\\
$\delta\rho$	&	$10^{-3}$	&	$10^{-3}$	\\
$\delta v_x$	&$1.97654\times10^{-3}+7.43855\times10^{-8} i$	&$1.97654\times10^{-3}+2.14211\times10^{-7}i$	\\
$\delta v_y$	&$-1.47061\times10^{-3}+1.37659\times10^{-7} i$	&$-1.47061\times10^{-3}+3.96547\times10^{-7} i$	\\
$\delta P 	$	&$1.00000\times10^{-3}+3.42168\times10^{-8} i$	&$1.00000\times10^{-3}+5.81556\times10^{-7}i$	\\
$\delta B_y$	&$2.25153\times10^{-3}-1.26062\times10^{-7} i$	&$2.25153\times10^{-3}-3.63108\times10^{-7} i$	\\
$\delta E_r$	&$-1.52464\times10^{-12}+1.26774\times10^{-8}i$&$2.54552\times10^{-9}+2.20203\times10^{-6} i$	\\
$\delta F_{r,x}$	&$-1.97654\times10^{-9}+2.53281\times10^{-12} i$	&$-1.97654\times10^{-7}+5.86240\times10^{-10}i$	\\
$\delta F_{r,y}$	&$-1.96050\times10^{-7}+2.45310\times10^{-9} i$	&$-1.96081\times10^{-7}+7.72241\times10^{-11}i$	\\
$\omega$		&$12.4190+4.67378\times10^{-4}i$	&$12.419+1.34593\times10^{-3}i$		\\
\hline\hline
\end{tabular}
\end{center}
\end{table}

\end{appendix}


 
\bibliographystyle{apj}
\bibliography{RMHD}

\end{document}